\documentclass[aps,prd,nofootinbib,superscriptaddress,showpacs,floatfix,a4paper]{revtex4-1}

\usepackage{amsfonts}
\usepackage{soul}
\usepackage{amsmath}
\usepackage{amssymb}
\usepackage{amsthm}
\usepackage{array}
\usepackage{booktabs}
\usepackage[usenames,dvipsnames]{xcolor}
\usepackage{comment} 
\usepackage{graphicx}
\usepackage{hyperref}
\usepackage{multirow}
\usepackage{slashed}
\usepackage{subfigure}
\usepackage{theorem}
\usepackage{textcomp}

\newcommand{\ttilde}{\raisebox{-3pt}{\textasciitilde{}}}

\allowdisplaybreaks[1]

\numberwithin{equation}{section}

\newcommand{\be}{\begin{equation}}
\newcommand{\ee}{\end{equation}}
\newcommand{\beq}{\begin{equation}}
\newcommand{\eeq}{\end{equation}}
\newcommand{\bea}{\begin{eqnarray}}
\newcommand{\eea}{\end{qnarray}}

\newcommand{\rr}[1]{\mathrm{#1}}

\def\<{\left\langle}
\def\>{\right\rangle}

\definecolor{brickred}{RGB}{150,25,14}

\begin{document}

\title{
Discovering $E_6$ SUSY models in gluino cascade decays at the LHC
}
	
\author{Alexander Belyaev}
\email[E-mail: ]{a.belyaev@soton.ac.uk}
\affiliation{School of Physics \& Astronomy, University of Southampton,
        Highfield, Southampton SO17 1BJ, UK}
\affiliation{Particle Physics Department, Rutherford Appleton Laboratory, 
       Chilton, Didcot, Oxon OX11 0QX, UK}
       
       \author{Jonathan~P.~Hall}
\email[E-mail: ]{halljp@indiana.edu}
\affiliation{Physics Department, Indiana University,
        Bloomington, IN 47405, USA}
	
\author{Stephen~F.~King}

\email[E-mail: ]{king@soton.ac.uk}
\affiliation{School of Physics \& Astronomy, University of Southampton, 
        Highfield, Southampton SO17 1BJ, UK}

\author{Patrik Svantesson}
\email[E-mail: ]{p.svantesson@soton.ac.uk}
\affiliation{School of Physics \& Astronomy, University of Southampton,
        Highfield, Southampton SO17 1BJ, UK}
\affiliation{Particle Physics Department, Rutherford Appleton Laboratory, 
       Chilton, Didcot, Oxon OX11 0QX, UK}

\begin{abstract}
We point out that the extra neutralinos and charginos generically appearing in a
large class of $E_6$ inspired models lead to distinctive signatures from gluino
cascade decays in comparison to those from the Minimal Supersymmetric Standard Model (MSSM). 
These signatures involve longer decay chains, more visible transverse energy,
higher multiplicities of jets and leptons, and less missing transverse energy than in the MSSM. 
These features make the gluino harder to discover for certain types of conventional analyses,
for example in the 6 jet channel we show that a 1~TeV MSSM gluino may resemble an 800~GeV $E_6$ gluino.
However, the enriched lepton multiplicity provides enhanced 3- and 4-lepton signatures,
which are much more suppressed in the case of the MSSM.
As an example we analyse various scenarios in the E$_6$SSM, demonstrating the
utility of a \texttt{CalcHEP} model that we have now made publicly available on HEPMDB (High Energy Physics Models DataBase). After extensive scans over the parameter space, we focus on representative
benchmark points. We perform detailed Monte Carlo analyses,
concentrating on the 3-lepton channel at the 7, 8, and 14 TeV Large Hadron Collider (LHC), and demonstrate that 
$E_6$ inspired models are clearly distinguishable from the MSSM in gluino cascade decays.
We emphasise to the LHC experimental groups that the distinctive features present in gluino cascade decays,
such as those discussed here, not only represents a unique footprint of a particular model but also may provide the key to an earlier discovery of supersymmetry.
\end{abstract}


\maketitle

\tableofcontents

\setcounter{footnote}{0}

\section{Introduction}

The general idea of softly broken supersymmetry (SUSY) provides a very
attractive framework
for physics models beyond the standard model (BSM), in which the hierarchy
problem is 
solved and the unification of gauge couplings can be realised
\cite{Chung:2003fi}. 
The simplest such extension, the Minimal Supersymmetric Standard Model (MSSM),
has been subjected to particularly close scrutiny at the CERN Large Hadron
Collider (LHC). 
For example, searches at ATLAS \cite{Aad:2011ib} and CMS
\cite{Chatrchyan:2011zy,Chatrchyan:2011qs}, under certain assumptions, constrain
the MSSM gluino mass to be greater than about 500--900 GeV. 
This mass limit range reflects the fact that the
precise limit depends crucially on the assumptions one makes about the rest of
the SUSY spectrum, in particular the mass spectrum of the squarks, charginos,
and neutralinos. As we emphasise in this paper, mass limits and the LHC
discovery potential
also depend on the particular SUSY  model under consideration and the quoted
mass limits strictly only apply to the constrained MSSM or simplified models of
SUSY\cite{Okawa2011} with very few particles. In other, well motivated,
non-minimal SUSY extensions of the SM, such as those discussed below, the
enriched particle content alters the SUSY discovery potential at the LHC and
this should be taken into account in the phenomenological and experimental
studies.

Despite its many attractive features, the MSSM 
suffers from the $\mu$ problem. The superpotential of the MSSM contains 
the bilinear term $\mu H_d H_u$, where $H_{d}$ and $H_{u}$ are the Higgs 
doublet superfields. In order to get the correct pattern of electroweak (EW)
symmetry breaking the parameter $\mu$ is required to be in the TeV region. 
At the same time, the incorporation of the MSSM into 
grand unified theories (GUTs) or string theory implies that $\mu$ could be of
the order of the
GUT or Planck scales, or possibly zero.
None of these possibilities is phenomenologically and theoretically acceptable.
The Next-to-Minimal Supersymmetric Standard Model (NMSSM)
\cite{Ellwanger:2009dp}
attempts to address this problem by postulating an additional gauge singlet
superfield $S$
with the interaction $\lambda S H_d H_u$ in the superpotential together with an
$S^3$ interaction in order
to break an accidental global $U(1)$ symmetry to a discrete $Z_3$ symmetry. 
At low energies ($\sim \mbox{TeV}$) the scalar component of $S$ acquires a
non-zero vacuum expectation 
value (VEV) $s$, giving an effective $\mu$ term.
However the resulting $Z_3$ discrete symmetry is broken at the same time,
leading to potentially dangerous
cosmological domain walls \cite{Ellwanger:2009dp}.

A solution to the $\mu$ problem which does not suffer from the problems of the
NMSSM arises within $E_6$ inspired SUSY models.
At high energies the $E_6$ gauge symmetry arising from GUTs or string theory in
extra dimensions
can be broken to the rank-5 subgroup
$SU(3)_c\times SU(2)_L\times U(1)_Y\times U(1)'$, where in general
\begin{equation}
U(1)'=U(1)_{\chi} \cos\theta+U(1)_{\psi} \sin\theta
\label{essm0}
\end{equation}
and the two anomaly-free $U(1)_{\psi}$ and $U(1)_{\chi}$ symmetries originate
from 
the breakings $E_6\to$ $SO(10)\times U(1)_{\psi}$, then $SO(10)\to$
$SU(5)\times$ $U(1)_{\chi}$. 
The extra $U(1)'$ gauge symmetry forbids $S^3$ interactions
since the SM-singlet superfield $S$ is now charged under $U(1)'$.
In addition the bilinear $\mu$ 
term is also forbidden if $\theta\ne 0$ or $\pi$, while the interaction $\lambda
S H_d H_u$ is allowed in the superpotential.

At low energies ($\sim \mbox{TeV}$) the scalar component of $S$ acquires a
non-zero VEV $s$ breaking $U(1)'$
and giving 
rise to a massive $Z'$ gauge boson together with an effective $\mu$ term.  
Within the class of rank-5, $E_6$ inspired SUSY models with an extra $U(1)'$
gauge 
symmetry, there is a unique choice of the extra Abelian gauge group that allows 
right-handed neutrinos to be uncharged. This is the $U(1)_{N}$ gauge symmetry
given by
$\theta=\arctan\sqrt{15}$ \cite{King:2005jy,King:2005my}. 
In this particular case, called the Exceptional Supersymmetric Standard Model 
(E$_6$SSM), based on the $SU(3)_c\times SU(2)_L\times U(1)_Y \times U(1)_N$
gauge group, the right-handed neutrinos may be superheavy, 
allowing a high scale see-saw mechanism in the lepton sector
\cite{King:2005jy,King:2005my}.

An elegant feature of the $E_6$ inspired models with an extra $Z'$ gauge 
boson at the TeV scale is that the conditions of anomaly cancellation 
may be satisfied by matter content filling out complete $27$ representations of
$E_6$ surviving to the TeV scale. One can therefore have three $27$
representations, with
each containing a generation of SM matter and fields with the quantum numbers
of Higgs doublets and SM-singlets.
Thus such models predict, in addition to three SUSY families of quarks and
leptons, also three SUSY families of Higgs
doublets of type $H_u$, three SUSY families of Higgs doublets of type $H_d$,
three SUSY SM singlets $S_i$,
and three SUSY families of charge $\mp 1/3$ colour triplet and anti-triplet
states $D$ and $\overline{D}$, which can get large mass terms due to the (third)
singlet VEV
$\langle S_3 \rangle=s/\sqrt{2}$ which is also responsible for the $\mu$ term.

In this paper we study gluino production and decay within the E$_6$SSM,
expected to be the main discovery channel for this model.
In fact the analysis also applies to a larger class of $E_6$ models
in which the matter content of three $27$ representations of $E_6$ survives to
the TeV scale. We study the E$_6$SSM as a concrete example and at the same time
demonstrate the use of the \texttt{E6SSM-12.02 CalcHEP} model that we have now
made publicly available on HEPMDB. After extensive scans over the parameter
space we create a number of representative benchmarks for different
E$_6$SSM scenarios, considering $Z'$ and Higgs boson physics,
the LSP dark matter relic density and direct detection
cross-section, and the perturbativity of dimensionless couplings.
We then analyse representative MSSM and E$_6$SSM benchmarks consistent with the
particle recently discovered at the LHC
\cite{ATLAS-CONF-2012-093,CMS-PAS-HIG-12-020} being the SM-like lightest Higgs boson.

The analysis is motivated by the fact that the gluinos are the expected to be
the lightest strongly interacting particles
in $E_6$ models \cite{Athron:2009ue}, so one should expect them to have the
largest production rate.
In particular, we are interested in gluino cascade decays
which not only provide observable signatures, but are also important for
distinguishing the $E_6$ inspired models from the MSSM. 
As we shall see, there are important differences between the two cases, which 
can affect the respective search strategies.
The differences arise because of the extra Higgsino and singlino states
predicted to be part of the $E_6$ matter content, as above.
In particular, in the $E_6$ inspired models, relative to the MSSM,
there are two extra families of Higgsinos, $\tilde{H}^{\alpha}_u$ and
$\tilde{H}^{\alpha}_d$, together with two extra singlinos, $\tilde{S}^{\alpha}$,
where $\alpha = 1,2$~\footnote{Note that the first and second family of Higgs
doublet and singlet fields
${H}^{\alpha}_u$, ${H}^{\alpha}_d$, and ${S}^{\alpha}$ predicted by the E$_6$SSM
do not develop VEVs and are called ``inert''.}.
There is also a third singlino, $\tilde{S}^{3}$, similar to the NMSSM singlino,
which mixes with the bino$'$, both states having a large mass of order the
effective $\mu$ parameter. The remaining extra Higgsinos and singlinos may be
lighter than the gluino. Indeed, it is possible to show that at least 
two linear combinations of the states $\tilde{H}^{\alpha}_u$, 
$\tilde{H}^{\alpha}_d$, $\tilde{S}^{\alpha}$
must be lighter than or of order $M_Z/2$. If these states mix with 
the usual neutralinos of the MSSM or NMSSM then the lightest supersymmetric
particle (LSP) will inevitably be
one of these states, leading to longer decay chains. For example, in regions of
MSSM parameter space where the bino is the LSP
the gluino typically undergoes a cascade decay to the bino. In the E$_6$
inspired models
the bino will mix with the extra Higgsinos and singlinos and the predominantly
bino state will subsequently decay into some lighter state 
having a mass of order $M_Z/2$, thereby typically giving a longer gluino cascade
decay chain
and producing less missing energy due to the lighter mass of the
LSP~\footnote{The decay to bino is expected to happen before the subsequent
decay into a lighter state since these lighter states are expected to have small
mixing to the MSSM-like sector, for reasons explained in section II.}.
For simplicity we shall assume that the $D$ and $\overline{D}$ states,
as well as the NMSSM type singlino $\tilde{S}^{3}$, are all heavier than the
gluino and so are irrelevant for gluino cascade decays.
Similarly we shall also assume all squarks and sleptons and Higgs scalars and
pseudoscalars (with the exception of the SM-like Higgs boson) to be heavier than
the gluino. These assumptions are motivated by the parameter space of the
constrained $E_6$ inspired models
\cite{Athron:2009ue,Athron:2009bs,Athron:2011wu}.

The main result of our analysis is that in $E_6$ inspired models the gluino
decays into the LSP with longer chains which involve more jets and leptons and
less missing energy than in the MSSM. This happens because the would-be LSP
(e.g. an MSSM-like bino dominated neutralino) undergoes further decays to the
extra light neutralinos and charginos predicted by the $E_6$ inspired models. 
As a result, the characteristics of the signal, such as lepton and jet
multiplicity, missing transverse momentum, effective mass, etc., are altered in
the E$_6$SSM as compared to the MSSM case even after the matching of the gaugino
masses in both models. Therefore, 
the search strategies designed for the MSSM need to be modified for the E$_6$SSM
case,
while one should stress that the gluino mass limits for the MSSM are not
applicable to the E$_6$SSM gluinos.

The layout of the remainder of this paper is as follows: In section II we
summarise the relevant features of the 
E$_6$SSM, its theoretical constraints and its experimental constraints not
relating to gluino detection. We also present the results of parameter scans and
discuss the viable parameter space, before going on to introduce a set of
benchmark points, including those which form the basis of our analysis. The
model implementation is discussed in section III (with a more in-depth
description of the publicly available code in Appendix B). We then present our
analysis of the different signatures and search prospects for different
strategies for each model in section IV. The conclusions are in section V.

\section{Model Setup and Parameter Space}

Before going on to consider the prospects
for the production and detection of gluinos, which are the main focus
of this paper, discussed in section IV, we must first determine the
limits on the E$_6$SSM from other experimental,
and cosmological, considerations.
We begin by giving an introduction to the E$_6$SSM (subsections \ref{sub2a} and
\ref{matrices}), explaining the important
features. We then discuss experimental
constraints relating to $Z'$ and Higgs boson physics and to exotic coloured
particles (subsection \ref{sub2ex}), before going on to
discuss dark matter considerations (subsection \ref{sub2dm}).
In light of these discussions we then show the results of some parameter
space scans (subsection \ref{sec:param}) and produce a set of benchmark points
(subsection \ref{sec:benchmarks}). These benchmarks example various
viable scenarios for the E$_6$SSM and we explain their features and issues.
Although the benchmarks presented look very different from each other
from many points of view, it turns out that they look very similar in terms
of their gluino decay signatures. In section IV we therefore
mostly show results
for two particular benchmarks that demonstrate the qualitative differences
between the MSSM and $E_6$ models for gluino searches.

\subsection{The E$_6$SSM}
\label{sub2a}
As discussed in the Introduction, the E$_6$SSM involves a unique 
choice for the extra Abelian gauge group that allows zero charges 
for right-handed neutrinos, namely $U(1)_{N}$.
To ensure anomaly cancellation the particle content of the E$_6$SSM at the TeV
scale is 
extended to include three complete fundamental $27$ representations of E$_6$,
apart from the three right-handed
neutrinos which are singlets and do not contribute to anomalies and so may be
much heavier.
The $27$ representations of $E_6$ decompose under the $SU(5)\times U(1)_{N}$
subgroup of E$_6$
as follows:
\begin{equation}
27_i\to (10,1 )_i+(5^{*},2)_i
+(5^{*},-3)_i +(5,-2)_i
+(1,5)_i+(1,0)_i\,.
\label{essm1}
\end{equation}
The first and second quantities in brackets are the $SU(5)$ representation and
extra $U(1)_{N}$ charge (where a GUT normalisation factor of
$1/\sqrt{40}$ is omitted \cite{King:2005jy,King:2005my}) respectively,
while $i$ is a family index that runs from 1 to 3.
An ordinary SM family, which contains the doublets of left-handed quarks $Q_i$
and
leptons $L_i$, right-handed up- and down-quarks ($u^c_i$ and $d^c_i$), and
right-handed charged leptons is assigned to
$(10,1)_i$ + $(5^{*},2)_i$.
Right-handed neutrinos $N^c_i$ should be associated with the last term in
Eq.~(\ref{essm1}),
$\left(1,0\right)_i$. The next-to-last term, $(1,5)_i$, 
represents SM-singlet fields $S_i$, which carry non-zero $U(1)_{N}$ charges and
therefore 
survive down to the EW scale. The pair of $SU(2)_L$-doublets ($H^d_{i}$ and
$H^u_{i}$) that 
are contained in $(5^{*},-3)_i$ and 
$(5,-2)_i$ have the quantum numbers of Higgs doublets.
They form either Higgs or inert Higgs
$SU(2)_L$ multiplets. Other components of these fundamental
$SU(5)$ multiplets form colour triplet and anti-triplet of exotic quarks
$D_i$ and $\overline{D}_i$
with electric charges $-1/3$ and $+1/3$ respectively. These exotic quark states
carry a 
$B-L$ charge $\mp2/3$, twice that of ordinary quarks.
In addition to the complete $27_i$ multiplets the low energy matter content of
the E$_6$SSM
can be supplemented by an $SU(2)_L$ doublet $L_4$ and anti-doublet
$\overline{L}_4$
from extra $27'$ and $\overline{27'}$ representations to preserve gauge coupling
unification \cite{King:2007uj}.
These states will typically be much heavier than the gluino and so will play no
role in the present analysis,
although they may play a role in leptogenesis \cite{King:2008qb}.

As in the MSSM the gauge symmetry in the E$_6$SSM does not forbid lepton and
baryon number 
violating operators that result in rapid proton decay, although the
situation is somewhat different.
The renormalisable superpotential of the E$_6$SSM automatically preserves
$R$-parity
because of the enlarged gauge symmetry,
however, although the $B-L$ violating operators of the $R$-parity violating MSSM
are forbidden by the gauge symmetry,
there are new $B$ and $L$ violating interactions involving the new exotic particles.
In the E$_6$SSM these terms are removed by imposing an exact $Z_2$ symmetry on
the superpotential.
There are two options for this symmetry under which the exotic quarks are either
interpreted as diquarks or leptoquarks, leading to $B$ and $L$ conservation.
Furthermore, the extra particles present in E$_6$
inspired SUSY models give rise to new Yukawa interactions that
in general induce unacceptably 
large non-diagonal flavour transitions. To suppress these effects in the
E$_6$SSM an 
additional approximate $Z^{H}_2$ symmetry is imposed. Under this symmetry all
superfields except one 
pair of $H_d^{i}$ and $H_u^{i}$ (say $H_d\equiv H_d^{3}$ and $H_u\equiv
H_u^{3}$) and one 
SM-type singlet field ($S\equiv S^3$) are odd. Ignoring $L_4$ and
$\overline{L}_4$,
the $Z^{H}_2$ symmetry reduces the structure 
of the Yukawa interactions to
\begin{eqnarray}
W_{\rm E_6SSM}&\simeq &  \lambda \hat{S} (\hat{H}_u \hat{H}_d)+
\lambda_{\alpha\beta} \hat{S} (\hat{H}_d^{\alpha} \hat{H}_u^{\beta})
+f_{u\alpha\beta} \hat{S}^{\alpha} (\hat{H}_d^{\beta}\hat{H}_u)
+f_{d\alpha\beta} \hat{S}^{\alpha} (\hat{H}_d \hat{H}_u^{\beta})
\nonumber\\[2mm]
&+&
h^U_{ij}(\hat{H}_{u} \hat{Q}_i)\hat{u}^c_{j} + h^D_{ij}(\hat{H}_{d}
\hat{Q}_i)\hat{d}^c_j
+ h^E_{ij}(\hat{H}_{d} \hat{L}_i)\hat{e}^c_{j}+ h_{ij}^N(\hat{H}_{u}
\hat{L}_i)\hat{N}_j^c\nonumber\\[2mm]
&+& \dfrac{1}{2}M_{ij}\hat{N}^c_i\hat{N}^c_j +\kappa_{i} \hat{S}
(\hat{D}_i\hat{\overline{D}}_i)\,,
\label{essm2}
\end{eqnarray}
where hats denote superfields, $\alpha,\beta=1,2$ and $i,j=1,2,3$.
The inert Higgs doublets and SM-singlets have suppressed couplings to matter,
whereas the third generation $SU(2)_L$ doublets $\hat{H}_u$ and 
$\hat{H}_d$ and SM-type singlet field $\hat{S}$, that are even under the
$Z^{H}_2$ 
symmetry, play the role of Higgs fields, radiatovely acquiring VEVs.
At the physical vacuum their scalar components develop VEVs
\begin{equation}
\langle H_d\rangle =\displaystyle\frac{1}{\sqrt{2}}\left(
\begin{array}{c}
v_1\\ 0
\end{array}
\right) , \qquad
\langle H_u\rangle =\displaystyle\frac{1}{\sqrt{2}}\left(
\begin{array}{c}
0\\ v_2
\end{array}
\right) ,\qquad
\langle S\rangle =\displaystyle\frac{s}{\sqrt{2}},
\label{essm21}
\end{equation}
generating the masses of the quarks and leptons. 
Instead of $v_1$ and $v_2$ it is more 
convenient to use $\tan\beta=v_2/v_1$ and
$v=\sqrt{v_1^2+v_2^2}=246\,\mbox{GeV}$.
The VEV of the SM-singlet field, $s$, breaks the extra $U(1)_N$ symmetry ,
generating exotic fermion masses and also inducing a mass for the $Z'$ boson.
Therefore the singlet field $S$ must acquire a large VEV in order to avoid
conflict with 
direct particle searches at present and past accelerators.

We shall assume that the $\kappa_i$ are sufficiently large that the exotic $D_i,
\overline{D}_i$
states are much heavier than the gluino and so will play no role in gluino
decays.
Similarly the right-handed neutrinos $N_i$ will be neglected since they are
assumed to be very heavy.

In this paper we shall be mainly concerned with the first line of
Eq.~(\ref{essm2}), namely the couplings
of the form $\hat{S}^i\hat{H}_d^j\hat{H}_u^k$, where $i,j,k=1,2,3$ label the
three families of Higgs doublet
and singlet superfields predicted in the E$_6$SSM. In particular we shall be
concerned with the resulting
chargino and neutralino mass terms coming from such
couplings involving one third-family scalar component and two fermion
components, i.e.
${S}\tilde{H}_d^j\tilde{H}_u^k$, $\tilde{S}^i{H}_d\tilde{H}_u^k$, and
$\tilde{S}^i\tilde{H}_d^j{H}_u$.
In the E$_6$SSM the presence of the extra
Higgsinos and singlinos $\tilde{H}^{\alpha}_u$, 
$\tilde{H}^{\alpha}_d$, and $\tilde{S}^{\alpha}$
means that the chargino and neutralino mass matrices are extended, as
discussed in the following subsection.

\subsection{Neutralino and chargino mass matrices}
\label{matrices}

In the MSSM \cite{Chung:2003fi} there are four neutralino interaction states,
the
neutral wino, the bino, and the two neutral Higgsinos. In the USSM
\cite{Kalinowski:2008iq}, a model similar to the NMSSM,
but where the $U(1)$ introduced by adding the singlet field $S$ is gauged
instead of reduced to a $Z_3$, two
extra states are added, the singlino and the bino$'$. In
the conventional USSM interaction basis we write the six neutralinos
as the column vector
\begin{equation}
\tilde{\chi}_\rr{int}^0 = (\begin{array}{cccc|cc} \tilde{B} & \tilde{W}^3 &
\tilde{H}_d^0 &
\tilde{H}_u^0 & \tilde{S} & \tilde{B}^\prime \end{array})^\rr{T}.
\end{equation}
Neglecting bino-bino$'$ mixing (assumed to be small as justified in
Ref.~\cite{Kalinowski:2008iq})
the USSM neutralino mass matrix in this basis becomes
\begin{equation}
M^n_\rr{USSM} = \left( \begin{array}{cccc|cc} M_1 & 0 & -m_Zs_Wc_\beta &
m_Zs_Ws_\beta & 0 & 0\\
0 & M_2 & m_Zc_Wc_\beta & -m_Zc_Ws_\beta & 0 & 0\\
-m_Zs_Wc_\beta & m_Zc_Wc_\beta & 0 & -\mu & -\mu_ss_\beta & g_1^\prime vc_\beta
Q_d^N \\
m_Zs_Ws_\beta & -m_Zc_Ws_\beta & -\mu & 0 & -\mu_sc_\beta &
g_1^\prime vs_\beta Q_u^N \\ \hline
0 & 0 & -\mu_ss_\beta & -\mu_sc_\beta & 0 & g_1^\prime sQ_s^N \\
0 & 0 & g_1^\prime vc_\beta Q_d^N & g_1^\prime vs_\beta
Q_u^N & g_1^\prime sQ_s^N & M_1^\prime \end{array}
\right),
\label{USSM}
\end{equation}
where $M_1$, $M_2$, and $M_1^\prime$ are the soft gaugino masses and $\mu_s =
\lambda v/\sqrt{2}$.  The variables $s_{\beta(W)}$ and $c_{\beta(W)}$ stand for $\sin$ and $\cos$ of
$\beta$(the Weinberg angle) and $Q^N_{(d,u,S)}$ are the $(H_d,H_u,S)$ $U(1)_N$
charges $(-3,-2,5)/\sqrt{40}$.

In the E$_6$SSM the neutralino sector is
extended to include the additional six neutral components of
$\tilde{H}^{\alpha}_u$, 
$\tilde{H}^{\alpha}_d$, and $\tilde{S}^{\alpha}$, where $\alpha =1,2$.
We then take the full list of twelve neutralino interaction states
into account in column vector
\begin{equation}
\tilde{\chi}_\rr{int}^0 = (\begin{array}{cccc|cc|ccc|ccc}
\tilde{B} & \tilde{W}^3 & \tilde{H}_d^0 & \tilde{H}_u^0 &
\tilde{S} & \tilde{B}^\prime & \tilde{H}_{d2}^0 & \tilde{H}_{u2}^0
& \tilde{S}_2 & \tilde{H}_{d1}^0 & \tilde{H}_{u1}^0 & \tilde{S}_1
\end{array})^\rr{T}.
\end{equation}
The first four states are the MSSM interaction states, the $\tilde{S}$
and $\tilde{B}^\prime$ are the extra states added in the USSM, and
the final six states are the extra inert doublet Higgsinos and Higgs singlinos
that come with the full E$_6$SSM.
Under the assumption that only the third generation Higgs doublets
and singlet acquire VEVs the full Majorana mass matrix is then
\cite{Hall:2009aj}
\begin{equation}
M^n_{\rr{E}_6\rr{SSM}} = \left( \begin{array}{ccc} M^n_\rr{USSM} & B_2 & B_1\\
B_2^\rr{T} & A_{22} & A_{21}\\
B_1^\rr{T} & A_{21}^\rr{T} & A_{11}\end{array} \right),\label{nmm}
\end{equation}
where the sub-matrices involving only the inert interaction states are given by
\begin{equation}
A_{\alpha\beta} = A^{\rr{T}}_{\beta\alpha} = -\frac{1}{\sqrt{2}} \left( \begin{array}{ccc}
0 & \lambda_{\alpha\beta} s & f_{u\beta\alpha}v\sin \beta\\
\lambda_{\beta\alpha} s & 0 & f_{d\beta\alpha}v\cos \beta\\
f_{u\alpha\beta}v\sin \beta & f_{d\alpha\beta}v\cos \beta & 0 \end{array}
\right)\label{AA}
\end{equation}
and the $Z_2^H$ breaking sub-matrices are given by
\begin{equation}
	B_\alpha = -\frac{1}{\sqrt{2}}\left(
	\begin{array}{ccc} 0 & 0 & 0 \\ 0 & 0 & 0 \\
0 & x_{d\alpha}s & z_\alpha v\sin \beta \\
x_{u\alpha}s & 0 & z_\alpha v\cos \beta \\
x_{u\alpha}v\sin \beta &
x_{d\alpha}v\cos \beta & 0 \\ 0 & 0 & 0
\end{array}
	\right)\label{BB}
\end{equation}
and involve the small $Z_2^H$ violating Yukawa couplings that were neglected in
Eq.~(\ref{essm2}), $x_{u\alpha}$, $x_{d\alpha}$, and $z_\alpha$.
Since these coupling are small,

the inert neutralino sector is only weakly coupled to the USSM sector, and may
be considered separately from
it to good approximation. However we emphasise that the $Z_2^H$ violating
couplings
are essential in order for the lightest neutralino from the USSM sector
to be able to decay into inert neutralinos and that these couplings are not
expected to be zero.
Exact $Z_2^H$ would also render exotic $D$ and $\overline{D}$ states stable.

Similarly, we take our basis of chargino interaction states to be
\begin{equation}
\tilde{\chi}_\rr{int}^{\pm} = \left( \begin{array}{c}
\tilde{\chi}_\rr{int}^+ \\ \tilde{\chi}_\rr{int}^- \end{array}
\right ), \nonumber\
\end{equation}
where
\begin{equation}
\begin{array}{ccc}
\tilde{\chi}_{\rr{int}}^+ = \left( \begin{array}{c} \tilde{W^+} \\
\tilde{H}_u^+ \\ \tilde{H}_{u2}^+ \\ \tilde{H}_{u1}^+ \end{array}
\right) & \rr{and} & \tilde{\chi}_{\rr{int}}^- = \left(
\begin{array}{c} \tilde{W^-} \\ \tilde{H}_d^- \\ \tilde{H}_{d2}^-
\\ \tilde{H}_{d1}^- \end{array} \right).
\end{array}
\end{equation}
The corresponding mass matrix is then
\begin{equation}
M^c_{\rr{E}_6\rr{SSM}} = \left( \begin{array}{cc} & C^\rr{T} \\ C
\end{array} \right), \nonumber
\end{equation}
where
\begin{equation}
C = \left( \begin{array}{cccc} M_2 & \sqrt{2}m_W\sin \beta & 0 & 0\\
\sqrt{2}m_W\cos \beta & \mu & \frac{1}{\sqrt{2}}x_{d2}s &
\frac{1}{\sqrt{2}}x_{d1}s \\
0 & \frac{1}{\sqrt{2}}x_{u2}s & \frac{1}{\sqrt{2}}\lambda_{22}s &
\frac{1}{\sqrt{2}}\lambda_{21}s \\
0 & \frac{1}{\sqrt{2}}x_{u1}s & \frac{1}{\sqrt{2}}\lambda_{12}s &
\frac{1}{\sqrt{2}}\lambda_{11}s \end{array} \right).
\label{cmm}
\end{equation}

It is clear that a generic feature of the E$_6$SSM is that the LSP is
usually (naturally) composed mainly of inert singlino and ends
up being typically very light. One can see this by inspecting the
new sector blocks of the extended neutralino mass matrix in Eq.~(\ref{nmm}),
such as $A_{11}$, and assuming a hierarchy of the form
$\lambda_{\alpha\beta} s \gg f_{u\alpha\beta}v, f_{d\alpha\beta}v$.
This is a natural assumption since we
already require that $s \gg v$ in order to satisfy the current experimental
limit
on the $Z'$ mass of around 2 TeV, as discussed below and for example in
Ref.~\cite{Accomando:2010fz}.

We emphasise again that for both the neutralinos and the charginos we see that
if the
$Z_2^H$ breaking couplings are exactly zero then the new part
of the E$_6$SSM neutralino mass matrix becomes decoupled from the USSM mass
matrix. However, although approximate decoupling is expected, exact decoupling
is not,
and will therefore not be considered.

\subsection{Experimental constraints}
\label{sub2ex}

The most recent limit on the $U(1)_N$ $Z'$ mass,
set by the CMS \cite{Chatrchyan:2012it}, searching for dilepton
resonances, is $m_{Z'} \gtrsim 1800$~GeV at a confidence level of 95\%.
Although the limit on the mass of the $Z'$ boson associated with the extra
$U(1)_N$ of the E$_6$SSM
can be inferred from this analysis,
this analysis neglects any other matter beyond that of the SM.
When decays of the $Z'$ boson into inert neutralinos
(inert Higgsino and singlino dominated mass eigenstates) are considered the $Z'$
width
tends to increase by a factor of about 2 (see for example
Ref.~\cite{Athron:2011wu},
although we confirm the result in our analysis). This then means that the
branching ratio
into leptons is decreased by a factor of about 2. Estimating the effect of
halving this expected branching ratio
on the analysis in Ref.~\cite{Chatrchyan:2012it} one can read off a 95\%
confidence level
lower bound of around 1600~GeV. This implies $s \gtrsim 4400$~GeV.

When the singlet VEV is this large, as required, the Higgs boson spectrum
typically becomes rather hierarchical with
a lightest mass eigenstate that participates in interactions as a SM-like Higgs
boson
and much heavier Higgs boson states that are approximately decoupled. 
Indeed the Higgs candidate recently discovered at the LHC can easily be
accommodated in the 
E$_6$SSM \cite{Athron:2012sq}.
Although in the MSSM large loop corrections are required in order for this limit
to be satisfied,
in the E$_6$SSM this is easier to achieve since there are extra contributions to
the SM-like Higgs mass
at tree-level due to $U(1)_N$ D-terms. 

If the exotic diquarks (or leptoquarks) are light enough

they would produce spectacular signatures
at the LHC and these exotic states already have strong limits on their masses.
The $E_6$ diquarks are excluded for masses below 3.5 TeV
\cite{Chatrchyan:2011ns}. 
Since these particles' masses must be so large they do not play a role in gluino
cascade decays and are excluded from our analysis.

\subsection{Dark matter considerations}
\label{sub2dm}

Stringent constraints on the E$_6$SSM inert parameter space come from
considerations relating to dark matter.
In the E$_6$SSM as described thus far the LSP is typically one of the two
necessarily light states from the inert neutralino sector.
As such this inert neutralino LSP 
 becomes a dark matter candidate.
The E$_6$SSM has been previously studied as a model attempting to explain
the observed amount of thermal relic cold dark matter
\cite{Hall:2009aj,Hall:2010ix}.
Unfortunately this dark matter scenario is now severely challenged by the most
recent 
XENON100 dark matter direct detection limits \cite{Aprile:2011hi}. The reason is
essentially as outlined in
the following paragraph and a more detailed analysis can be found in
Ref.~\cite{Hall:2010ix}.

In the E$_6$SSM the LSP is generically singlino dominated, a situation
which arises from  the
extended neutralino mass matrix in Eq.~(\ref{nmm}) under the
condition $s \gg v$.
One can show that \cite{Hall:2009aj,Hall:2010ix} if there is no
hierarchy in the Yukawa couplings then the LSP would not annihilate very
efficiently
at the time of thermal freeze-out and would therefore lead to an unacceptable 
overabundance of dark matter in the Universe.
On the other hand, by allowing the largest $f_{d\alpha\beta}$ and
$f_{u\alpha\beta}$ couplings
to be significantly larger than the largest $\lambda_{\alpha\beta}$ couplings
and $\tan\beta$ to be less than about 2
the observed amount of dark matter can be predicted.
In this case the LSP is heavier and,
although still inert singlino dominated, has {substantial} inert Higgsino
admixtures and can
annihilate efficiently enough in the early universe via an s-channel $Z$ boson.
The largest $\lambda_{\alpha\beta}$ coupling
cannot be too small, otherwise the inert charginos would be too light to have
so far escaped detection. At the same time the largest $f_{d\alpha\beta}$ and
$f_{u\alpha\beta}$
couplings cannot be too large if it is required that perturbation theory remains
valid up to the GUT scale.
This being the case, the LSP and the NLSP cannot be made much larger than about
60~GeV \cite{Hall:2010ix}.
In this dark matter scenario there should be some suppression of the coupling of
the LSP to the $Z$ boson,
by partial cancellation between the up-type and down-type inert Higgsino
components,
in order to be consistent with the precision measurement of the invisible $Z$
boson width from LEP,
if the LSP mass is below half of the $Z$ mass.
For given Yukawa couplings increasing $\tan\beta$ has the effect of both
suppressing the LSP mass and
increasing the coupling to the $Z$ boson, lessening this cancellation.
Although this partial cancellation can occur
in the coupling of the LSP to the $Z$ boson, the coupling of the LSP to the
SM-like Higgs boson
is necessarily large if the LSP is to produce the observed amount of dark matter
(due to its inert Higgsino admixtures).
This in turn means a large spin-independent direct detection cross-section,
larger than is now consistent with experiment. However, if the relic abundance
is less than the observed value, then the direct detection constraint can be
avoided, and that is the strategy that we follow for benchmark points of this
kind, as we now discuss in more detail.

It is important to note that for both the MSSM and the standard E$_6$SSM we
analyse points where less than the
observed amount of dark matter is predicted and assume that the majority of dark
matter
is not made up of MSSM/E$_6$SSM neutralinos. 
This choice of the parameter space is actually
dictated by limits from direct detection experiments: if less than the total
amount of the dark matter in the universe is
made up of  LSPs, then the expected number of events for a direct detection
experiment
for a given  LSP direct detection cross-section will be correspondingly smaller.
In the E$_6$SSM such points require the previously discussed hierarchy of Yukawa
couplings appearing
in the inert block of the neutralino mass matrix and in this case if the LSP
mass is below half of the $Z$ mass then
$\tan\beta$ still cannot be too large in order
for the LSP not to contribute too much to the invisible width of the $Z$ boson,
as measured at LEP.
Large values of $\tan\beta$ mean that the up-type inert Higgsino admixture in
the LSP
greatly outweighs the down-type inert Higgsino admixture, necessarily leading to
a too large coupling to the $Z$ boson.

Alternatively, in a variation of the model known as the EZSSM \cite{Hall:2011zq}
the
situation is quite different and the model may be responsible for all dark
matter
and consistent with all experiment. Here an extra discrete symmetry $Z_2^S$ is
imposed that
forbids the terms in the superpotential involving the inert singlet superfields,
i.e. the $f_{u\alpha\beta}$ and $f_{d\alpha\beta}$ and $Z_2^H$ violating
$x_{u\alpha}$ and $x_{d\alpha}$ Yukawa couplings are forced to be zero.
This then means that the inert singlinos are exactly massless and decoupled from
the
rest of the neutralinos. In this variation of the model the lightest
non-inert-singlino
LSP is absolutely stable.

If this stable particle is the bino and there are a
pair of inert Higgsinos close by in mass then the bino can be responsible for
all of
dark matter \cite{Hall:2011zq}. The massless inert singlinos themselves slightly
increase the
expansion rate of the universe prior to nucleosysnthesis in agreement with
observation of the $^4$He relic abundance and in this scenario there would
currently be
a cosmic inert singlino background slightly colder than the cosmic neutrino
background \cite{Hall:2011zq}.
The phenomenology of this scenario as regards the gluino cascade decay is
essentially identical to MSSM one and we do not make of point of trying to
distinguish this type of scenario from the MSSM in this study.

Finally we shall consider a scenario where the two lightest (predominantly inert
singlino) neutralinos
are both very light, with one around a GeV, and one much lighter, in principle
in the keV range.
In this case, interesting phenomenoly can emerge in the gluino cascade decays as
in the usual case
where the lightest neutralino states are around half the Z mass. However, unlike
that case, the 
lightest neutralino is not subject to direct detection limits. Moreover, it is
possible to arrange for the
correct relic abundance in such a scenario, where the lightest neutralino in the
keV mass range is stable and constitutes Warm Dark Matter (WDM)
\cite{King:2012wg}. The idea is that both the light neutralinos are thermally
produced in the early Universe due to their couplings to the $Z$ and $Z'$ gauge
bosons, but the GeV state decays 
late (due to its weak couplings) after both of the neutralinos have gone out of
thermal equilibrium, and reheats the Early Universe, effectively diluting the
number density of the stable keV neutralinos, such that they are responsible for
the observed relic abundance. It is very interesting to compare the gluino
cascade decays in this case to that where the lightest neutralinos are around
half the $Z$ mass, in order to provide an experimental ``confirmation'' of
keV dark matter at the LHC.

\subsection{Parameter space under study}

\label{sec:param}

In this study we consider a pattern of low energy soft gaugino masses that is
consistent with
$E_6$ grand unification \cite{King:2005jy}. This typically implies that at the
EWSB scale
$M_2 \approx 2M_1$ and if $M_1 = 150$~GeV the physical gluino mass is around
800~GeV.
In order to have a direct comparison, $M_1$ is made equal in both the MSSM and
E$_6$SSM, 150~GeV
in the following analysis. We also fix the physical gluino mass to be equal,
800~GeV in
the following analysis. For large $\mu$ in the MSSM, and given that the
effective $\mu$ in the
E$_6$SSM is large due to the limit on $s$ coming from the limit on the $Z'$
mass,
there will be a neutralino that is almost the bino with a mass very close to
$M_1$.
For lower values of $\mu$ in the MSSM bino-Higgsino mixing occurs.

We also consider large squark masses, with all squarks heavier than the gluino.
This is motivated by the GUT constrained E$_6$SSM, where such large squark
masses
are a feature, although it should be noted that although the EZSSM scenario
is consistent with such GUT constraints it has not been shown that
a GUT constrained standard E$_6$SSM scenario consistent with dark matter
observations exists.
Nonetheless we assume heavy squarks for all scenarios.

\begin{figure}[t]
	\centering
	\includegraphics[width=.8\columnwidth]{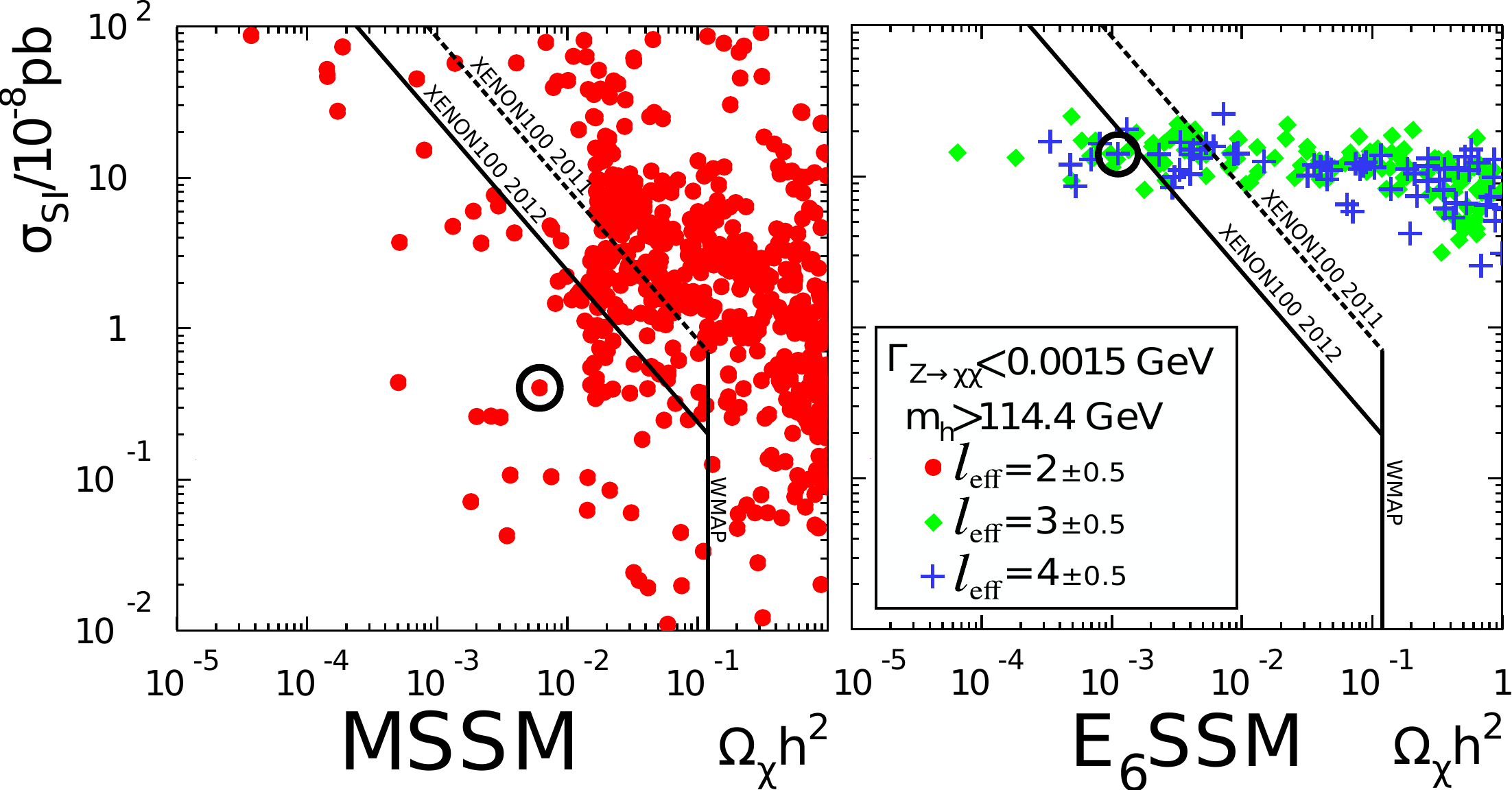}
	\caption{The scanned regions of the parameter spaces projected onto the
plane spanned by the spin-independent cross-section, $\sigma_{SI}$, and the
relic density of the LSP, $\Omega_{\tilde\chi_1^0} h^2$. The area right of the
vertical solid line is excluded by WMAP \cite{Komatsu:2010fb} and the area above
the diagonal line is excluded by XENON100, where the LSP direct
detection cross-section exclusion gets weighted by its relic density.
The 90\% confidence level limit on the spin-independent
LSP-nucleon cross section
for a weakly interacting LSP with mass around 50 GeV and which makes up all of
the observed amount of DM has been pushed down from
$0.7\times10^{-44}$cm$^2=0.7\times10^{-8}$pb in 2011 \cite{Aprile:2011hi} to
$0.2\times10^{-44}$cm$^2=0.2\times10^{-8}$pb in 2012 \cite{XENON100:2012}.
The LEP constraints on chargino masses ($m_{\tilde \chi_1^\pm}>103$ GeV) and
invisble $Z$ width ($\Gamma(Z\rightarrow\tilde\chi^0\tilde\chi^0)<1.5$~MeV) have been applied.
The constraint on the Higgs mass is also taken from LEP since it holds
for invisible Higgs decays, a common feature among the E$_6$SSM points.
The colours/shapes represent the effective gluino decay chain length
$l_{\mathrm{eff}}=\sum_l l\cdot P(l)$ for each point, where $P(l)$ is the
probability for a chain length of $l$, as defined in Fig.~\ref{fig:chain}. The
benchmarks entitled MSSM and E$_6$SSM-I, which are consistent
with the particle recently discovered at the LHC being the SM-like
lightest Higgs boson, are encircled.}
\label{fig:scan}
\end{figure}

In Fig.~\ref{fig:scan} we present the results of scans over the MSSM and the
E$_6$SSM
parameter space in the $(\Omega_\chi h^2,\sigma_{\rr{SI}})$ plane.
The parameter space scanned over is shown in Tabs.~\ref{tab:mssm-region}
and~\ref{tab:e6ssm-region}
and points are linearly distributed over these ranges.
The ranges chosen are motivated from the discussions in the previous two
subsections.
For both the MSSM and E$_6$SSM the points are shown as long as they are
consistent with the LEP limit on the SM-like Higgs boson mass,
applicable even if the Higgs has large invisible branching fractions,
but we also highlight benchmark points that are consistent with the particle
recently discovered at the LHC being the SM-like lightest Higgs boson.
For the E$_6$SSM, where
the LSP mass
may be less than half of the $Z$ mass, points are only shown if they are
consistent with
LEP limits on the invisible $Z$ width, contributing less than 1-sigma.
For both scans squark soft masses are set to 2~TeV, although the effects of
squark mixing
on the mass eigenstates are included. The physical gluino mass is set to 800~GeV
and
the $U(1)_Y$ gaugino mass is set to 150~GeV in both cases. We do not consider
scenarios
in which any squarks are less massive than the gluino.

 \begin{table}[htb]
\begin{minipage}[h]{.45\linewidth}
	\centering
	\begin{tabular}{rrrr}
		parameter	&	min	&	max\\
		\hline
		\hline
		$\tan\beta$	&	2	&	60\\
		\hline
		$A_t=A_b=A_\tau=A_\mu$		&	-3 	&	3&
\multirow{3}{*}{\rotatebox{-90}{[TeV]}}\\
		$M_A$		&	0.1  	&	2\\
		$\mu$		&	-2	&	2\\
		\hline
	\end{tabular}
	\caption{The MSSM scanning region. A common squark and slepton mass scale
was fixed to $M_{S}=2$~TeV. The gaugino masses were fixed to $M_1=150$~GeV,
$M_2=285$~GeV, and $M_3=619$~GeV, providing a gluino mass close to 800 GeV.}
	\label{tab:mssm-region}
\end{minipage}
\begin{minipage}[h]{.45\linewidth}
	\centering
	\begin{tabular}{rrrr}
		parameter	&	min	&	max\\
		\hline
		\hline
		$\tan\beta$	&	1.4	&	2\\
		$|\lambda|$	&	0.3	&	0.7\\
		$\lambda_{22}$	&	0.0001	&	0.01\\
		$\lambda_{21}$	&	0.01	&	0.1\\
		$\lambda_{12}$	&	0.01	&	0.1\\
		$\lambda_{11}$	&	0.0001	&	0.01\\
		$f_{d21}$	&	0.0001	&	0.01\\
		$f_{d21}$	&	0.1	&	1\\
		$f_{d12}$	&	0.1	&	1\\
		$f_{d11}$	&	0.0001	&	0.01\\
		$f_{u22}$	&	0.0001	&	0.01\\
		$f_{u21}$	&	0.1	&	1\\
		$f_{u12}$	&	0.1	&	1\\
		$f_{u11}$	&	0.0001	&	0.01\\
		$x_{d2}$		&	$10^{-4}$&	$10^{-2}$\\
		$x_{d1}$		&	$10^{-4}$&	$10^{-2}$\\
		$x_{u2}$		&	$10^{-4}$&	$10^{-2}$\\
		$x_{u1}$		&	$10^{-4}$&	$10^{-2}$\\
		$z_1$		&	$10^{-3}$&	$10^{-1}$\\
		$z_2$		&	$10^{-3}$&	$10^{-1}$\\
		\hline
		$A_t=A_b=A_\tau$&	-3 	&	3&
\multirow{3}{*}{\rotatebox{-90}{[TeV]}}\\
		$M_A$		&	1	&	5\\
		$s$		&	3.7	&	8\\
		\hline
	\end{tabular}
	\caption{The E$_6$SSM scanning region. A common squark and slepton mass
scale was fixed to $M_{S}=2$~TeV. The gaugino masses were fixed to $M_1=150$~GeV, $M_1'=150$~GeV, $M_2=300$~GeV, and $M_{\tilde{g}}=800$~GeV.}
	\label{tab:e6ssm-region}
\end{minipage}
\end{table}

\begin{figure}[htb]
\centering
	\includegraphics[width=.8\textwidth]{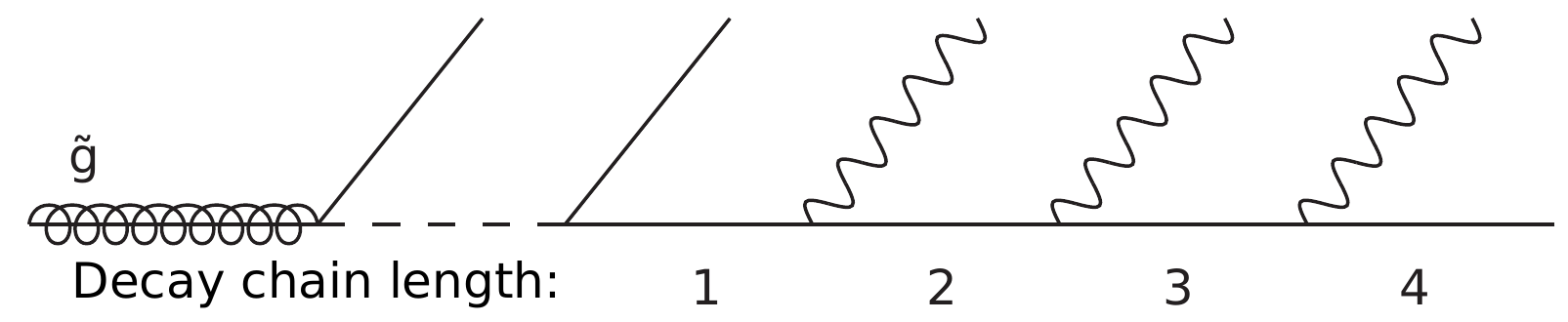}
\caption{In the first step in a gluino decay chain the gluino decays into a
quark and a squark (in our scenario it will be a virtual squark) which in turn
decays into a second quark and a neutralino or chargino. This is the shortest
possible gluino decay chain, which we define as having length $l=1$. The
neutralino or chargino can then decay into lighter neutralinos or charginos by
radiating $W$, $Z$, or Higgs bosons, which typically decay into pairs of
fermions. For each such decay the decay chain length is taken to increase by
one. The radiated bosons could be on-shell or off-shell depending on the mass
spectrum of the model.
Light squarks or leptons could appear further along in the decay
chain, leading to radiation of SM fermions without intermediate $W$, $Z$, or
Higgs bosons, but in our study squarks and sleptons are heavy so this is not
relevant.}
\label{fig:chain}
\end{figure}

\begin{figure}[htb]
\centering

\subfigure[The distribution of the effective chain length, $l_{\rr{eff}} =
\sum_{l}l\cdot P(l)$, where $P(l)$ is the probability of having a decay chain of
length $l$ for a point in parameter space.]{
	\includegraphics[width=.45\linewidth]{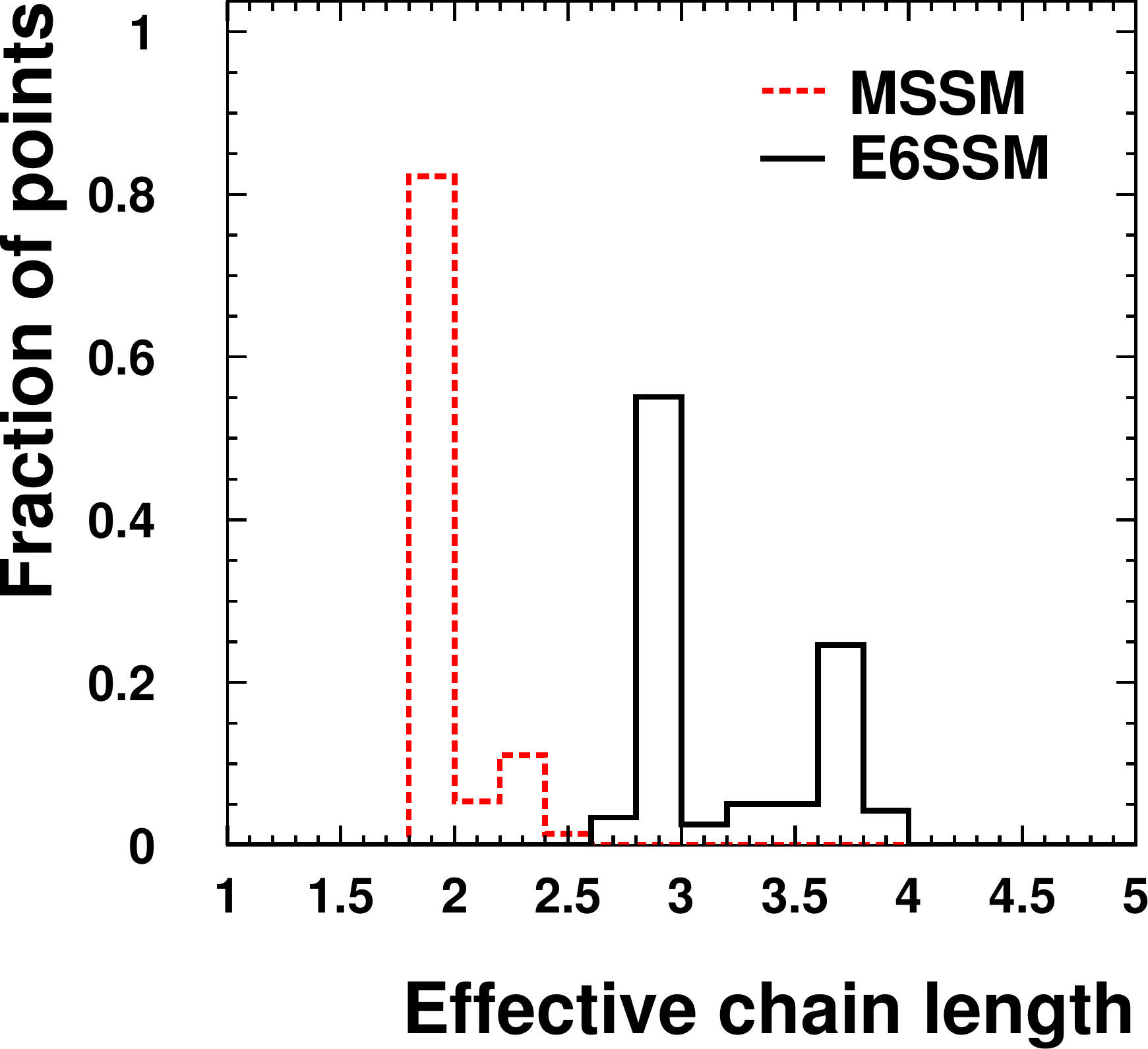}%
\label{fig:efflen}
}
\subfigure[The probability for a certain gluino decay chain length, averaged
over all points in the parameter scan, satisfying dark matter and collider
constraints.]{
	\includegraphics[width=.45\linewidth]{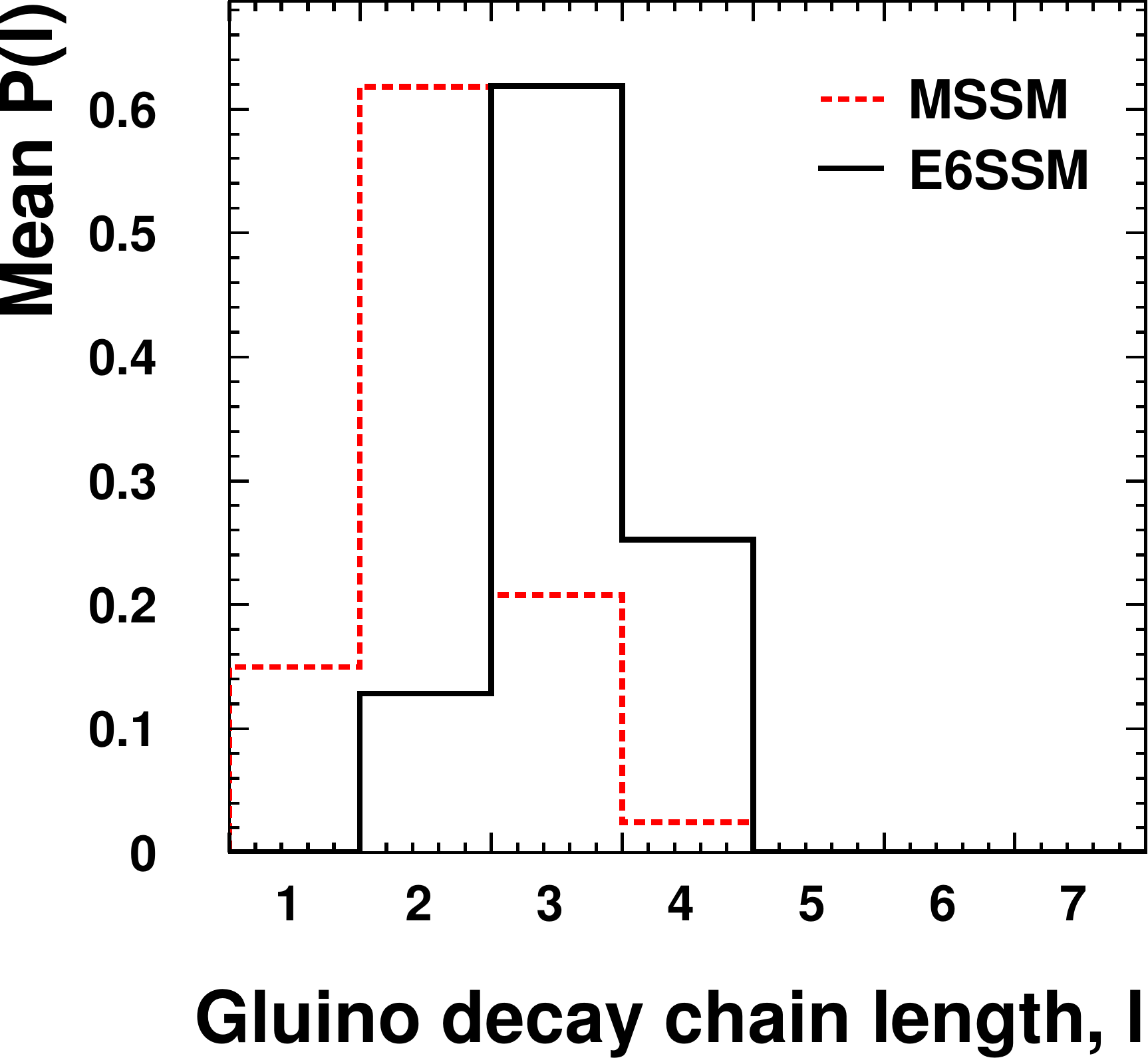}%
\label{fig:lengthprobmean}
}
\caption{Statistical properties of the gluino decay chain length in the scanned
parameter space. Figure \ref{fig:efflen} shows how the effective gluino decay
chain length evaluated at each point is distributed over the scanned parameter
space. Most points have an effective decay chain length close to an integer
indicating that there usually is a largely dominant decay chain length. The two
peaks just below 3 and 4 clearly show how the E$_6$SSM generally introduces one
or two extra steps in the chain. Figure \ref{fig:lengthprobmean} show the
average probabilities of different gluino decay chain lengths for both models'
parameter spaces. Again, the E$_6$SSM is shown to shift the probabilities to
longer decay chain lengths.}
\end{figure}

We define the length of a gluino decay chain to be the number of decays after
the virtual squark as in
Fig.~\ref{fig:chain}. We then also define an effective chain length for each
point in parameter space
\begin{equation}
l_{\rr{eff}} = \sum_{l}l\cdot P(l),
\end{equation}
where $P(l)$ is the probability of having a decay chain of length $l$ for that
point.
Intervals of effective chain length are colour/shape coded
in Fig.~\ref{fig:scan}.
These scans
indicate that in the E$_6$SSM these decay chains are typically longer due to the
bino decaying
into the lower mass inert states. The distribution of effective chain lengths
for these scans are also plotted
in Fig.~\ref{fig:efflen}.

In the MSSM, for the parameters chosen, the typical decay length is 2, with
the lightest chargino being initially produced from the virtual squark decay
before subsequently itself decaying to the bino LSP.
When the magnitude of $\mu$ is small the mixing between the gauginos (bino and
wino) and the Higgsinos increases
and the entire neutralino and chargino spectrum
is pushed down. Specifically the heavier neutralinos and chargino are brought
down below the gluino mass causing extra steps in the gluino decay chain.

In the E$_6$SSM the effective decay length is typically either 3 or 4. Initially
an either charged or neutral wino
is produced and this subsequently decays to the bino. The bino then decays into
either of the two light inert neutralinos that
are the LSP and NLSP. Which of these the bino preferentially decays into depends
on the values of the $Z_2^H$ violating couplings
in blocks $B_\alpha$ in the neutralino mass matrix in Eq.~(\ref{nmm}).
Therefore in the E$_6$SSM we typically expect the gluino cascade decays to be
either one or two steps longer than in the MSSM.

\subsection{{Benchmarks}}
\label{sec:benchmarks}

\begin{figure}[ht]
	\flushleft
	MSSM:\\
	\includegraphics[height=.2\linewidth]{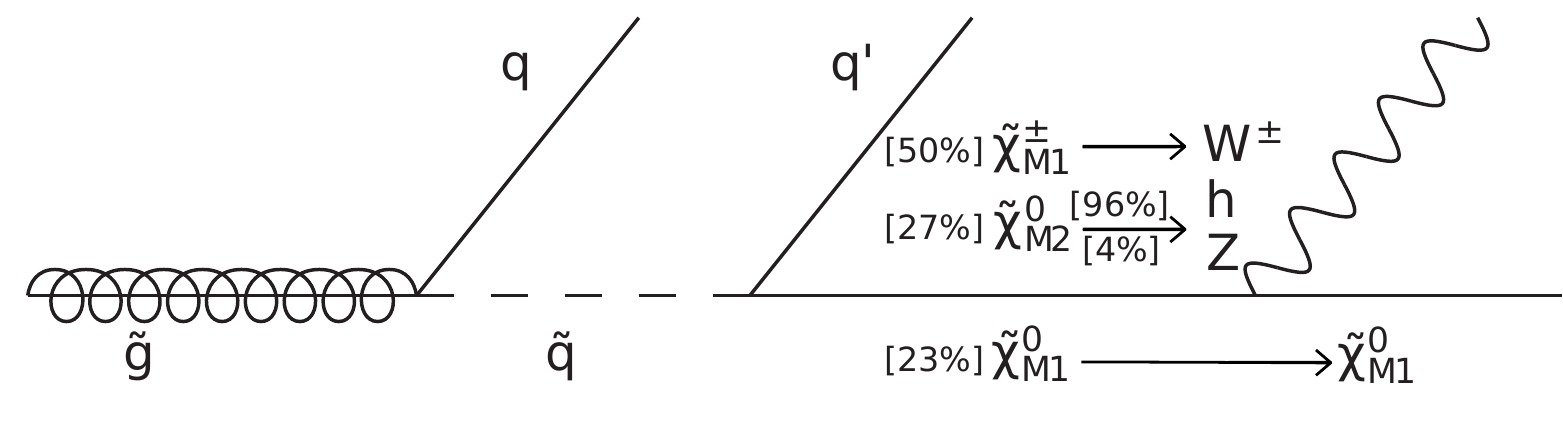}\\
	E$_6$SSM-I:\\
	\includegraphics[height=.2\linewidth]{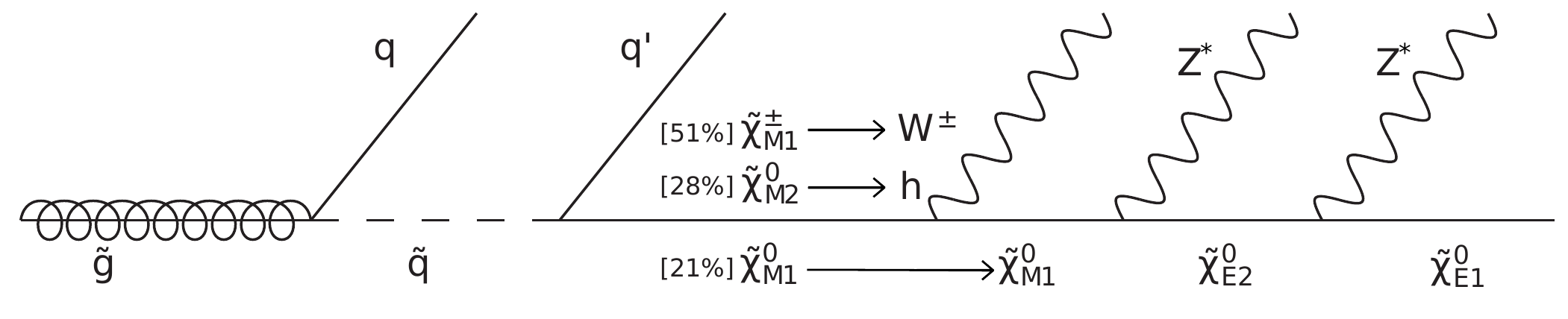}\\
	\caption{Feynman diagrams for the leading gluino decay chains for our two main benchmarks, MSSM and E$_6$SSM-I. The branching ratios for production of particles are denoted in brackets. The decay chains for both benchmarks are essentially the same up to the first two steps, with just a slight difference in branching ratios. 
	The essential difference is that, in the case of the E$_6$SSM, the lightest MSSM-like neutralino is no longer stable and decays in two steps to the lightest E$_6$SSM neutralino.
	About 20\% of the gluino decays go directly into the lightest MSSM-like neutralino implying a chain length $l=1$ for the MSSM and $l=3$ for the E$_6$SSM. On the other hand, about 80\% of the gluino decays  into a heavier neutralino or chargino, which subsequetly decays into the lightest MSSM-like neutralino state, giving the MSSM a chain length $l=2$ and the E$_6$SSM $l=4$.}
	\label{fig:diagrams}
\end{figure}

Below follow descriptions of the main features of the chosen benchmarks. The
details of their spectrum and parameter values are given in
Tab.~\ref{tab:bm-table}. The diagrams for the main decay channels of the gluinos
are
shown in Fig.~\ref{fig:diagrams}
for two main benchmarks, MSSM and E$_6$SSM-I
discussed below. The branching ratios for production of particles are denoted in brackets. The decay chains for both benchmarks are essentially the same up to the first two steps, with just a slight difference in branching ratios.
The essential difference is that, in the case of the E$_6$SSM, the lightest MSSM-like neutralino is no longer stable and decays in two steps to the lightest E$_6$SSM neutralino.
About 20\% of the gluino decays go directly into the lightest MSSM-like neutralino implying a chain length $l=1$ for the MSSM and $l=3$ for the E$_6$SSM. On the other hand, about  80\% of the gluino decays are into a heavier neutralino or chargino, which subsequetly decays into the lightest MSSM-like neutralino state giving the MSSM a chain length $l=2$ and the E$_6$SSM $l=4$. More complete diagrams showing how the gluinos
decay for the different benchmarks (below) are shown in Fig.~\ref{fig:gluino-decays} in
Appendix \ref{ap:A}.

The following benchmarks provide the main focus of our study:
\begin{itemize}

	\item MSSM:\\
		In this benchmark we have an LSP with a mass of 150 GeV. The low dark
matter relic density is achieved via LSP resonance annihilation through the
CP-odd and heavy Higgs bosons, $A$ and $H$. The lightest Higgs boson, $h$, has a
mass of 124.4 GeV and can be produced in gluino decay chains. This happens when
the gluino decays via the next to lightest neutralino $\tilde\chi_2^0$. The
gluino decay chain length is dominantly $l=2$ for this benchmark as for all the
MSSM points scanned over.
	\item E$_6$SSM-I:\\
		In this benchmark the LSP and NLSP annihilate efficiently through the
Higgs
		boson resonance leaving a relic density less than the observed relic
		density of dark matter. The lightest Higgs mass is around 125 GeV and
the two lightest
		inert neutralino state masses are slightly above half of the Higgs mass.
		In this case the Higgs is SM-like in both its composition and its
decays,
		since only decays into SM final states are kinematically allowed.
		If the LSP mass
		was be slightly below half of the Higgs mass then the Higgs boson would
decay invisibly.
		In order for the two lightest inert neutralino states to be heavy enough
		some of the $f_{u\alpha\beta}$ and $f_{d\alpha\beta}$
		are required to be large enough such that Yukawa coupling running
		becomes non-perturbative on the way up to the GUT scale. We estimate
that
		here the Yukawa couplings remain purtubative up to an energy scale of
		order $10^{12}$~GeV.
		Compared to the MSSM benchmark there are typically two extra
		steps in the decay chain as the bino-like neutralino decays into first
the NLSP which
		subsequently decays into the LSP.
		The two extra steps in the chain make the 
		most common total gluino decay chain length $l=4$.
		The decay of the bino into the NLSP is preferred
		over the decay directly into the LSP because of the structure of the
		$Z_2^H$ breaking trilinear Higgs Yukawa couplings given in
Tab.~\ref{tab:e6-yukawa}.
		
	\item E$_6$SSM-II:\\
		This benchmark represents a typical scenario, with decay 
		chain features similar to in the E$_6$SSM-I benchmark, where the most
		common decay length is 4. Yukawa coupling running remains perturbative
		up to the GUT scale, as it does in all of the following benchmarks.
		Just as in previous benchmark the gluino decay chain length here is
typically two 
		steps longer than in the MSSM because of the two extra light neutralino
states, but here their masses
		are smaller.
		The LSP mass is below half of the $Z$ boson mass, but decays of the
		$Z$ boson into LSP-LSP contribute to the effective number of neutrinos
		as measured at LEP less than 1-sigma.
		Because the LSP and NLSP have masses not very far away from half of the
Z mass they are able to 
		annihilate relatively efficiently via an s-channel $Z$ boson in the
early Universe 
		and the LSP contributes much less than the observed relic density of
dark matter.
		This benchmark has a rather heavy lightest Higgs with a mass around 134
GeV. 
		This Higgs decays dominantly invisibly into a pair LSPs and is ruled out
if the boson candidate 
		discovered at the LHC is interpreted as a SM-like Higgs boson. 
		
	\item E$_6$SSM-III:\\
		In contrast to previous benchmarks this benchmark represents the other
typical E$_6$SSM scenario 
		where the bino-like neutralino decays straight to the LSP (not via the
NLSP). In this case there 
		is only one extra step compared to the MSSM and the most
		common decay length is 3. Points with this shorter decay length are
slightly
		more common when one scans over the $Z_2^H$ breaking Yukawa couplings.
		In the same way as in E$_6$SSM-II the LSP annihilates efficiently via an
s-channel
		$Z$ boson, even though it is farther away from resonance,
		and contributes much less than the observed relic density of dark
matter.
		In this benchamark the lightest Higgs has a mass around 116 GeV, much
lighter than in E$_6$SSM-II.
		As above, this is ruled out if the boson candidate 
		discovered at the LHC is interpreted as a SM-like Higgs boson.
		This benchmark represent the slightly more common scenario in the
		parameter space where the typical decay chain has one extra step
		after the 150~GeV bino-like neutralino.
		
	\item E$_6$SSM-IV (EZSSM-I):\\
		This benchmark represents an EZSSM dark matter scenario as described in
		Ref.~\cite{Hall:2011zq}. Here the bino-like neutralino is stable and
		makes up all of the observed dark matter relic density. A low enough
		relic density is achieved via the bino-like neutralino upscattering
		into the inert Higgsino pseudo-Dirac pair $\tilde{\chi}^0_{E3}$ and
		$\tilde{\chi}^0_{E4}$. 
		Since the bino is stable the gluino decays of this benchmark are
essentially
		identical to those of the MSSM benchmark
		and also have the same dominant decay chain length $l=2$.
		The bino-like neutralino's
		direct detection cross-section is small.

		This benchmark looks very similar to the MSSM benchmark 
		and cannot be distinguished from it purely by analysis of gluino cascade
decays. However, if a heavy
		or CP-odd Higgs around 300 GeV was to be excluded then the MSSM, but not
the EZSSM,
		benchmark would be excluded. This relies on the fact that the EZSSM can
have a stable bino at 150 GeV without requiring
		resonance annihilation through heavy Higgs boson states, which in
contrast is required in the MSSM.
		The Higgs boson has a mass around 125 GeV and is SM-like in its
composition
		and decays.

	\item E$_6$SSM-V (EZSSM-II):\\
		This is another benchmark of the EZSSM, but here the dark matter relic
density
		is not explained. Here the inert Higgsino pseudo-Dirac pair have masses
around 120 GeV,
		well below the bino-like neutralino. These states co-annihilate
efficiently
		in the early universe and contribute a dark matter relic density less
than the
		total observed dark matter relic density. The effective decay chain
length is
		about 4. 
		The complete decay modes for the gluino
		are shown in Fig.~\ref{fig:E6SSM-V-decays} in Appendix \ref{ap:A}. 
		This benchmark provides an example of heavier lightest inert neutralino
masses
		(excluding the decoupled inert singlinos) without requiring
non-perturbatively
		large Yukawa couplings as in the non-EZSSM benchmark E$_6$SSM-I. Because
the 
		lightest inert-Higgsino-like neutralinos have masses much larger
		than half of
the lightest Higgs mass, which is 126 GeV, 
		the Higgs will decay very SM-like.
	\item E$_6$SSM-VI (approximate $Z_2^S$):\\
		In this benchmark the inert singlino decoupling is not exact and the
		inert singlino-like LSP mass is not zero, but has been pushed down to
the 100 keV scale.
		This point represents the scenario where the lightest neutralino in the
keV mass range is stable and           constitutes Warm Dark Matter (WDM)
\cite{King:2012wg}, as described earlier, although we do not
calculate the relic density here.


		The approximate decoupling leads to a quite long-lived bino (with a
width of order $10^{-11}$~GeV)
		and the step in the decay after the bino appears as a displaced vertex
at the order of 0.1~mm
		from the previous step.
		The decay chain length is typically 4. The complete decay modes for the
gluino is shown in Fig.~\ref{fig:E6SSM-VI-decays} in Appendix \ref{ap:A}.
		As remarked, the observed relic density could be achieved by pushing the
LSP mass down to the
		keV scale. In this case the last steps of the gluino decay
		would be likely to occur outside the detector and one would be left with
something that looks like the MSSM. The lightest Higgs has a mass of 126~GeV and
decays to the LSP are 
		very suppressed due to its small mass. The Higgs could decay to the GeV
scale NLSP, which in turn would decay to the LSP outside the detector since its
width is of the order $10^{-20}$~GeV, leading to invisible Higgs decays. The
branching ratio for $h\rightarrow \tilde\chi_2^0\tilde\chi_2^0 $ is however
around 6\% and is not excluded by current Higgs data.
\end{itemize}

In the following analysis, in section IV,
we use MSSM and E$_6$SSM-I as our main benchmarks.
With the exception of E$_6$SSM-IV (EZSSM-I) the results obtained for each of the
E$_6$SSM benchmarks are very similar. (E$_6$SSM-IV on the other hand looks very
similar to the MSSM, since the bino is stable.) We therefore mainly give just
the results for MSSM and E$_6$SSM-I,
demonstrating the qualitative difference between the MSSM and $E_6$ models.
We also include some results for the E$_6$SSM-VI
to demonstrate the effects of having
an even less compact spectrum and also to show how little our conclusions
depend on the exact spectrum of the $E_6$ model.

\begin{table}
	\centering
	\begin{tabular}{cccccccccc}
	                &	MSSM	&	 E$_6$SSM-I 	&	E$_6$SSM-II	&
E$_6$SSM-III	&	E$_6$SSM-IV	&	E$_6$SSM-V	&	E$_6$SSM-VI	&	\\
\hline	      
\hline
$\tan\beta$	        &	10	&	1.5		&	1.42		&	1.77		&	3
&	1.42		&	1.42		&	\\
$\lambda$	        &	-	&	0.497		&	0.598		&	-0.462		&
-0.4		&	0.598		&	0.598		&	\\
\hline	        
$s$		        &	-	&	5180		&	5268		&	5418		&
5500		&	5268		&	5268		&
\multirow{8}{*}{\rotatebox{-90}{[GeV]}}\\
$\mu$		        &	1578	&	(1820)		&	(2228)		&	(−1770)
&	 (-1556)	&	(2228)		&	(2228)		&	\\
$A_t=A_b=A_\tau$        &	-2900	&	-3110		&	-3100		&	476.2
&	4638		&	-2684		&	-2684		&	\\
$M_A$		        &	302.5	&	3666		&	4365		&	2074
&	4341		&	4010		&	4000		&	\\
$M_1$		        &	150	&	150		&	150		&	150		&	150		&
150		&	150		&	\\
$M_2$		        &	285	&	300		&	300		&	300		&	300		&
300		&	300		&	\\
$M_{1'}$	        &	-	&	151		&	151		&	151		&	151		&
151		&	151		&	\\
$m_{\tilde g}$	        &	800.2	&	800.0		&	800.0		&	800.0
&	800.0		&	800.0		&	800.0		&	\\
\hline	

$m_{\tilde \chi^0_{M1}}$&	148.7	&	148.9		&	149.1		&	151.2
&	150.6		&	149.1		&	149.1		&\\
$m_{\tilde \chi^0_{M2}}$&	302.2	&	296.1		&	296.8		&	303.7
&	301.7		&	296.8		&	296.8		&	\\
$m_{\tilde \chi^0_{M3}}$&	1582	&	1763		&	2233		&	1766
&	1557		&	2233		&	2233		&	\\
$m_{\tilde \chi^0_{M4}}$&	1584	&	1823		&	2246		&	1771
&	1558		&	2246		&	2246		&	\\
$m_{\tilde \chi^\pm_{M1}}$&	302.2	&	299.0		&	299.2		&	300.9
&	300.4		&	299.2		&	299.2		&	\\
$m_{\tilde \chi^\pm_{M2}}$&	1584	&	1822		&	2229		&	1771
&	1557		&	2229		&	2229		&	\\

$m_{\tilde \chi^0_{U1}}$&	-	&	1878		&	1835		&	1909
&	1937		&	1835		&	1835		&
\multirow{2}{*}{\rotatebox{-90}{$\!\!\!\!\!$[GeV]}}\\
$m_{\tilde \chi^0_{U2}}$&	-	&	1973		&	2003		&	2062
&	2087		&	2003		&	2003		&	\\

$m_{\tilde \chi^0_{E1}}$&	-	&	62.7		&	43.5		&	45.2
&	0		&	0		&	0.00011		&\\
$m_{\tilde \chi^0_{E2}}$&	-	&	62.8		&	48.6		&	53.2
&	0		&	0		&	1.53		&	\\
$m_{\tilde \chi^0_{E3}}$&	-	&	119.8		&	131.3		&	141.6
&	164.1		&	119.9		&	120.1		&	\\
$m_{\tilde \chi^0_{E4}}$&	-	&	121.0		&	163.6		&	187.4
&	164.1		&	119.9		&	122.8		&	\\
$m_{\tilde \chi^0_{E5}}$&	-	&	183.0		&	197.0		&	227.8
&	388.9		&	185.8		&	185.8		&	\\
$m_{\tilde \chi^0_{E6}}$&	-	&	184.4		&	224.3		&	265.6
&       388.9		&	185.8		&	187.0		&	\\
$m_{\tilde \chi^\pm_{E1}}$&	-	&	109.8		&	119.9		&	122.7
&	164.1		&	119.9		&	119.9		&	\\
$m_{\tilde \chi^\pm_{E2}}$&	-	&	117.7		&	185.8		&	225.1
&	388.9		&	185.8		&	185.8		&	\\

$m_{h}$		        &	124.4	&	125.4		&	133.8		&	116.3
&	124.7		&	126.1		&	125.8		&\\

\hline
$P(l=1)$	        &	0.188	&	$<10^{-9}$      &	$<10^{-5}$	&
$<10^{-5}$	&	0.1727		&	$<10^{-8}$ 	&	$<10^{-12}$	&	\\
$P(l=2)$	        &	0.812	&	$<10^{-4}$	&	0.01524		&	0.1723
&	0.8273		&	0.01	        &	$<10^{-5}$	&	\\
$P(l=3)$	        &	0	&	0.1746		&	0.2336		&	0.7986		&
$<10^{-6}$	&	0.2	        &	0.1721		&	\\
$P(l=4)$	        &	0	&	0.8196		&	0.7512		&	0.02915		&
$<10^{-15}$	&	0.8	        &	0.8280		&	\\
$P(l=5)$	        &	0	&	0.0058		&	$<10^{-7}$	&	0		&	0
&	 $<10^{-15}$	&	0		&	\\
\hline	
$\Omega h^2$	        &	0.00628	&	0.00114		&	0.0006842	&
0.0006937	&	0.101		&	$0.00154$	&			&	\\
$\sigma_{SI}$ 	        &	$0.401\times10^{-9}$		&
$15.34\times10^{-8}$&	$9.35\times10^{-8}$&	$16.35\times10^{-8}$&
$3.75\times10^{-11}$	&	$3.98\times10^{-13}$ 	&	 	&	[pb]\\
\hline
	\end{tabular}
	\caption{Benchmarks motivated by the parameter scans presented in
	Fig.~\ref{fig:scan} and Tabs.~\ref{tab:mssm-region} and \ref{tab:e6ssm-region}. From
top to bottom the classes of parameters are dimensionless input parameters;
dimensionful input parameters; neutralino, chargino, and lightest Higgs masses
(in absolute values); probabilities for certain gluino decay chain lengths; and
finally dark matter properties. The $\tilde\chi^{0(\pm)}_{Mi}$ are MSSM-like
states, the $\tilde\chi^{0}_{Ui}$ are USSM-like states, being mainly mixtures of
$\tilde{S}$ and $\tilde{B}'$. The $\tilde\chi^{0(\pm)}_{Ei}$ are states
introduced by the inert sector of E$_6$SSM. The scale for squark and slepton
masses is $M_S=2$~TeV in all benchmarks.}
	\label{tab:bm-table}
\end{table}

\begin{table}
	\centering
		\setlength{\tabcolsep}{8pt}
	\begin{tabular}{ccccccc}
				&E$_6$SSM-I		&E$_6$SSM-II		&E$_6$SSM-III
&E$_6$SSM-IV 		&E$_6$SSM-V 	&E$_6$SSM-VI \\
		\hline
		\hline\\[-6pt]
		
		$\lambda$	& $3.93\times10^{-1}$	& $5.98\times10^{-1}$	&
$-4.77\times10^{-1}$	& $-4.0\times10^{-1}$	& $5.98\times10^{-1}$	&
$5.98\times10^{-1}$\\

		$\lambda_{22}$	& $-3.57\times10^{-4}$	& $-4.48\times10^{-3}$	&
$5.14\times10^{-3}$	& $1.0\times10^{-1}$	& $-4.48\times10^{-3}$ 	&
$-4.48\times10^{-3}$ \\
		$\lambda_{21}$	& $3.0\times10^{-2}$	& $-4.34\times10^{-2}$	&
$-8.77\times10^{-2}$	& 0			& $-4.34\times10^{-2}$ 	&
$-4.34\times10^{-2}$ \\
		$\lambda_{12}$	& $3.21\times10^{-2}$	& $-3.83\times10^{-2}$	&
$6.40\times10^{-2}$	& 0			& $-3.83\times10^{-2}$& $-3.83\times10^{-2}$  \\
		$\lambda_{11}$	& $7.14\times10^{-4}$	& $-1.25\times10^{-2}$	&
$-3.36\times10^{-2}$	& $4.22\times10^{-2}$	& $-1.25\times10^{-2}$ 	&
$-1.25\times10^{-2}$ \\
		$f_{d22}$	& $1\times10^{-3}$	& $-9.02\times10^{-3}$	&
$7.06\times10^{-3}$	& 0			& 0		& $2.0\times10^{-1}$\\
		$f_{d21}$	& $6.844\times10^{-1}$	& $-3.48\times10^{-1}$	&
$6.10\times10^{-1}$	& 0			& 0		& $-3.48\times10^{-3}$\\
		$f_{d12}$	& $6.5\times10^{-1}$	& $6.92\times10^{-1}$	&
$-7.64\times10^{-1}$	& 0			& 0		& $6.92\times10^{-3}$\\
		$f_{d11}$	& $1\times10^{-3}$	& $6.17\times10^{-3}$	&
$9.26\times10^{-3}$	& 0			& 0		& $6.17\times10^{-5}$\\
		$f_{u22}$	& $1\times10^{-3}$	& $-6.77\times10^{-3}$	&
$3.93\times10^{-3}$	& 0			& 0		& $1.0\times10^{-1}$\\
		$f_{u21}$	& $6.7\times10^{-1}$	& $-7.86\times10^{-1}$	&
$-8.56\times10^{-1}$	& 0			& 0		& $-7.86\times10^{-3}$\\
		$f_{u12}$	& $6.4\times10^{-1}$	& $2.52\times10^{-1}$	&
$-2.71\times10^{-1}$	& 0			& 0		& $2.52\times10^{-3}$\\
		$f_{u11}$	& $1\times10^{-3}$	& $8.59\times10^{-3}$	&
$-2.24\times10^{-3}$	& 0			& 0		& $8.59\times10^{-5}$\\
		$x_{d2}$	& $7.14\times10^{-4}$	& $4.04\times10^{-3}$	&
$2.35\times10^{-4}$	& $4.04\times10^{-5}$	& $4.04\times10^{-5}$&
$4.04\times10^{-5}$	\\
		$x_{d1}$	& $7.14\times10^{-4}$	& $5.11\times10^{-4}$	&
$2.96\times10^{-4}$	& $5.11\times10^{-6}$	& $5.11\times10^{-6}$	&
$5.11\times10^{-6}$\\
		$x_{u2}$	& $7.14\times10^{-4}$	& $-2.01\times10^{-3}$	&
$9.04\times10^{-4}$	& $-2.01\times10^{-5}$	& $-2.01\times10^{-5}$ &
$-2.01\times10^{-5}$ \\
		$x_{u1}$	& $7.14\times10^{-4}$	& $1.01\times10^{-3}$	&
$-2.21\times10^{-3}$	& $1.01\times10^{-5}$	& $1.01\times10^{-5}$	&
$1.01\times10^{-5}$\\
		$z_{2}$		& $1\times10^{-3}$	& $6.02\times10^{-2}$	&
$4.16\times10^{-3}$	& 0			& 0		& $6.02\times10^{-4}$\\
		$z_{1}$		& $1\times10^{-3}$	& $2.63\times10^{-3}$	&
$1.18\times10^{-2}$	& 0			& 0		& $2.63\times10^{-5}$\\
		\hline
	\end{tabular}
	\caption{Trilinear Higgs
	Yukawa couplings in the E$_6$SSM benchmarks. The couplings
$\lambda_{ijk}$ come from the terms
$\lambda_{ijk}\hat{S}_i\hat{H}_{dj}\hat{H}_{uk}$ in the superpotential. Here
$\lambda_{333}=\lambda$,
		 $\lambda_{3\alpha\beta}=\lambda_{\alpha\beta} $,
		 $\lambda_{\alpha3\beta}=f_{d\alpha\beta} $,
		 $\lambda_{\alpha\beta3}=f_{u\alpha\beta} $,
		 $\lambda_{33\alpha}=x_{d\alpha} $,
		 $\lambda_{3\alpha3}=x_{u\alpha} $, and
		 $\lambda_{\alpha33}=z_{\alpha} $.}
	\label{tab:e6-yukawa}
\end{table}

\section{Model Implementation}
To scan parameter spaces of models and generate Monte Carlo events the models have to be transferred from paper to computer. There are various ways of doing this. Many implementations of the MSSM have already been created, but for the E$_6$SSM there are no available sources. We have chosen to use the software package \verb+LanHEP+ \cite{Semenov2010} to calculate the Feynman rules for the E$_6$SSM. \verb+LanHEP+ finds the interactions and mass mixings between the particle states in the model and writes an output which can be read by a Feynman diagram calculator or Monte Carlo event generator. \verb+LanHEP+ can output in several formats. We have been using the format used by \verb+CalcHEP+ since that is the software we are using for calculating Feynman diagrams and generating events.

In the \verb+LanHEP+ implementation of the model the particle content and Lagrangian is specified. We have used a slightly stripped down version of the E$_6$SSM, suitable for our purposes. What is not included from the three families of 27 representations of the $E_6$ group are the exotic coloured states~\footnote{These can be diquarks or leptoquarks depending on the model definition
.}, their superpartners, and the inert Higgs doublets and SM-singlets, from the two first families. However, one should note that we are including the superpartners of these inert Higgs and singlet states which, as described in section \ref{matrices}, extends the neutralino and chargino sectors. The diagonalisation of large mass matrices appearing in the model, e.g.~ the $12\times12$ neutralino mass matrix, is performed with routines available from the \verb+SLHAplus+ \cite{Belanger2010} package which is well integrated with \verb+LanHEP+ and \verb+CalcHEP+.

The $Z_2^H$ violating Yukawa couplings, $x_u$, $x_d$, and $z$, connecting the inert neutralino sector with the USSM sector, have been included in the \verb+LanHEP+ model. Turning on these couplings causes the neutralino mass matrix to leave its block-diagonal form and acquire non-zero off-block-diagonal elements.
The model is parameterised such that the dimensionless input parameters are the Yukawa couplings $\lambda_{ijk}$ from the $\hat{S}\hat{H}_u\hat{H}_d$-terms  in the superpotential, the ratio of the Higgs doublet VEVs, $\tan \beta$, and the gauge couplings. Dimensionful input parameters of the model are the third generation soft trilinear scalar A-couplings, the soft masses of the squarks and sleptons, and the soft gaugino masses at the electro-weak scale. One should note that we use the notation of the physical gluino mass, $m_{\tilde g}$
(\verb+MGo+) instead of $M_3$.
The soft trilinear coupling $A_\lambda$ associated with the $SH_uH_d$-term is exchanged for the pseudo-scalar Higgs mass $M_A$.

The details and notations of the \verb+CalcHEP+ model \verb+E6SSM-12.02+ described above can be found in Appendix \ref{ap:model} and the model files are accessible from the High Energy Physics Model DataBase (HEPMDB)\cite{HEPMDB}.

\section{Gluino Production and Decays in the MSSM and E$_6$SSM}

The most important processes for supersymmetry searches at hadron colliders are
the production of gluinos and squarks, {provided that they are not much heavier than charginos and neutralinos}. In this paper we consider the case where
all of the squarks are heavier than the gluino,
which is motivated by the GUT constrained E$_6$SSM as discussed in section \ref{sec:param},
which makes the pair-production of gluinos the most attractive
process for E$_6$SSM search and discovery. 

\subsection{Production cross-sections} 

\begin{figure}[htb]
\centering
\includegraphics[width=.8\linewidth]{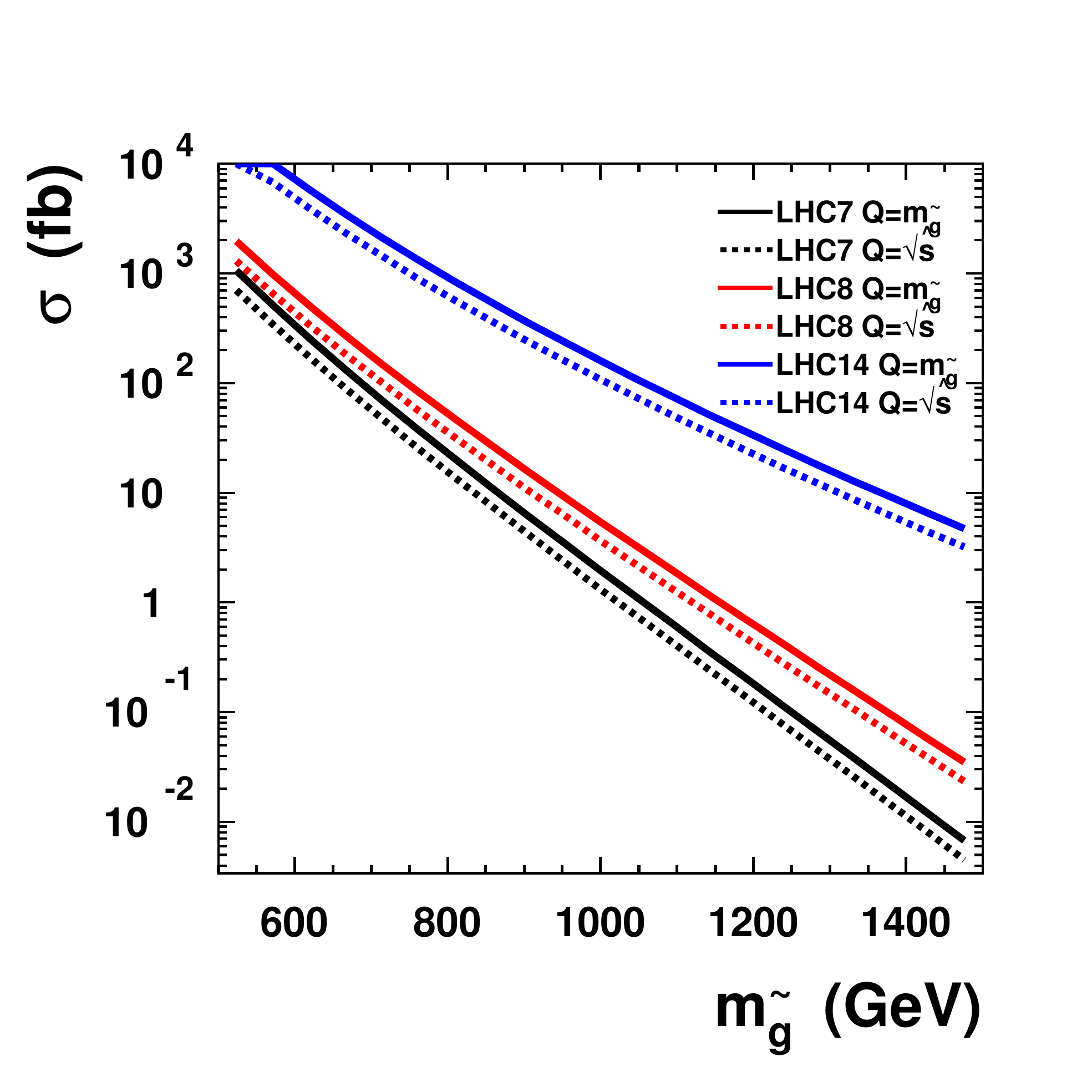}
\caption{The tree-level cross section for gluino pair production as a function of the gluino mass, $m_{\tilde{g}}$. The solid (or dashed) lines represent, from bottom to top, the LHC at 7 TeV (black), 8 TeV (red), and 14 TeV (blue). The CTEQ6LL set is used for PDFs. The QCD scale, $Q$, is set to the gluino mass, $Q=m_{\tilde g}$, for the solid lines and to the centre of mass energy, $Q=\sqrt{\hat s}$, for the dashed lines. 
The scale dependence of the cross section is an effect of the uncertainty of the leading order calculation. 
Including NLO corrections is known to bring up the cross section by at least a factor 2 \cite{Beenakker:1996ch}, so we are underestimating the production rate for gluinos slightly with this leading order calculation.}
\label{fig:gluinoproduction}
\end{figure}

The tree-level cross section for gluino pair-production at the LHC at 7, 8, and 14~TeV is shown as a function of the gluino mass in Fig.~\ref{fig:gluinoproduction}. The CTEQ6LL \cite{Pumplin2002} PDFs are used and the cross section is evaluated at the QCD scales  $Q=\sqrt{\hat s}$ and
$Q=m_{\tilde g}$~\footnote{Both renormalisation and factorisation scales were chosen to be equal.}.

In Fig.~\ref{fig:gluinoproduction} one can see a large scale dependence of the cross section due to the uncertainty in the leading order calculation, which is substantially reduced at NLO level
\cite{Beenakker:1996ch,Kulesza:2009kq,Beenakker:2009ha,Beenakker:2011fu}.
At $Q=m_{\tilde g}$, for which the cross section is about 50\%
larger than at  $Q=\sqrt{s}$ at tree level, the product of 
NLO and NLL K-factors is in the 2.5--5  range for $\sqrt{s}=7$~TeV for  the 500--1500~GeV mass range \cite{Beenakker:2011fu}.
To be on the conservative side we use LO cross sections evaluated at $Q=m_{\tilde g}$
in our analysis.
For our benchmarks with a gluino mass of 800~GeV, 
the production cross sections are 20.6~fb, 47.5~fb, and 839~fb for
the $\sqrt{s}=7$, 8 and 14~TeV respectively. So, at the present $\sqrt{s}=8$~TeV  LHC energy 
and expected  integrated luminosity of 20 fb$^{-1}$ by the end of 2012
about 1000 gluino pairs would have been produced.

\subsection{Signatures and distributions}

Since the E$_6$SSM introduces new neutralinos, naturally lighter than the MSSM-like LSP, the gluino decay chains will be longer than in the MSSM in general. This is confirmed and illustrated by the parameter scans in 
Fig.~\ref{fig:scan} and the benchmarks defined in Tab.~\ref{tab:bm-table}. 
In order to study the LHC phenomenology of gluino cascade decays of the  
E$_6$SSM, and the MSSM for comparison, we have performed  
Monte Carlo analyses using the
\texttt{CalcHEP} \cite{calchep} package with CTEQ6LL \cite{Pumplin2002} PDFs.
With the exception of the multi-jet analysis in section \ref{sec:6jet8}, we restrict ourselves to a parton-level analysis. We do however take into account a realistic electromagnetic energy resolution, given by
$0.15/\sqrt{E \mbox{(GeV})}$, typical for the ATLAS and CMS detectors, as well
as their typical hadronic energy resolution of $0.5/\sqrt{E\mbox{(GeV)}}$
and perform the respective Gaussian smearing for leptons and quarks. We define leptons (jets) by requiring $p_T>10$ GeV ($20$ GeV) and $|\eta|<2.5$ (4.5) and a lepton isolation of $\Delta R(\mathrm{lepton,jet})>0.5$.

\begin{figure}[htb]
\includegraphics[width=.48\linewidth]{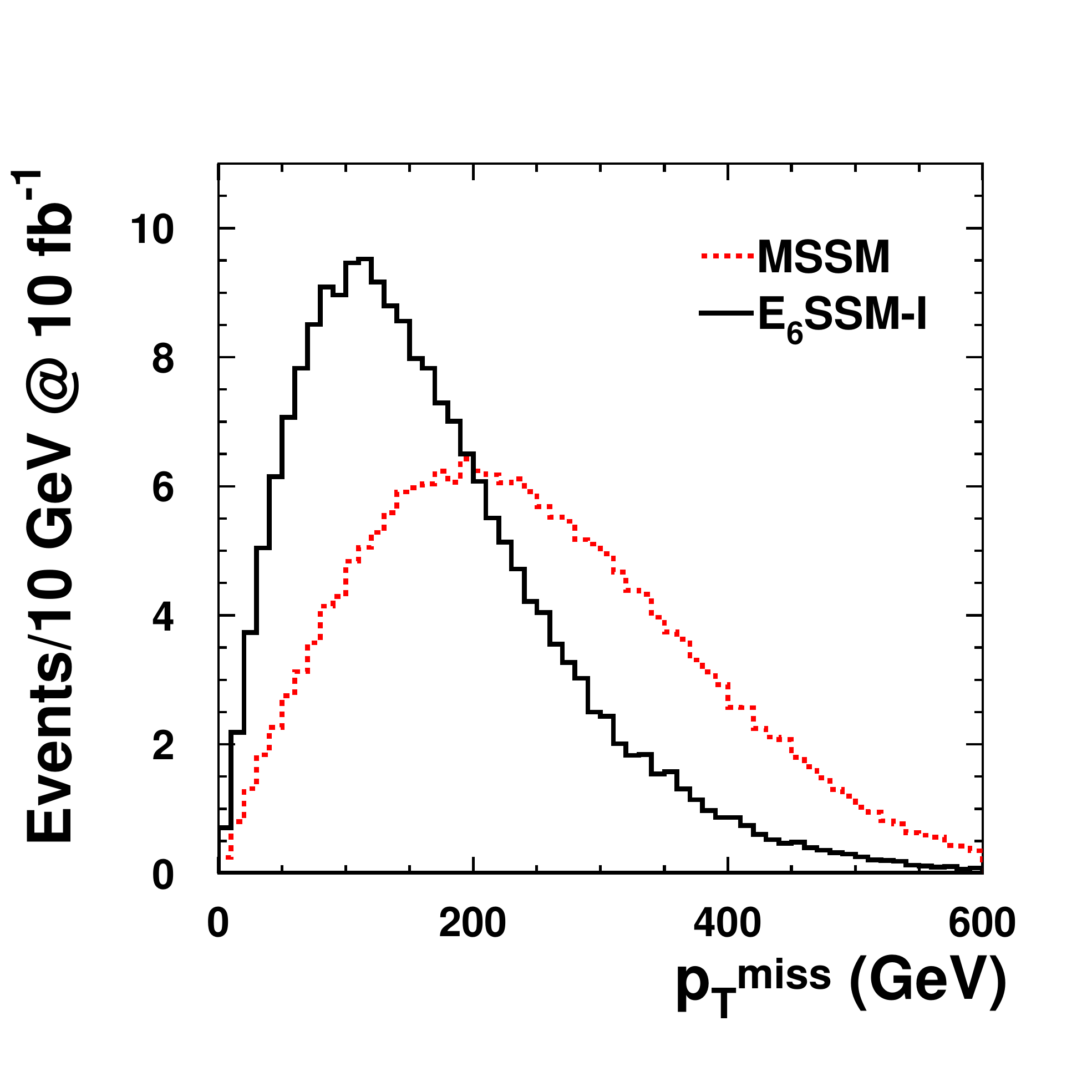}%
\includegraphics[width=.48\linewidth]{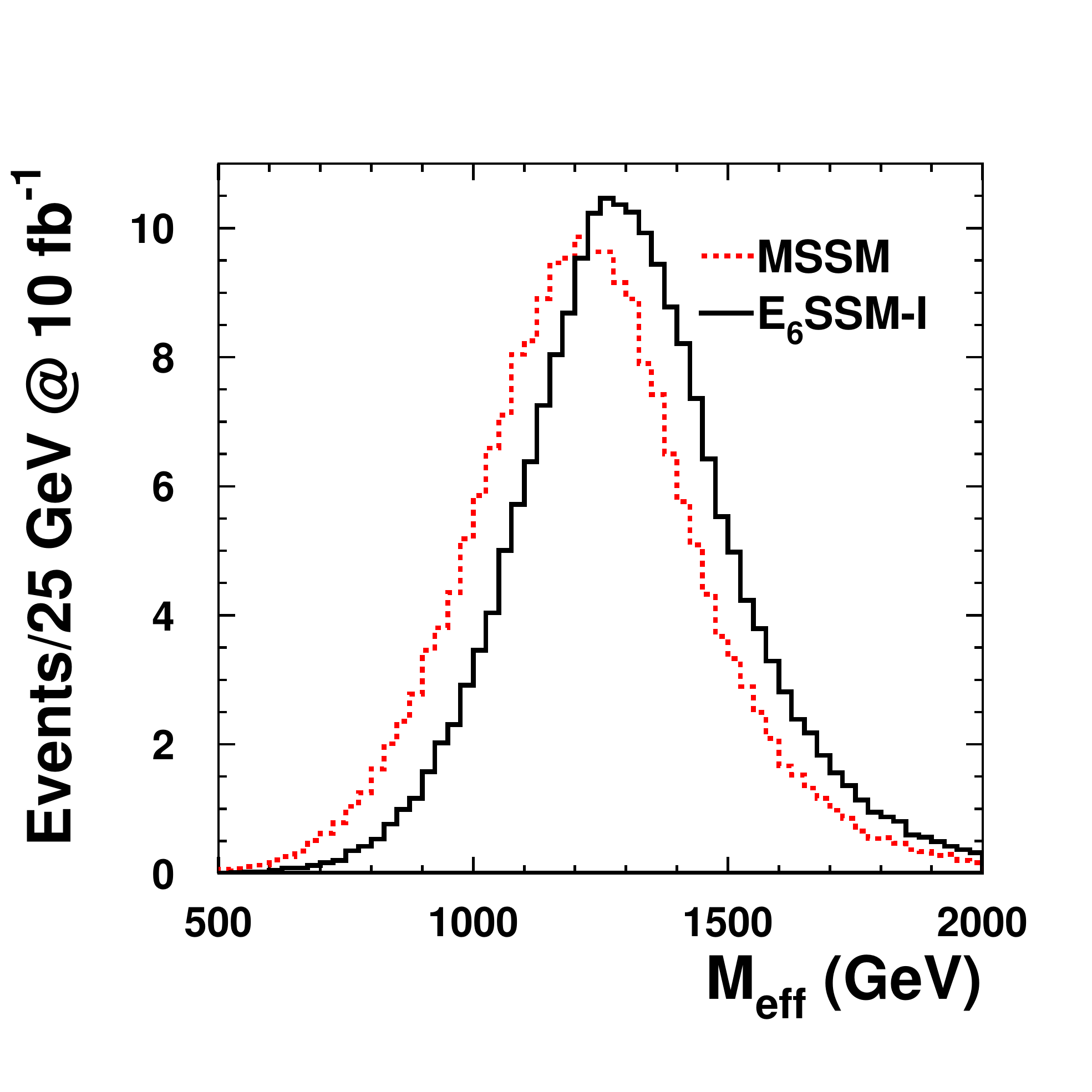}
	\caption{Missing transverse momentum (left) and the effective mass (right) before cuts for the MSSM and E$_6$SSM-I benchmark with $m_{\tilde g}=800$~GeV at $\sqrt{s}=7$~TeV. The E$_6$SSM predicts significantly less missing transverse momentum and slightly larger effective mass compared to the MSSM. The longer gluino decay chains of the E$_6$SSM, with a lighter LSP in the end, provide less missing and more visible transverse momentum. The effective mass does not distinguish the features of these models since it is a sum of visible and missing transverse momenta.}
	\label{fig:ptmiss}
\end{figure}

The longer decay chains of gluinos lead to less missing momentum, $p_T^{\mathrm{miss}}$,
and larger effective mass $M_{\mathrm{eff}}=p_T^{\mathrm{miss}} + \sum_{\mathrm{visible}}|p_T^{\mathrm{visible}}|$, 
as measured in the detector as one can see in 
Fig.~\ref{fig:ptmiss} which presents the respective distributions
for our main benchmarks (MSSM and E$_6$SSM-I). 
Although the $p_T^{\mathrm{miss}}$ distribution 
is quite different,
one should note that the effective mass distribution is not significantly different between the models. 
This happens because the effect from the suppressed  missing momentum in the case of the E$_6$SSM is partially  canceled by the effect of the increase of visible momentum, due to the longer gluino cascade decay.
There is a slight overall increase of the effective mass due to the fact that visible momenta are added up as magnitudes while the missing momentum is a vectorial sum. The reduced amount of missing transverse momentum in the E$_6$SSM makes it less discoverable, compared to the MSSM, in typical SUSY searches which focus on all-hadronic events with large missing momentum.

\begin{figure}[htb]
	\includegraphics[width=.48\linewidth]{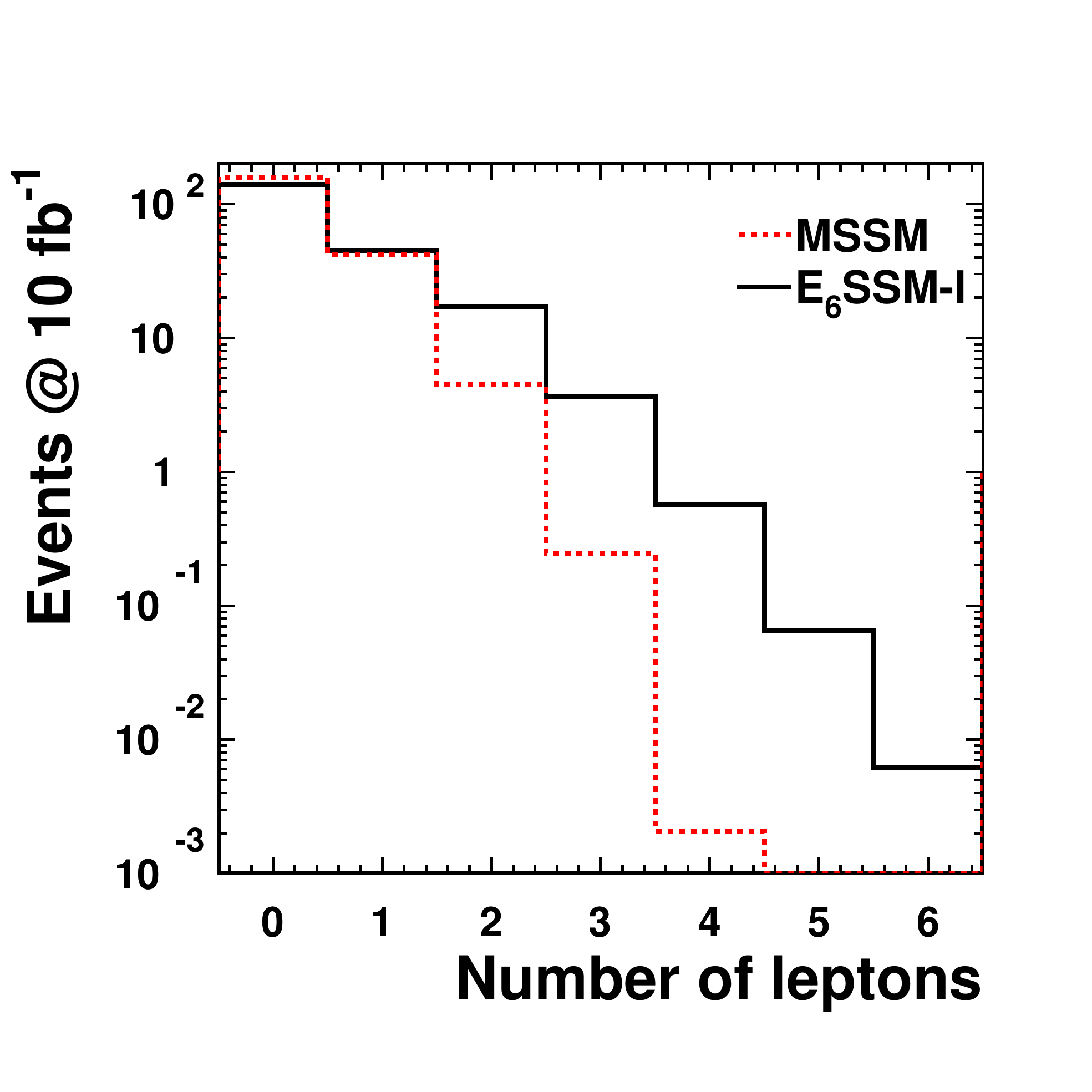}%
	\includegraphics[width=.48\linewidth]{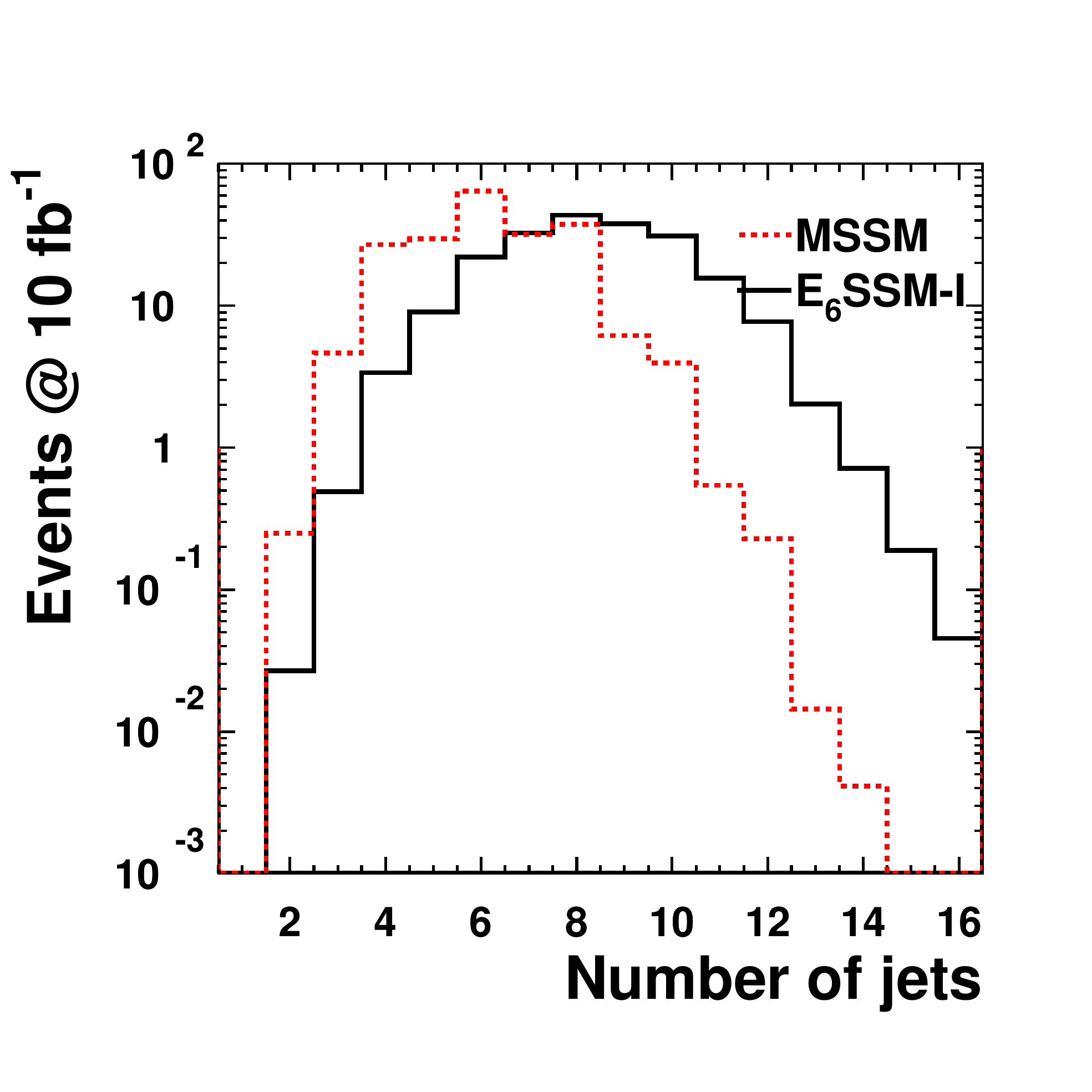}
	\caption{Lepton multiplicity (left) and jet multiplicity (right), requiring $p_T>10$ GeV, $|\eta|<2.5$, and $\Delta R(\mathrm{lepton,jet})>0.5$ for leptons and $p_T>20$ GeV and $|\eta|<4.5$ for jets. The benchmarks considered are the MSSM and E$_6$SSM-I as presented in Tab.~\ref{tab:bm-table} with $m_{\tilde g}=800$~GeV. The LHC setup is used with $\sqrt{s}=7$~TeV and normalised to
	10~fb$^{-1}$ of integrated luminosity. Due to the longer gluino decay chains of the E$_6$SSM it predicts many more visible particles in collider experiments, both leptons and jets. This suggests that searches for E$_6$SSM gluinos should be more favourable in multi-lepton and multi-jet searches.}
	\label{fig:multiplicity}
\end{figure}

Another important feature of the long decay chains of the E$_6$SSM is the increase in lepton as well as jet multiplicity, as shown in Fig.\ \ref{fig:multiplicity}, again for the benchmarks MSSM and E$_6$SSM-I. This feature allows us to rely on multi-lepton requirements for background reduction rather than cuts on missing energy. There is a significant loss of statistics by using this strategy, however it turns out to be a very important channel for discovery of gluinos with long decay chains and indeed a channel in which the E$_6$SSM is largely dominant compared to the MSSM. This makes the multi-lepton channels essential for distinguishing the models.

\subsection{Searches at $\sqrt{s}=7$ TeV LHC}
There has not been any indications of SUSY from the LHC during its run at $\sqrt{s}=7$ TeV. We have investigated different SUSY search channels at this energy to understand the status of our benchmarks and what limits can be put on the E$_6$SSM and which channels we expect to be the most favourable for discovery and distinguishing the models. We compare our signals with published backgrounds used by CMS and ATLAS at this energy. We have scaled all the channels to an integrated luminosity of 10 fb$^{-1}$ for comparison, which is approximately the amount of 7 TeV data acquired by the two experiments. Benchmarks with an 800 GeV gluino mass are considered here. 

\subsubsection{0 leptons} 
The long gluino cascade decays with less missing momentum would be less visible in  the main SUSY searches based on jets and missing energy (see e.g.\ \cite{Aad:2011ib} and \cite{CMS:2011-jets}) which provide the best statistics and strongest exclusions for MSSM. In these searches the E$_6$SSM parameter space
is less constrained as compared to the MSSM and the
acquired exclusions do not hold for this model. 
The main reason for this is the hard cuts on missing energy and 
its ratio to the effective mass since the distributions for these variables are significantly different for MSSM versus E$_6$SSM as we demonstrate here.

The effective mass distribution for our benchmarks for an 800~GeV gluino mass is plotted on the top of the backgrounds from ATLAS and CMS in 
Figs.~\ref{fig:meff-ATLAS-bg-1fb} and \ref{fig:meff-CMS-bg-1fb} after all cuts have been applied except for the final selection cut on the effective mass itself. The signal from E$_6$SSM is suppressed as compared to the MSSM and, more importantly, both benchmarks are well below the background, 
illustrating the difficulty of discovering SUSY at the 7 TeV LHC in the case where the gluino mass is around 800 GeV, assuming the squarks to be much heavier.

\newcommand{\picturewidth}{0.38}
\begin{figure}[ht]
\centering
\subfigure[0 leptons, 4 jets: Backgrounds from ATLAS \cite{Aad:2011ib}.]{
	\includegraphics[width=\picturewidth\linewidth]{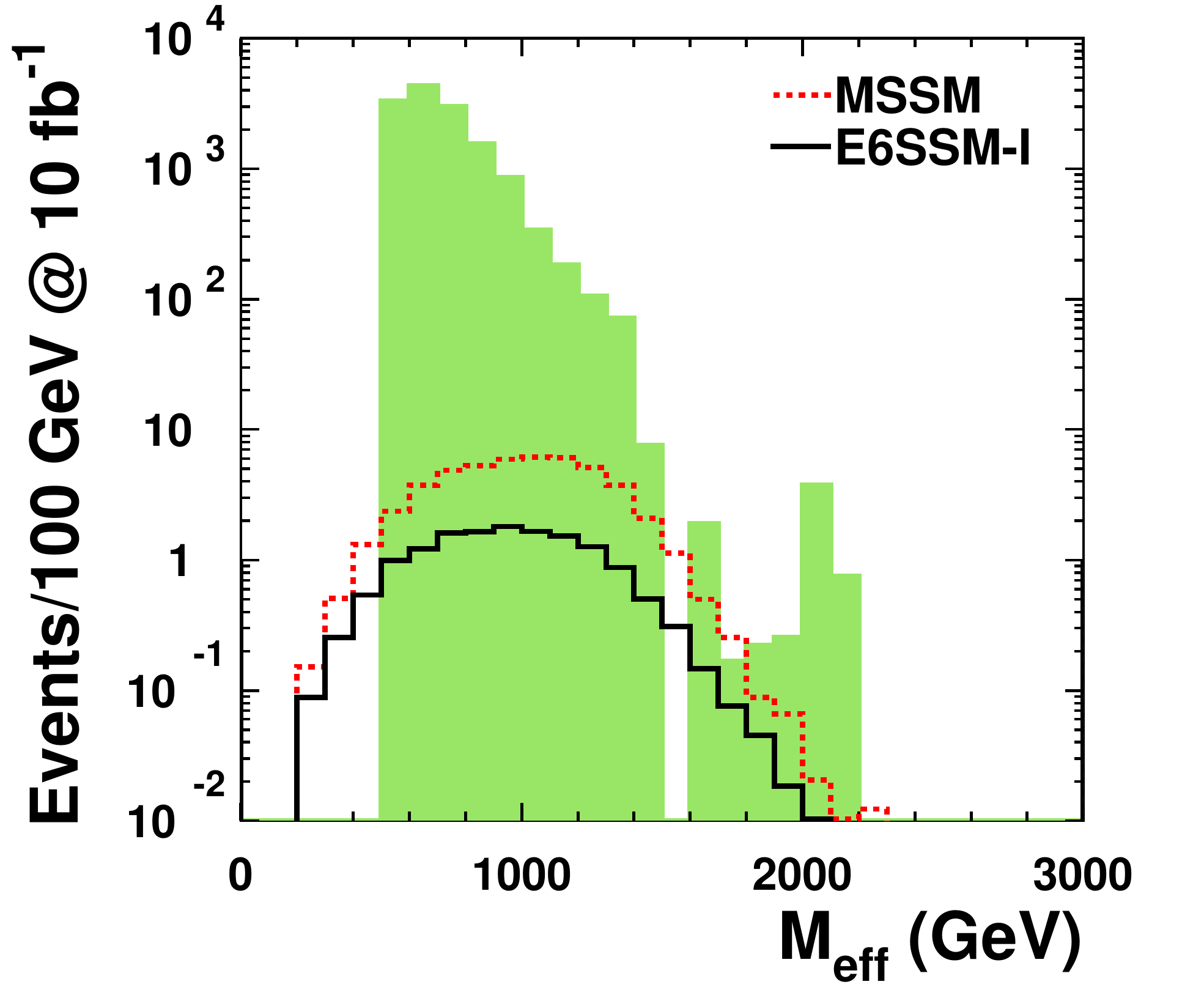}
\label{fig:meff-ATLAS-bg-1fb}
}
\subfigure[0 leptons, 3 jets: Backgrounds from CMS\cite{CMS:2011-jets}. ]{
	\includegraphics[width=\picturewidth\linewidth]{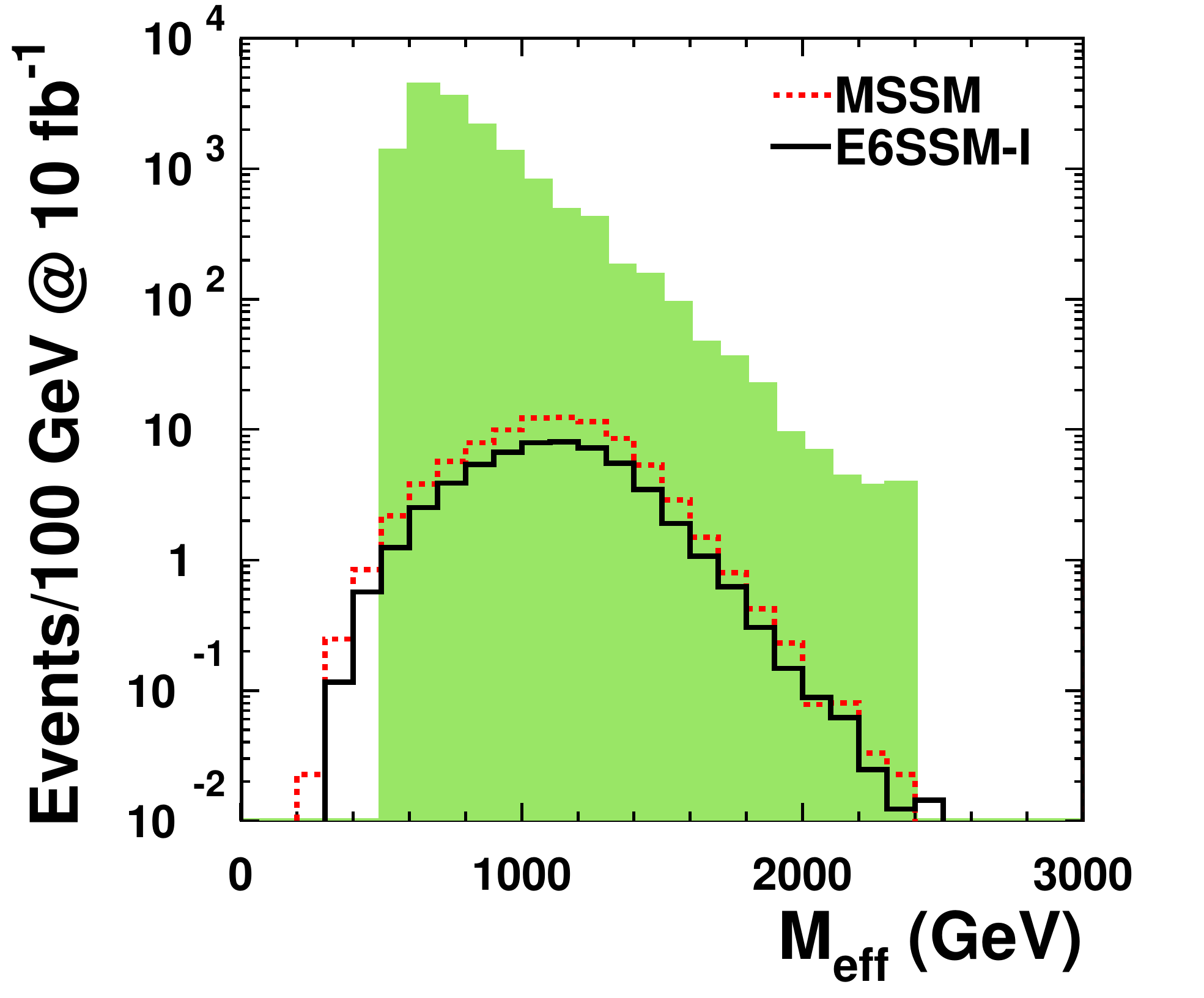}
\label{fig:meff-CMS-bg-1fb}
}
\subfigure[1 muon, 4 jets:  Backgrounds from ATLAS \cite{ATLAS:2011ad}]{
	\includegraphics[width=\picturewidth\linewidth]{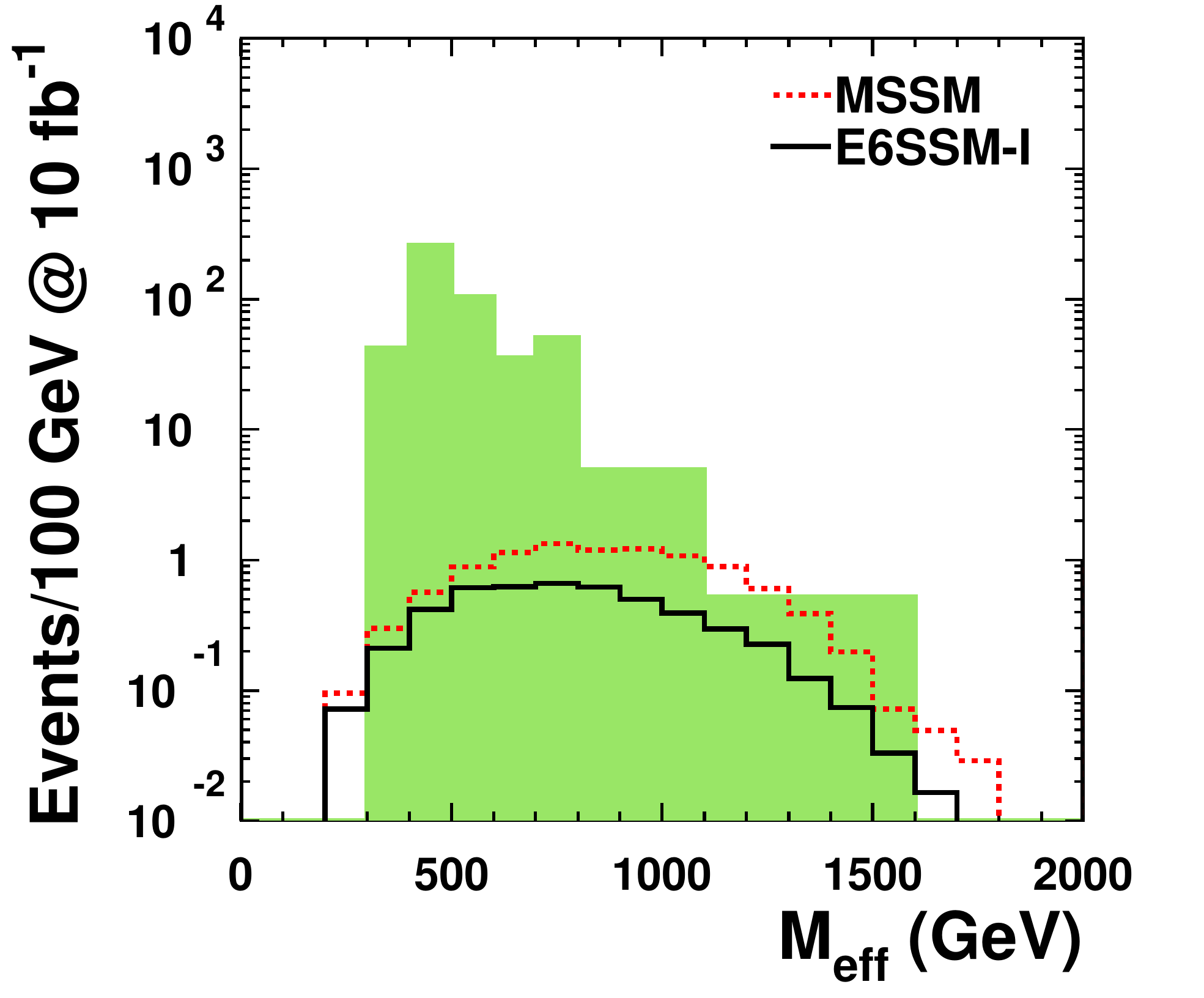}
\label{fig:meff-atlas-muon}
}
\subfigure[2SS leptons: SS-SR2. Backgrounds from ATLAS  \cite{Aad:2011cwa}]{
	\includegraphics[width=\picturewidth\linewidth]{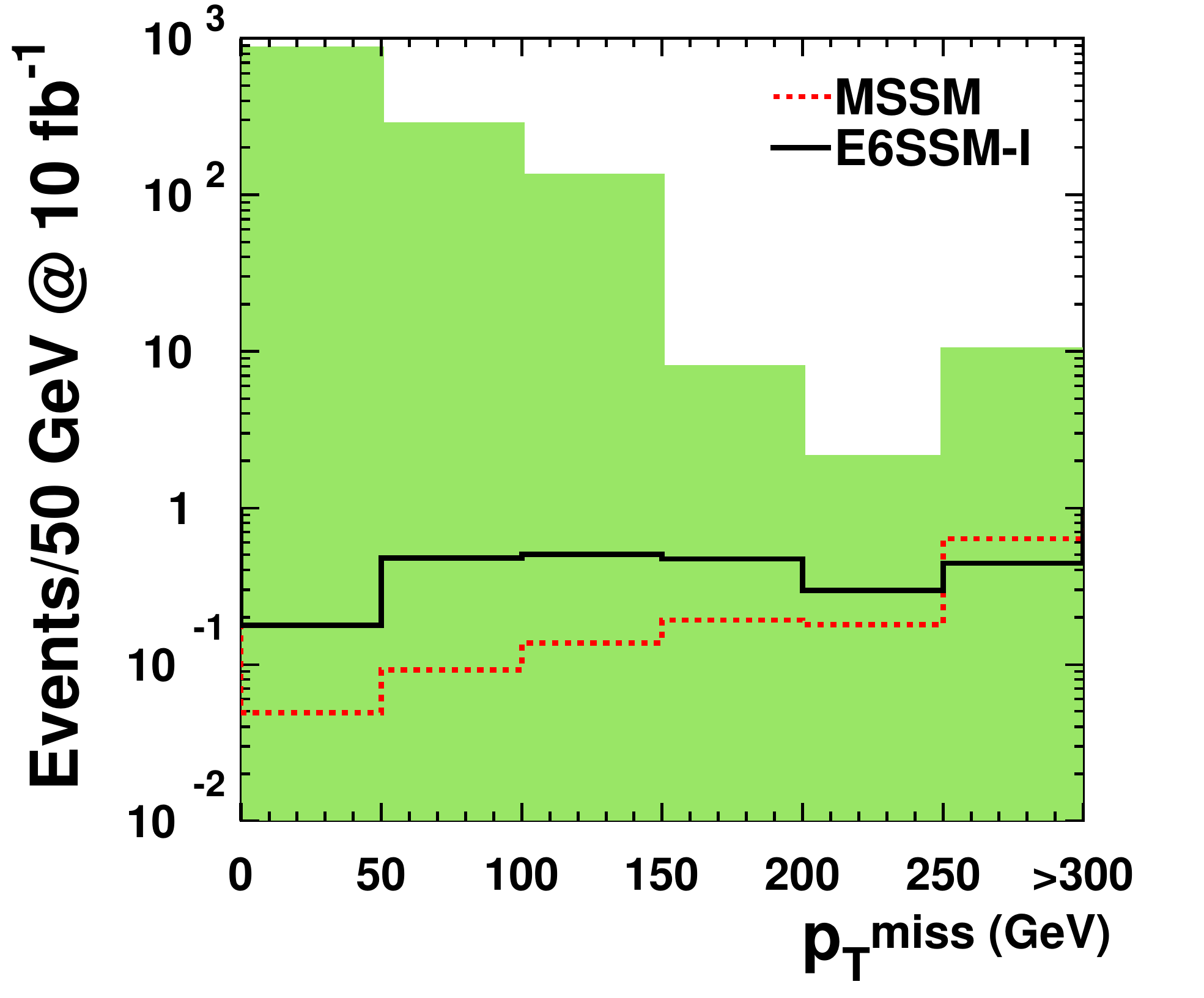}
\label{fig:ptmiss-atlas-2sslep}
}
\subfigure[3 leptons: Backgrounds from CMS \cite{CMS-PAS-SUS-11-013}]{
	\includegraphics[width=\picturewidth\linewidth]{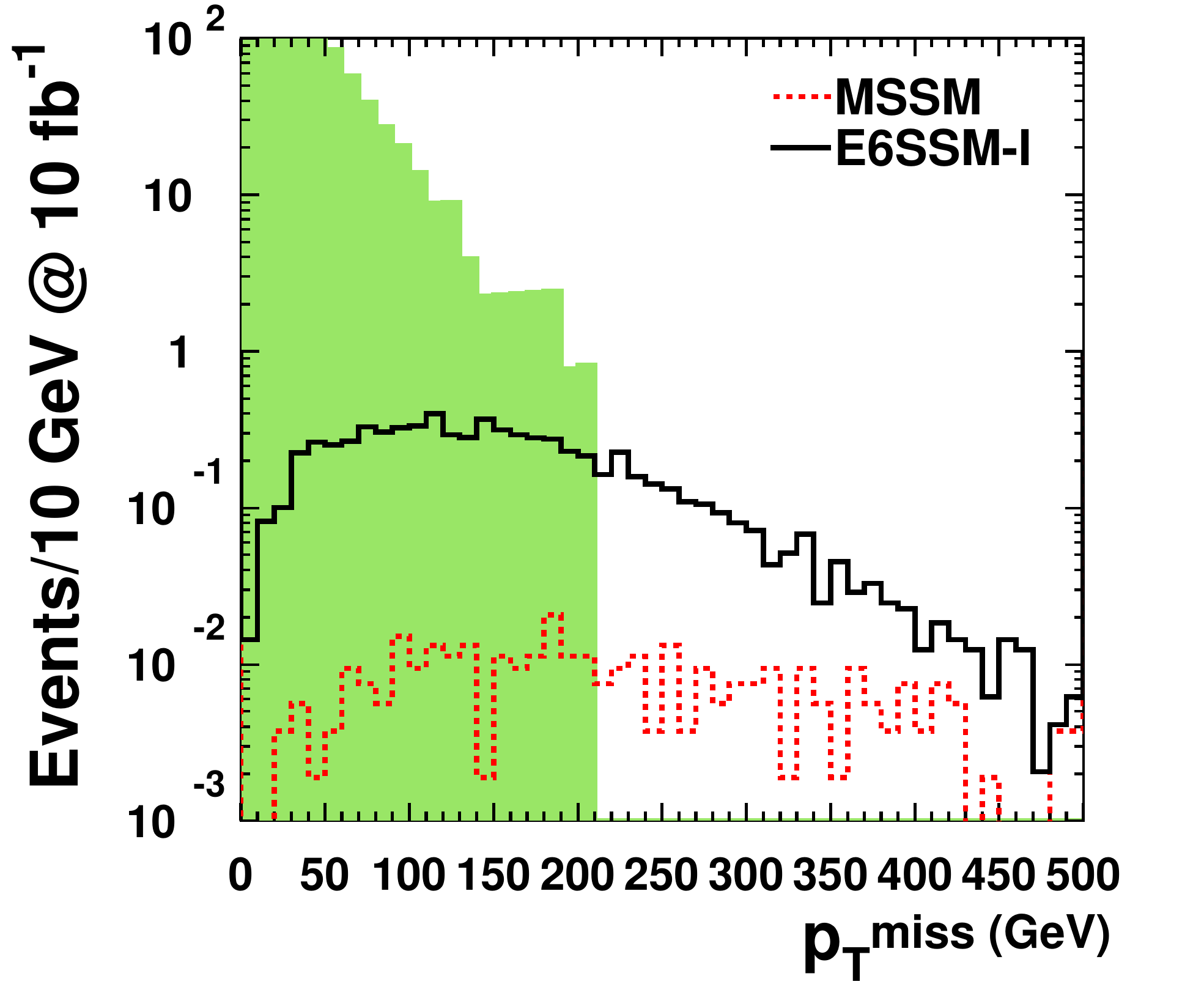}
\label{fig:ptmiss-3lep7-stolen}
}
\subfigure[4 leptons: Backgrounds from ATLAS \cite{ATLAS-CONF-2012-001}]{
	\includegraphics[width=\picturewidth\linewidth]{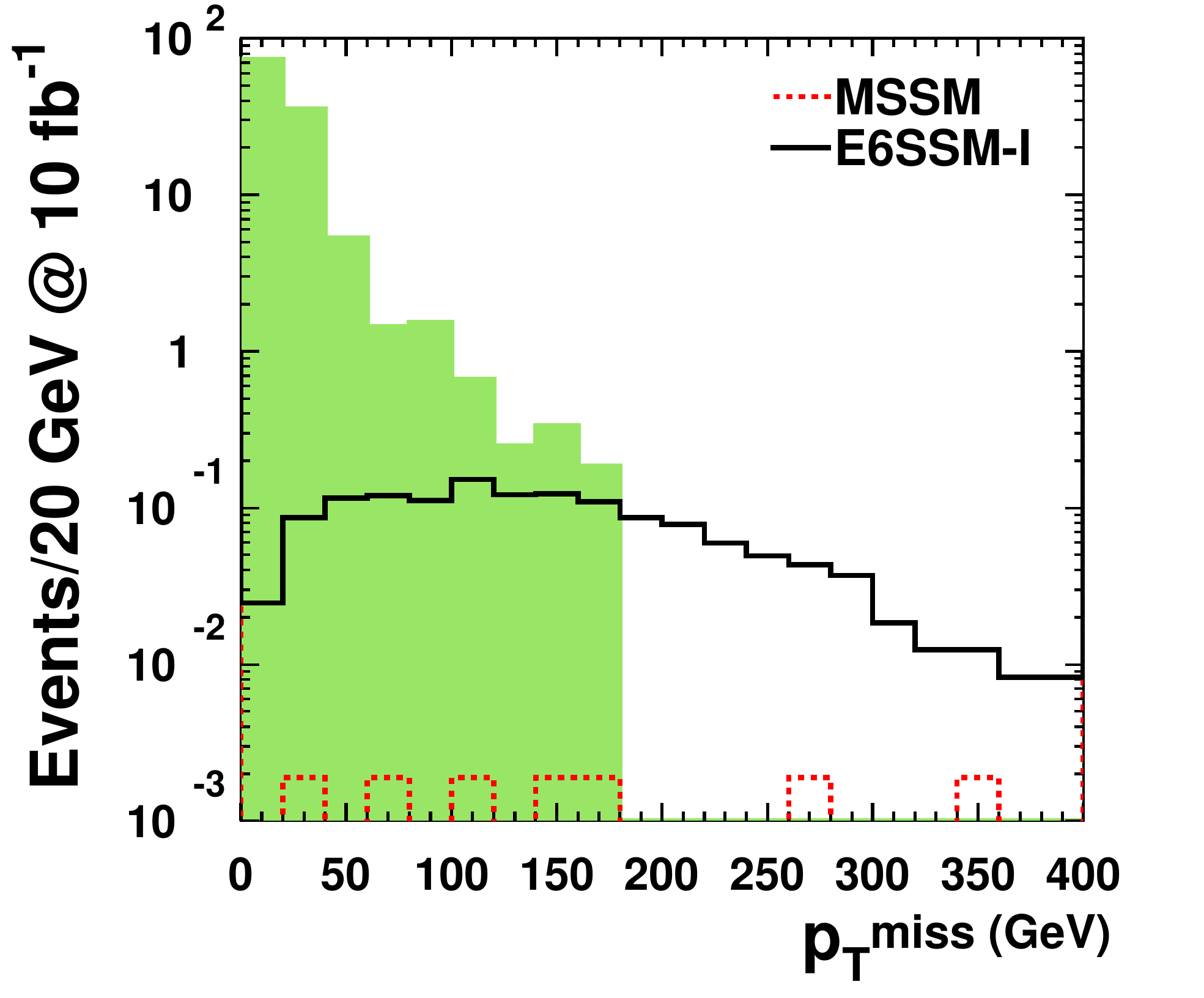}
\label{fig:ptmiss-ATLAS-4lep-stolen}
}
\caption{Distributions for published 0--4 lepton searches at 7~TeV, scaled to  10~fb$^{-1}$ for comparison. 
The signal contributions from the MSSM and E$_6$SSM-I benchmarks with  $m_{\tilde g}=800$~GeV are plotted on top of published backgrounds.
The distributions shown are after all cuts except the final selection cut on the plotted distribution.
0 lepton searches (with jets$\leq$4) give bad background suppression and favour MSSM. Requiring two or more leptons makes MSSM more suppressed. 
For multi-lepton searches the signal to background ratio for the E$_6$SSM is better but the signal statistics is low.
}
\label{fig:meff-bg}
\end{figure}
Even though the $0$-lepton signature with a jet multiplicity of about four is not favoured for E$_6$SSM hunting, the $0$-lepton signature
for this model could be still interesting
for the cases of larger jet-multiplicity. 
For multi-jet channels, analyses beyond the parton level are essential.
We discuss this in detail for the case of $\sqrt{s}=8$~TeV in section~\ref{sec:6jet8}
where we perform one example of a beyond-the-parton level
analysis. Apart from this, we restrict ourselves here by the parton level analysis
and in particular we shall focus on the tri-lepton signature.

\subsubsection{1--2 leptons}
The selection of events with leptons provides easy triggering and efficient background suppression at the cost of worse statistics. To exemplify this we compare
the signal distributions for
the benchmarks versus the backgrounds for two ATLAS searches, a single lepton search in Fig.~\ref{fig:meff-atlas-muon} and a two same-sign lepton search in Fig.~\ref{fig:ptmiss-atlas-2sslep}. One can see that the signal-to-background ratio is
better in these leptonic searches compared to the all-hadronic searches. 
In the effective mass distribution of the 1 muon channel from ATLAS's 1 lepton plus 4 jets search shown in  Fig.~\ref{fig:meff-atlas-muon} one can see that the signal level is not extremely far below the background level in the high effective mass region. The E$_6$SSM signal is still suppressed as compared to the MSSM in this one lepton search, however, considering the $p_T^{\mathrm{miss}}$ distribution for the two same-sign lepton search by ATLAS in Fig.~\ref{fig:ptmiss-atlas-2sslep}, one sees how the E$_6$SSM signal overtakes the MSSM benchmark's by requiring one more lepton. This is due to the fact that two same-sign leptons in the final state become more likely in the E$_6$SSM than in the MSSM, simply because it predicts more leptons in general. 
Even though the E$_6$SSM signal has got stronger than the MSSM signal in this 2SS channel compared to the one lepton channel, the signal-to-background ratio now looks worse. This is because the choice of using the missing momentum instead of the effective mass as the variable to define the signal region is not favourable for the E$_6$SSM since the respective signal is very vulnerable to hard cuts on this variable. Also for higher multiplicity searches, using the effective mass to define the signal region allows for the signal-to-background ratio to be improved as we will show below.

\subsubsection{3--4 leptons}
Requiring additional leptons makes the statistics even worse, 
but it allows the signal to appear above the background enough 
to allow a reasonable signal significance in order to test the
models under study.
Comparison of the benchmark signals with the background from a 3 lepton search by CMS is shown in Fig.~\ref{fig:ptmiss-3lep7-stolen} and from a 4 lepton search by ATLAS in Fig.~\ref{fig:ptmiss-ATLAS-4lep-stolen}. The three lepton channel  defined by the cuts
\begin{align}
	\begin{split}
	p_T(l_1) &> 20 \mbox{ GeV}\\
	p_T(l_2) &> 10 \mbox{ GeV}\\
	p_T(l_3) &> 10 \mbox{ GeV}\\
	\end{split}
	\label{3lepcut}
 \end{align}
 where $l=\mu$ or $e$ with $|\eta|<2.5$, and $\Delta R(\mathrm{lepton,jet})>0.5$,
seems promising, showing a possible excess in the high missing transverse momentum region. The background used by CMS is not evaluated for large enough missing transverse momentum however. To explore the large missing transverse momentum region and to expand the analysis further we have produced backgrounds for this channel, including several processes, using \verb+CalcHEP+.

The dominant backgrounds come from $ZWj$ and $t\bar t V$. Other important contributions come from $ZW$ and $t\bar t$. We also considered backgrounds such as $ZWjj$, $ZZ$ and $ZZZ$   which we found to be subleading. {Our background predictions at $\sqrt{s}=7$~TeV agree well with backgrounds used in the multi-lepton searches by CMS \cite{CMS-PAS-SUS-11-013} and ATLAS \cite{ATLAS-CONF-2012-001}.}
They only major difference is in the very low end of the $p_T^{\mbox{\footnotesize miss}}$ distribution where the \verb+CalcHEP+ generated backgrounds are suppressed. 

This difference in the transverse missing momentum distribution 
between our results and the results from the full detector simulation
occurs because we do not simulate any source of instrumental missing energy in our analysis. 
However, this difference does not affect the our results 
since we are not using  missing  $p_T$ information directly, similar to approach of \cite{baer2008}, and, moreover, this difference 
effectively vanishes after the cut is applied on the effective mass variable as used at last stage of our analysis.

We would like to stress that parton level analysis
are quite accurate to the tri-lepton signature we study in this paper.
Signatures with lower lepton multiplicity require considering QCD backgrounds
from  jets faking leptons and instrumental defects which could affect the low $p
_T^{\mathrm{miss}}$ region. This background  is difficult to take into account at the parton-level event generator and even 
at the level  fast detector simulation. Therefore analysis of signatures with the lepton
multiplicites below three are outside of the scope of our paper.
In case we  compare signal versus background for these signatures for the illustration purposes
we chose to rely on published backgrounds for those channels in this paper.

\begin{figure}[h!]
\centering
\subfigure[ ]{
	\includegraphics[width=.45\linewidth]{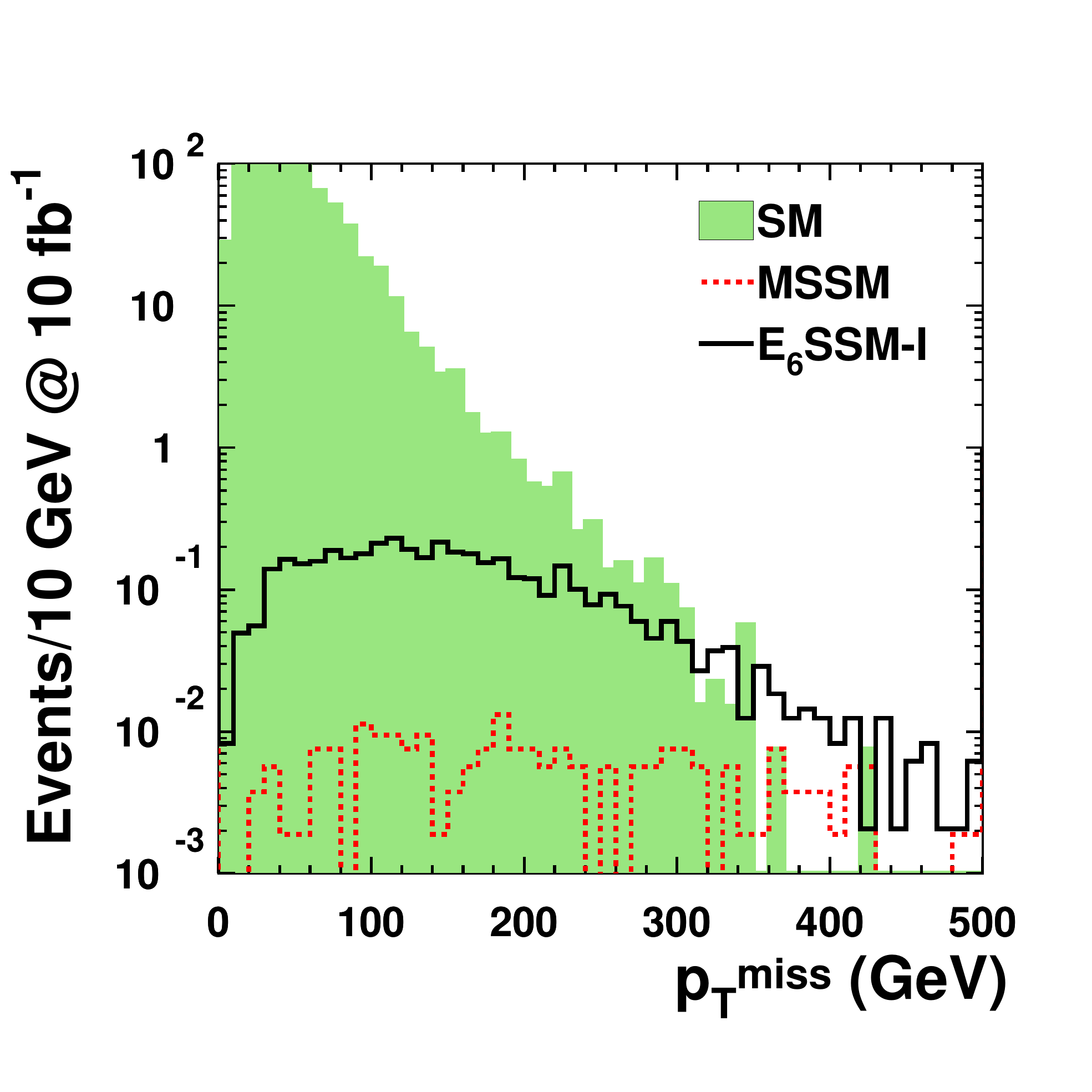}
\label{fig:ptmiss-3lep7}
}
\subfigure[]{
	\includegraphics[width=.45\linewidth]{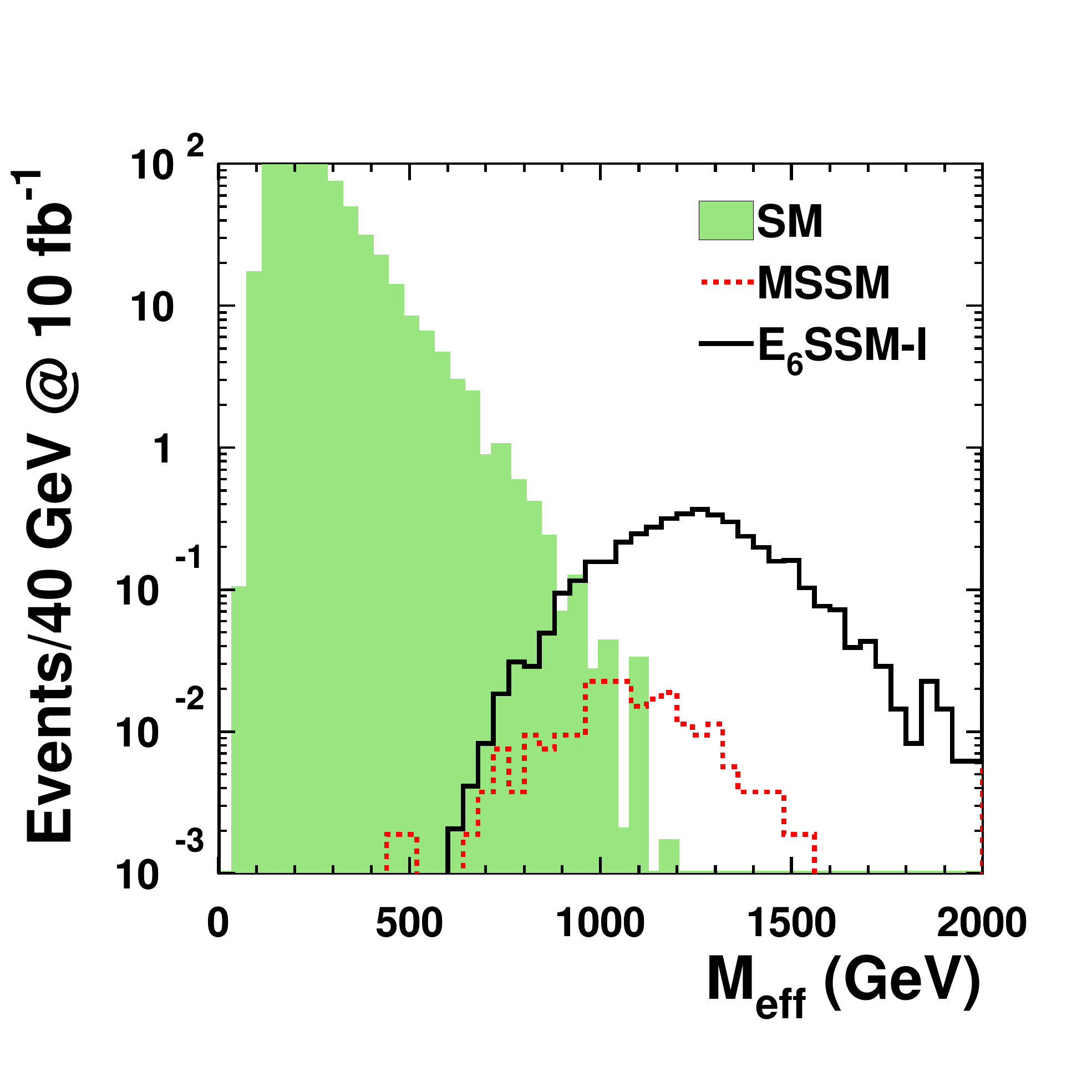}
\label{fig:meff-3lep7}
}
\caption{{Distributions for $p_T^{\mathrm{miss}}$ (a) and $M_{\mbox{\tiny eff}}$ (b) at the LHC with 10~fb$^{-1}$ at 
$\sqrt{s}=7$ TeV after requiring at least three leptons. The benchmarks both have a gluino mass $m_{\tilde g}=800$~GeV.
Backgrounds have been generated by
CalcHEP
and agree well with published ones such the ones by CMS, shown in Fig.~\ref{fig:ptmiss-3lep7-stolen}. The E$_6$SSM-I benchmark is shown to present a signal larger than the background for large missing momentum (a) even though it is a model that predicts quite small amounts of missing momentum.
The signal presents itself more strongly in the effective mass variable where there is no need for a cut on the missing transverse momentum.
}
}
\label{fig:3lep7}
\end{figure}

The result for the $p_T^{\mbox{\footnotesize miss}}$ distribution is shown in Fig.~\ref{fig:ptmiss-3lep7}. One can see that the E$_6$SSM signal is now at the same level as the background and maybe a little higher for large $p_T^{\mbox{\footnotesize miss}}$. 
If one instead considers the effective mass as a selection variable
for the three lepton channel,
the situation looks much more promising, at least for the E$_6$SSM. This can be seen from the effective mass distribution presented in Fig.~\ref{fig:meff-3lep7}.
CMS has not been using the effective mass to define the signal region for this channel but uses the missing transverse energy and the  hadronic transverse energy instead. Our way of defining the signal region by the effective mass is on the other hand much closer to the way presented by ATLAS in \cite{ATLAS-CONF-2012-001} or as suggested in \cite{baer2008}.
A cut on $M_{\mbox{\tiny eff}}$ at 950~GeV gives $S=11.5$ signal events for the 800 GeV gluino mass E$_6$SSM-I benchmark and $B=0.4$ background events, providing an expected $5.7\sigma$ excess at 10~fb$^{-1}$, using the definition of statistical significance $S_{12}=2(\sqrt{S+B}-\sqrt{B})$ valid for small statistics \cite{Bartsch:824351,Bityukov2007}. 

\subsection{Searches at $\sqrt{s}=8$ TeV LHC}

\subsubsection{6 jets}
\begin{figure}[h!]
	\centering
	\subfigure[]{
	\includegraphics[width=.48\linewidth]{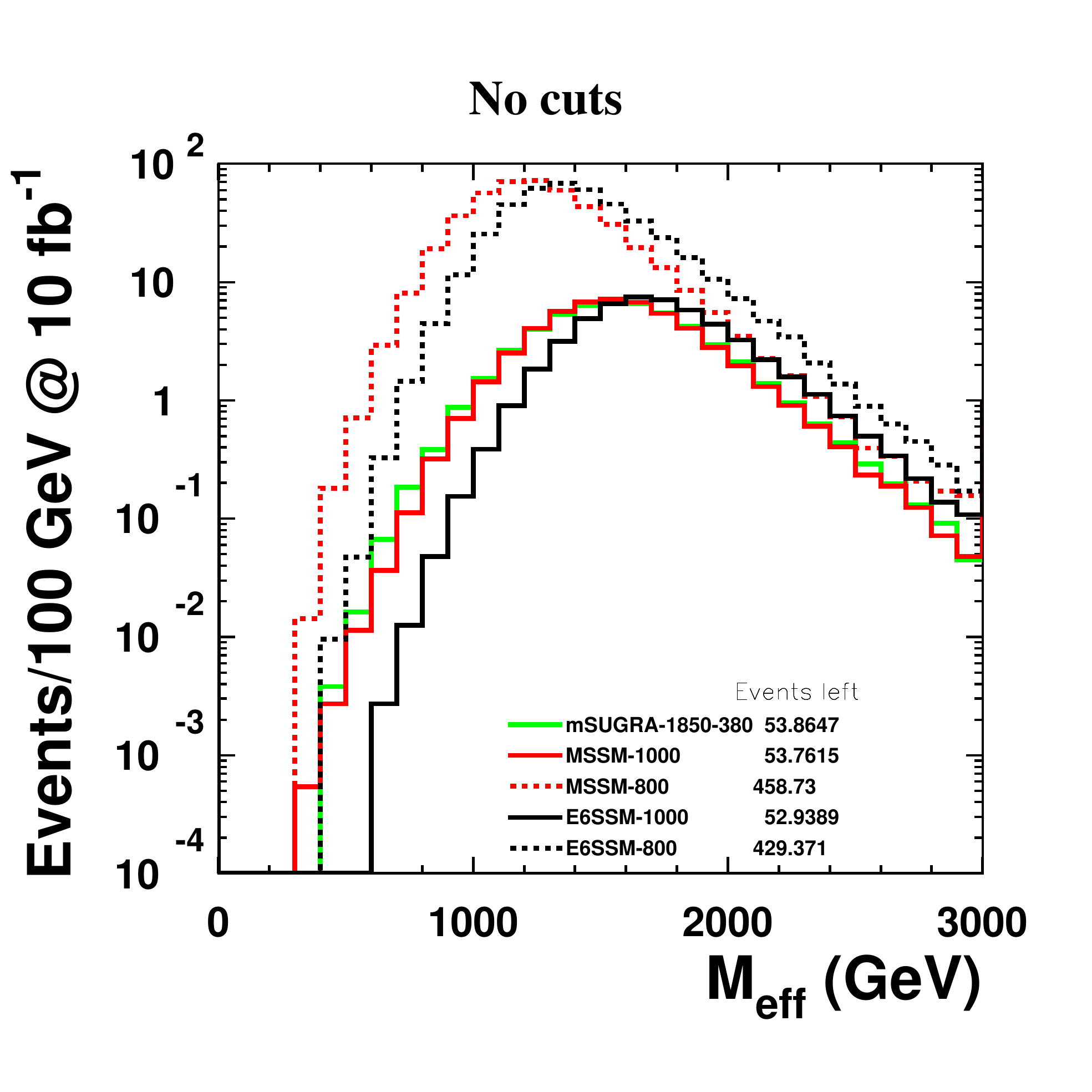}
	}
	\subfigure[]{
	\includegraphics[width=.48\linewidth]{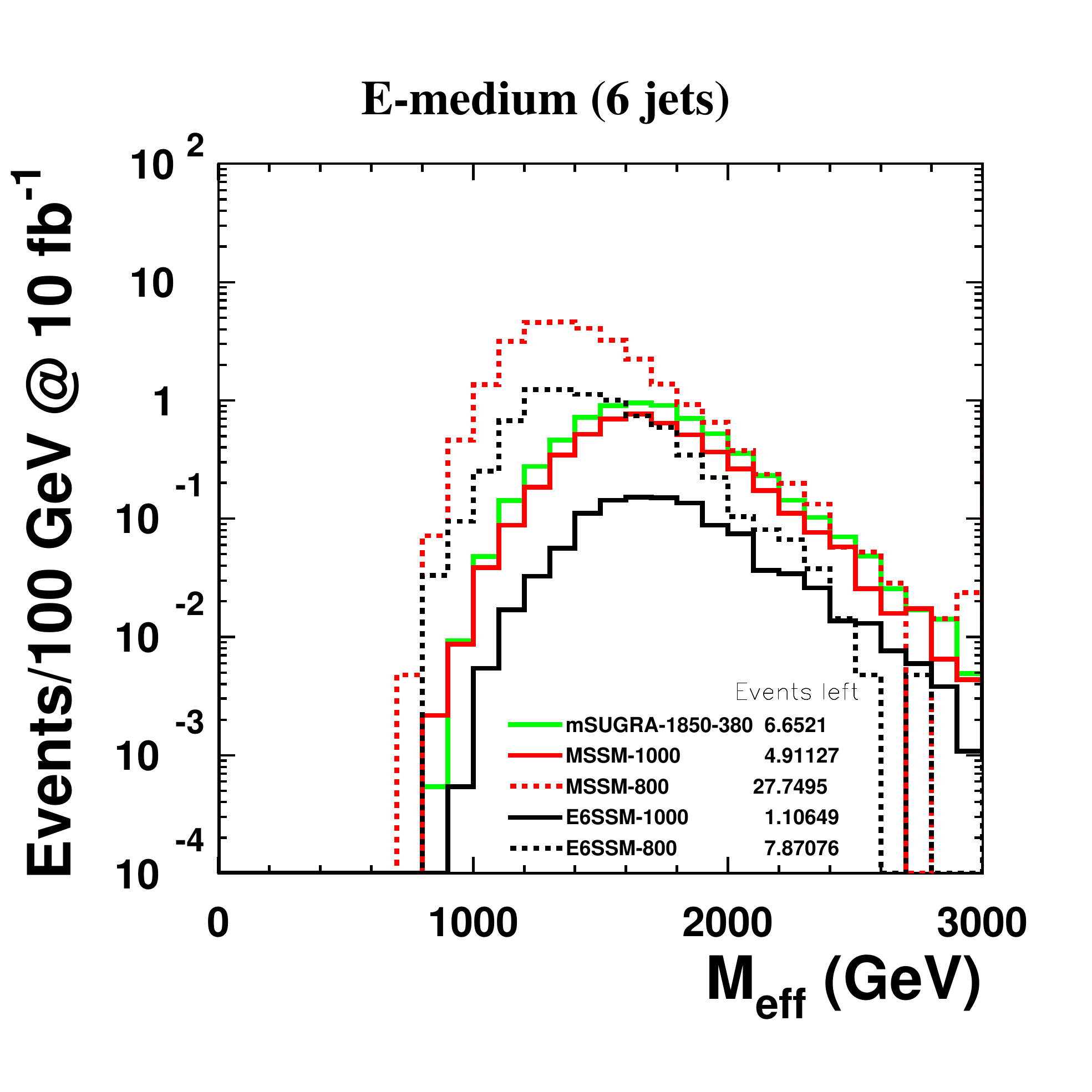}
	}
	\caption{Comparison between mSUGRA, MSSM, and E$_6$SSM benchmarks in the 6 jet channel, E-medium, used by ATLAS \cite{ATLAS-CONF-2012-109}. The effective mass distribution for the gluino signal are plotted before (a) and after (b) cuts at $\sqrt{s}=8$~TeV and 10~fb$^{-1}$ of integrated luminosity. The events left after the last signal region cut on the effective mass, $M_{\mathrm{eff}}>1300$~GeV, are given in Tab. \ref{tab:6jet8}. The benchmarks with solid lines all have a gluino mass of 1 TeV while the benchmarks with dashed lines have a gluino mass of 800~GeV. After cuts (b) the E$_6$SSM benchmark with an 800~GeV gluino mass has a distribution not very different from the mSUGRA point with a 1~TeV gluino mass.}
	\label{fig:6jet8}
\end{figure}

Since the multi-jet channels could provide good prospects of discovery of and differentiation between benchmarks we investigate the effect 
of the cuts,
\begin{align}
	\begin{split}
	E_T^{\mathrm{miss}}	&>	160 	 \mbox{ GeV}\\
	p_T(j_1)		&>	130	 \mbox{ GeV}\\
	p_T(j_2)		&>	60 	 \mbox{ GeV}\\
	p_T(j_3)		&>	60 	 \mbox{ GeV}\\
	p_T(j_4)		&>	60 	 \mbox{ GeV}\\
	p_T(j_5)		&>	60 	 \mbox{ GeV}\\
	p_T(j_6)		&>	60 	 \mbox{ GeV}\\
	\Delta\phi(\mbox{jet},p_T^{\mathrm{miss}})_{\mathrm{min}}	&>	0.4 (i=\{1,2,3\}), 0.2 (p_T>40\mbox{ GeV jets})\\
	E_T^{\mathrm{miss}}/m_{\mathrm{eff}}	&> 0.25 (6j)\\
	m_{\mathrm{eff}}			&> 1300 \mbox{ GeV},
	\end{split}
	\label{ecuts}
\end{align}
used by ATLAS \cite{ATLAS-CONF-2012-109}
for 6-jets analysis
applied to our benchmarks. To perform this kind of multi-jet analysis we are forced to go beyond our parton-level analysis. The need for a more dedicated analysis, with hadronisation and detector effects, comes mainly from the importance of initial and final state radiation that plays an important role when requiring more than four jets in the gluino decays. This is because the first four jets are well approximated by our parton level analysis since they typically are the ones that originate straight from the gluino decay. To generate the events at the level of fast detector simulation we fed events from \verb+CalcHEP+ to the \verb+PGS+ \cite{Conway} package. We also compare the signal from the MSSM and E$_6$SSM with an mSUGRA point which is excluded, but very close to the limit. The result for the effective mass distribution for the signals before and after cuts are shown in Fig.~\ref{fig:6jet8}. 

Our result, produced with \verb+CalcHEP+  and  \verb+PGS+, is in good agreement with the ATLAS result and is consistent with the experimental data.

\label{sec:6jet8}
Our analysis shows that, also in the multi-jet channels, the E$_6$SSM will be suppressed compared to the MSSM and mSUGRA models, mostly because of its small missing transverse momentum. The larger jet multiplicity of the 
E$_6$SSM at the parton level
turns to be not such a great discriminator between two models, at least for this signature,
because the initial and final state radiation allows most of the signal from the
MSSM model to pass the 6-jets selection requirement.
Even though it might still be a good discovery channel for 
different selection/analysis for
the E$_6$SSM, we note that, for the current 
selection, the E$_6$SSM signal from an 800~GeV gluino is about 
at the same level as the signal from the mSUGRA model with a 1~TeV gluino.
This is the very important point we would like to convey in this paper:
the MSSM limits are not quite applicable to the E$_6$SSM, for example 
the gluino mass limit could easily differ by about 200 GeV 
between two models in the heavy squark limit as we demonstrate above.
Therefore the status of the E$_6$SSM is quite different even from generic MSSM one 
for the current LHC analysis, which needs to be tuned in order to explore E$_6$SSM parameter space.

\begin{table}[h]
	\centering
	\begin{tabular}{lcr}
						&	$m_{\tilde g}$/TeV	& Events\\
						\hline
						\hline
		mSUGRA (1850,380)		&	1			&6.18\\
						\hline
		\multirow{2}{*}{MSSM}		&	1			&5.59\\
						&	0.8			&18.16\\
						\hline
		\multirow{2}{*}{E$_6$SSM}	&	1			&1.05\\
						&	0.8			&5.58\\
						\hline
	\end{tabular}
	\caption{The events left at 8 TeV and 10 fb$^{-1}$ for benchmarks from three models after the E-medium set of cuts, including the final cut on the effective mass, $M_{\mathrm{eff}}>1300$ GeV. Here the E$_6$SSM benchmark with a 800 GeV gluino mass is left with less events than the mSUGRA benchmark with a 1 TeV gluino mass.}
	\label{tab:6jet8}
\end{table}

\subsubsection{3 leptons}
\label{sec:3lep8}
\begin{figure}[htb]
\centering
\subfigure[$m_{\tilde g}=700$ GeV]{
	\includegraphics[width=.4\columnwidth]{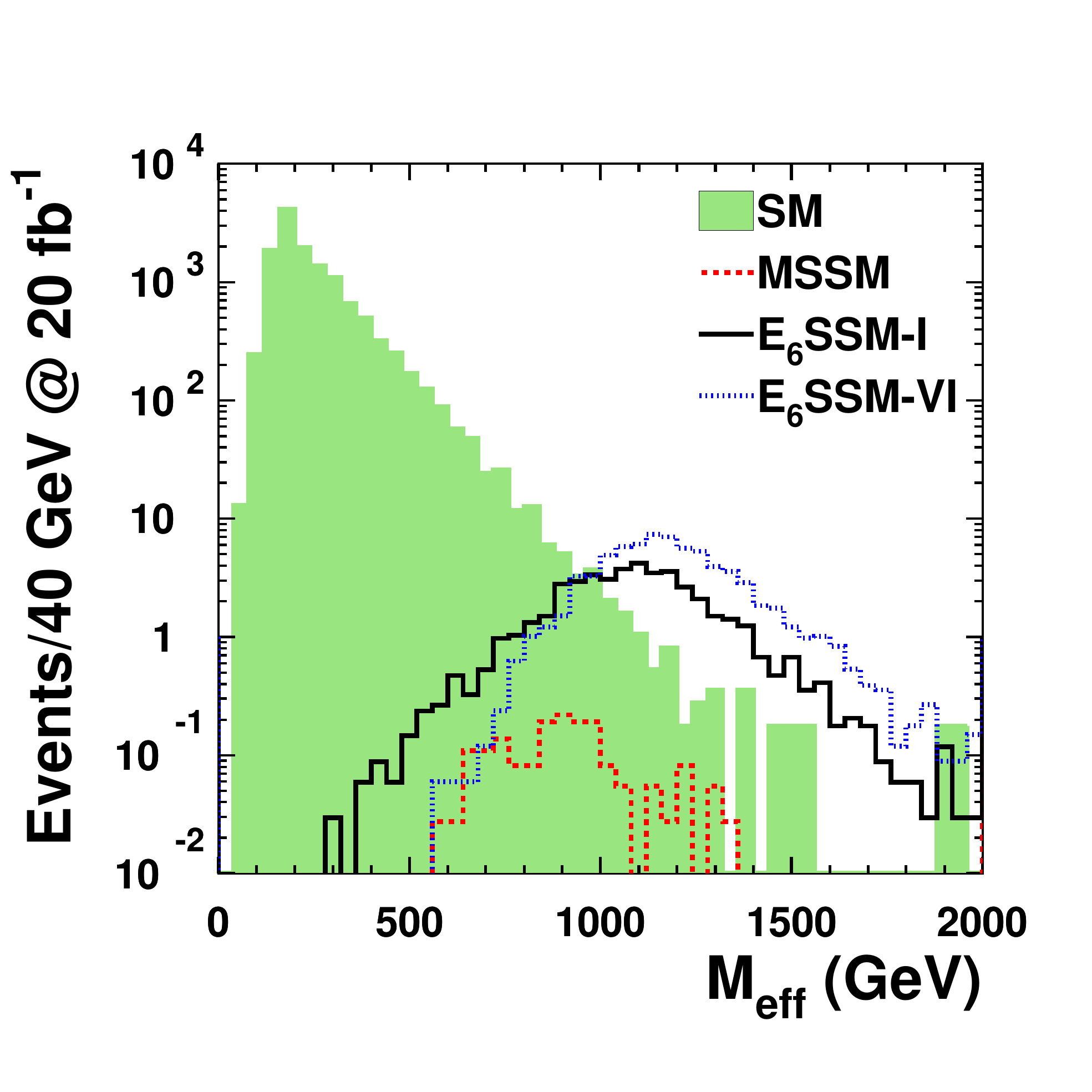}
\label{fig:meff-3lep700}
}
\subfigure[$m_{\tilde g}=800$ GeV]{

	\includegraphics[width=.4\columnwidth]{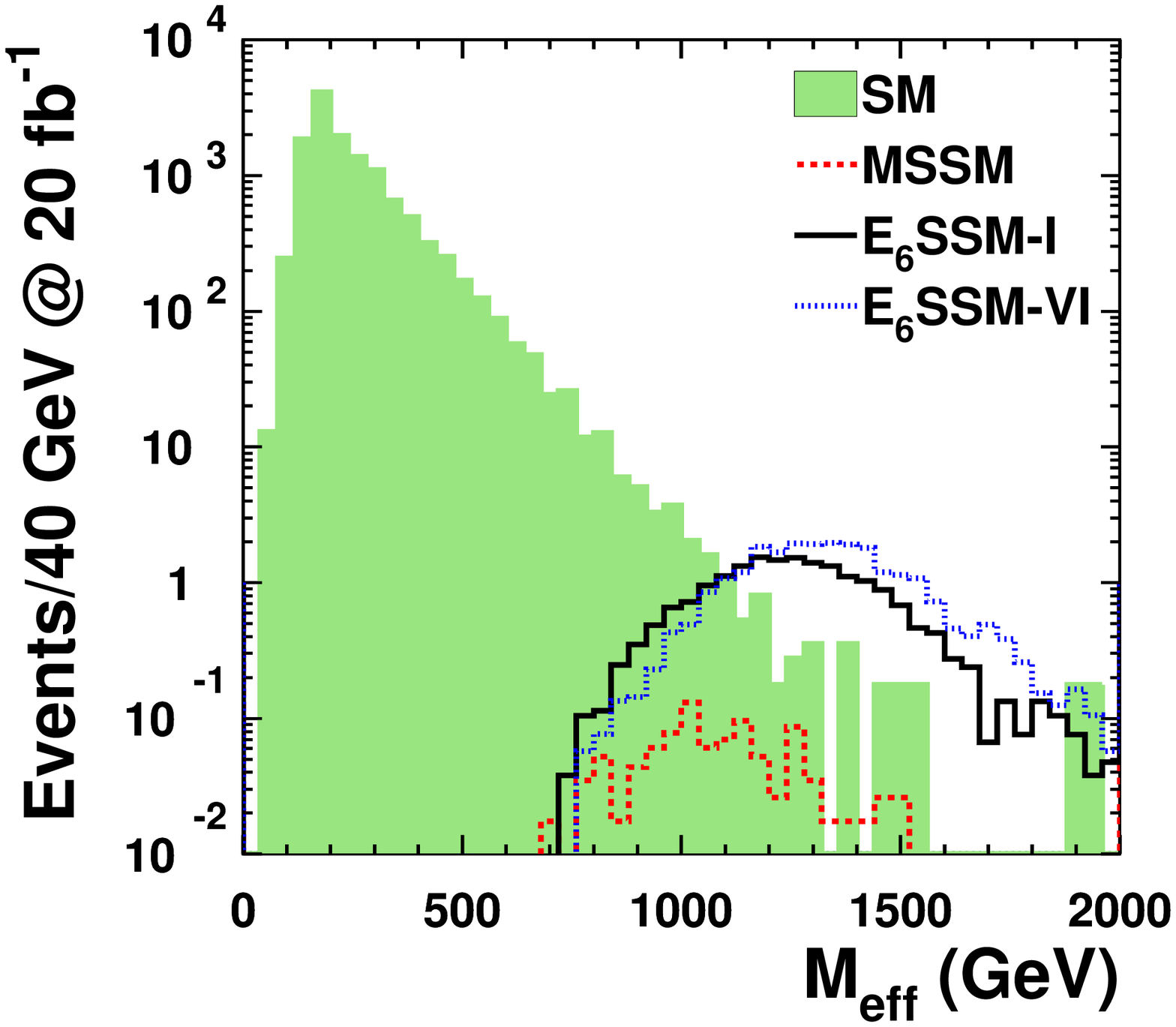}
\label{fig:meff-3lep800}
}
\subfigure[$m_{\tilde g}=900$ GeV]{

	\includegraphics[width=.4\columnwidth]{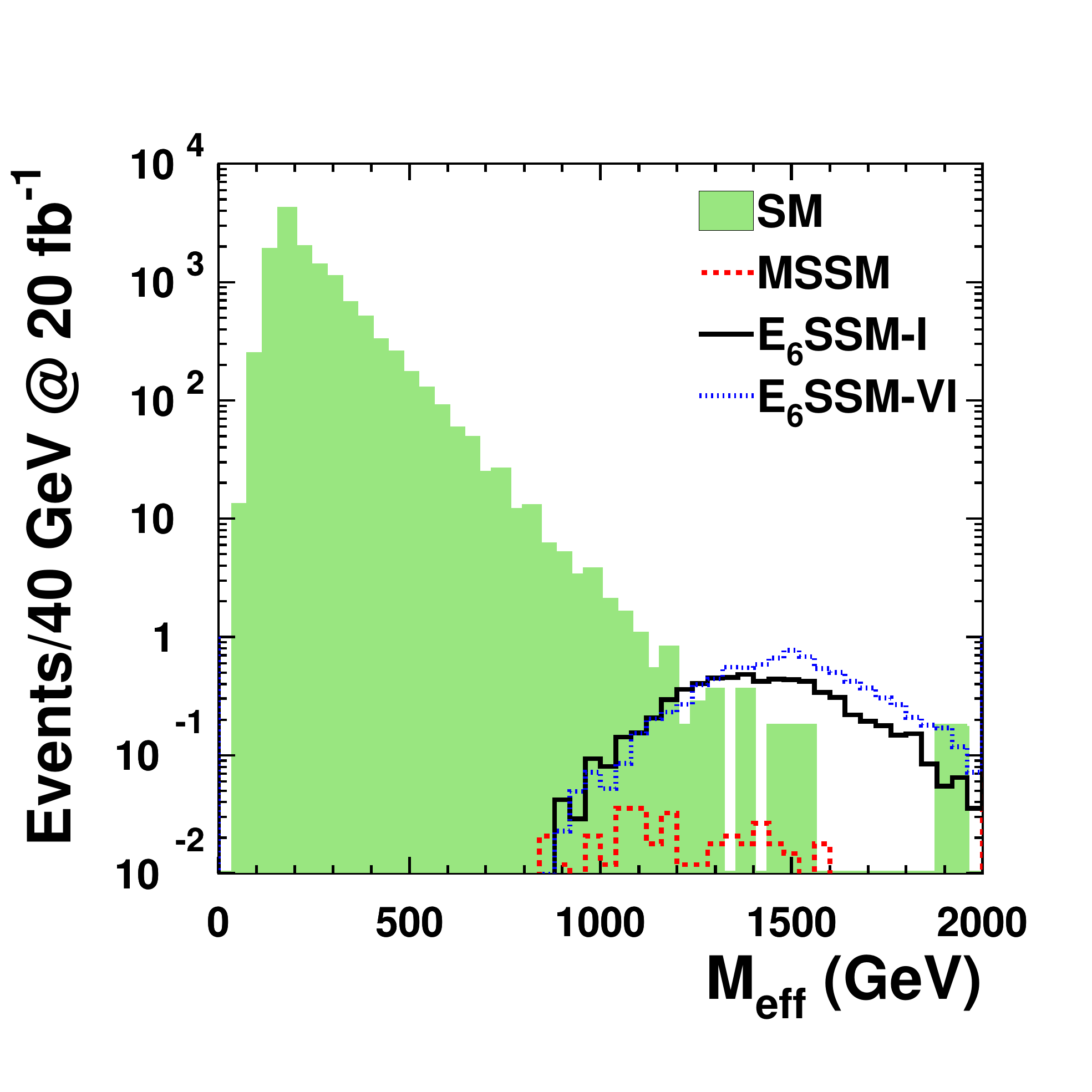}
\label{fig:meff-3lep900}
}
\subfigure[$m_{\tilde g}=1000$ GeV]{

	\includegraphics[width=.4\columnwidth]{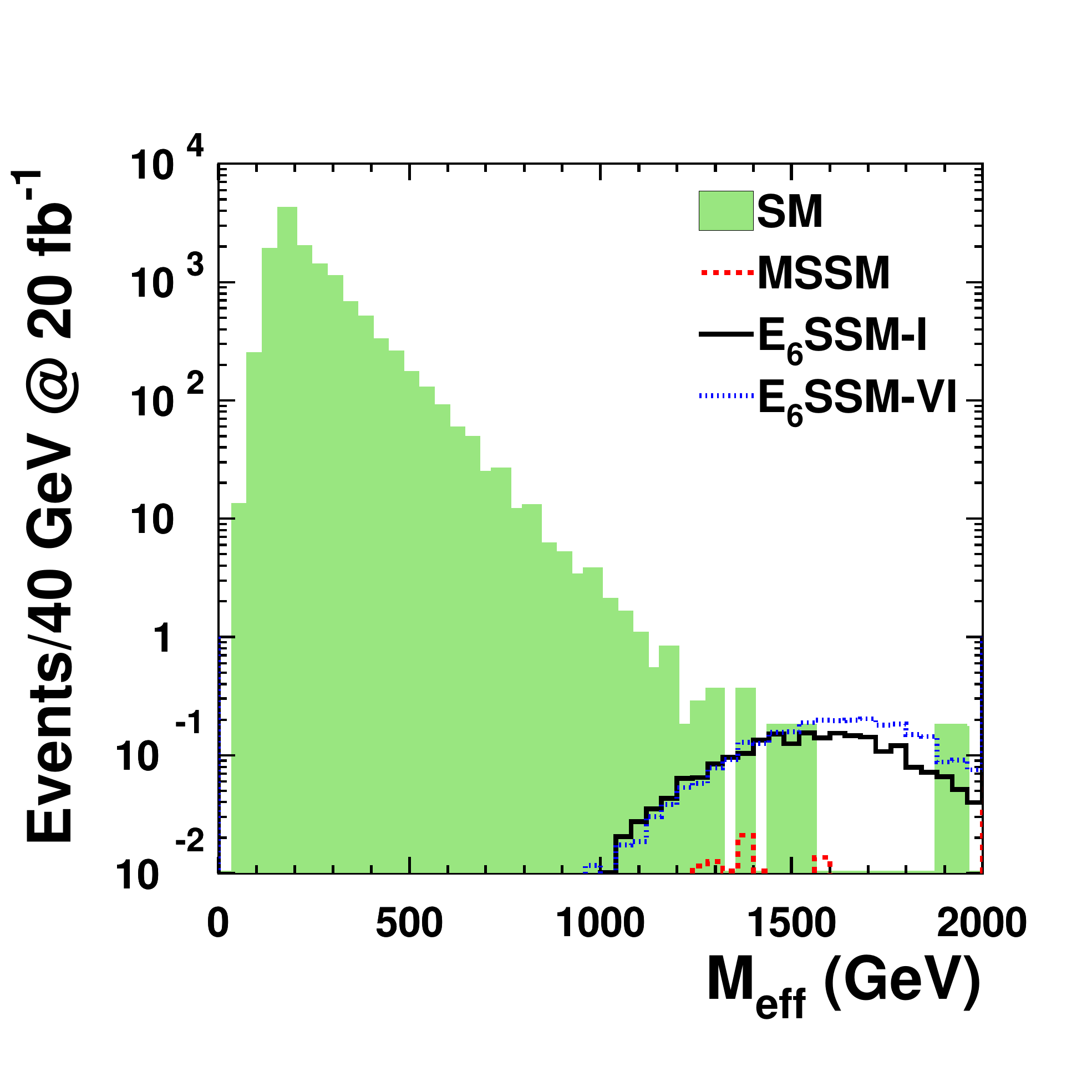}
\label{fig:meff-3lep1000}
}

\caption{
{
Plots of 
$M_{\mbox{\tiny eff}}$ after 
requiring at least 3 leptons at LHC at $\sqrt{s}=$8~TeV. The integrated luminosity is taken to be 20~fb$^{-1}$. The different subfigures show the signal distributions for the MSSM, E$_6$SSM-I, and E$_6$SSM-VI benchmarks for different values of the gluino mass. The E$_6$SSM-VI benchmark is similar to E$_6$SSM-I, but the lighter LSP mass, and thus larger mass gap between it and the bino-like neutralino, causes the signal to be stronger since higher $p_T$ leptons are more likely to be produced. The distributions for gluino masses of 700~GeV (a), 800~GeV (b), 900~GeV (c), and 1000~GeV (d) show that the signal to background ratio is not affected much as the gluino mass increases, but the statistics become bad since the cross section gets small.
}
}
\label{fig:3lep}
\end{figure}

Performing the same analysis for the promising 3 lepton channel at the LHC with $\sqrt{s}=8$~TeV as was done for the case with $\sqrt{s}=7$~TeV we are able to calculate the discovery prospects for our benchmarks. Again, the 3 lepton signature is defined by the cuts in \ref{3lepcut}.
In Fig.~\ref{fig:3lep} the effective mass distributions are plotted for our benchmarks and SM backgrounds for 
20~fb$^{-1}$ of integrated luminosity. The different plots show how the effective mass distributions change for the benchmarks if the gluino mass is varied. 

In these plots we have included another benchmark, E$_6$SSM-VI, for comparison with the E$_6$SSM-I and MSSM benchmarks. 
This benchmark gives a signal slightly above that of the E$_6$SSM-I. This is because the E$_6$SSM-VI benchmark has a long gluino decay chain with an even less compact spectrum. The larger mass difference between the MSSM-like lightest neutralino and the inert singlinos implies that higher $p_T$ leptons can be radiated from that step in the decay chain. This causes an increase in the number of lepton surviving the lepton identification cuts.

A final cut, defining the signal region, is made on the effective mass. 
We let the signal region cut depend on the gluino mass to enhance the expected significance and define it as $M_{\mathrm{eff}}>1.4m_{\tilde g}$.

Using the definition of statistical significance, $S_{12}=2(\sqrt{S+B}-\sqrt{B})$, valid for small statistics \cite{Bartsch:824351,Bityukov2007}, we calculate the expected excess for different gluino masses using our mass dependent signal region cut. 

\begin{figure}[h!]
\centering
\subfigure[Expected significance at 20~fb$^{-1}$, $\sqrt{s}=8$~TeV]{
	\includegraphics[width=.4\columnwidth]{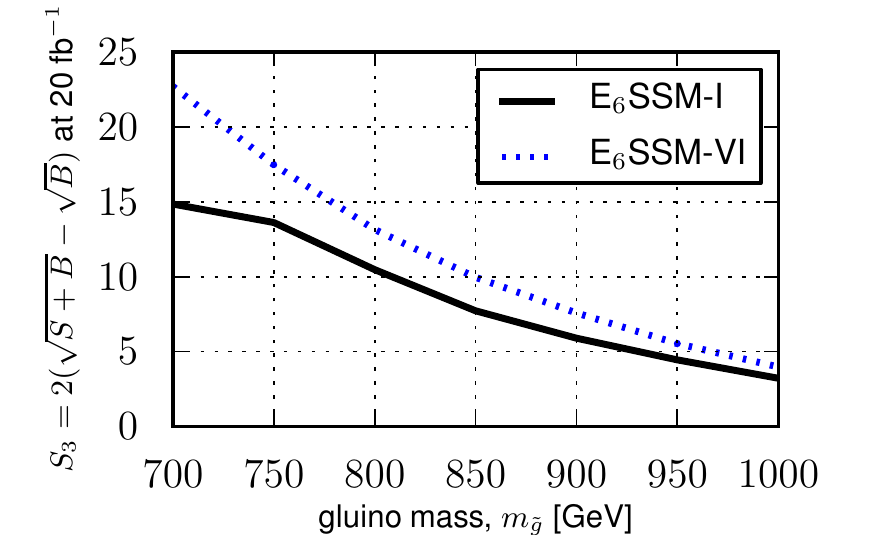}
\label{fig:3lep-significance-8TeV}
}
\subfigure[Luminosity required for 5$\sigma$ discovery and 2$\sigma$ exclusion, $\sqrt{s}=8$TeV]{
	\includegraphics[width=.4\columnwidth]{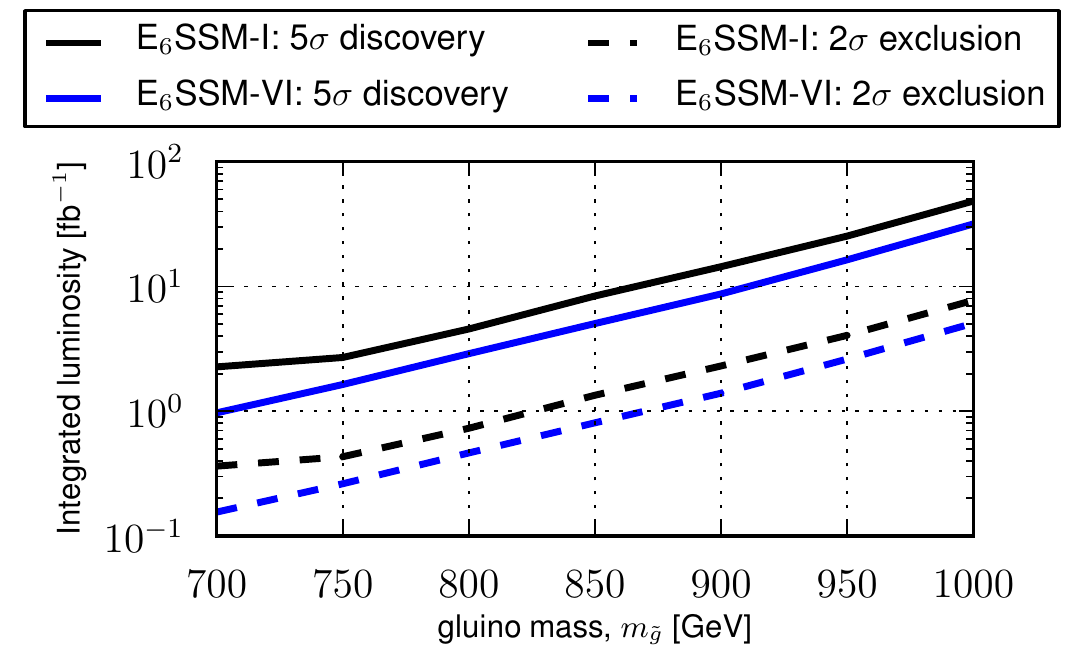}
\label{fig:3lep-discolumi-8TeV}
}

\caption{The gluino mass reach at $\sqrt{s}=$8~TeV for the three lepton channel. The gluino mass is varied for the benchmarks E$_6$SSM-I and E$_6$SSM-VI to estimate the expected significance for different gluino masses. The significance is calculated with the events remaining after a selection cut requiring $M_{\mathrm{eff}}>1.4m_{\tilde g}$. A K-factor of 3 has been applied to the signal. The E$_6$SSM-VI benchmark (shown in blue) is more accessible for exclusion or discovery than benchmark E$_6$SSM-I since it has a bigger mass gap between the bino-like and inert-singlino-like neutralinos, providing higher $p_T$ leptons.}
\label{fig:3lep-reach}
\end{figure}

The significance is plotted as a function of the gluino mass in Fig. \ref{fig:3lep-significance-8TeV} where a K-factor of 3 has been applied to the signal.
{The expected number of events for the E$_6$SSM-I and E$_6$SSM-VI benchmarks with gluino mass $m_{\tilde g}=900$ GeV and the background before and after the final cut on the effective mass ($M_{\mathrm{eff}}>1.4m_{\tilde g}=1260$ GeV) are presented in Tab.~\ref{tab:3lepeff}. The table also lists the expected significances with and without a K-factor of 3 for the two benchmarks.} 
The integrated luminosity needed for discovery and exclusion of a particular gluino mass in the E$_6$SSM in the 3 lepton channel is shown in Fig.~\ref{fig:3lep-reach}, where again a K-factor of 3 has been applied to the signal. The plot shows that a 2$\sigma$ exclusion of gluino masses below 1~TeV is possible with data acquired by ATLAS or CMS at the end of the year 2012. The MSSM benchmark is still well below the background at this stage and is therefore not included in these plots.

\begin{table}[h]
\centering
\begin{tabular}{lcccc}

& $N_{\mathrm{lep}}\geq3$	&	 $M_{\mathrm{eff}}>1.4m_{\tilde g}$	& $\sigma_{K=1}$ & $\sigma_{K=3}$ \\
\hline
\hline

E$_6$SSM-I	& 6.72			& 5.08	&	2.80	&	5.88	\\	
E$_6$SSM-VI	& 8.95			& 7.60	&	3.71	&	7.57	\\
BG		& $8.66\times10^{3}$	& 1.25	&		&		\\
\hline
\end{tabular}
\caption{The expected number of events after the first and second cuts in the 3 lepton analysis for the E$_6$SSM-I and E$_6$SSM-VI benchmarks with $m_{\tilde g}=900$~GeV and SM background at 20~fb$^{-1}$ and 8~TeV. Also the significance, based on signal and background events after the second cut, with K-factors of 1 and 3 applied to the signal.}
	\label{tab:3lepeff}
\end{table}

\subsection{3 lepton searches at $\sqrt{s}=14$ TeV LHC}
\begin{figure}[htb]
\centering
\subfigure[$m_{\tilde g}=900$ GeV]{
	\includegraphics[width=.4\columnwidth]{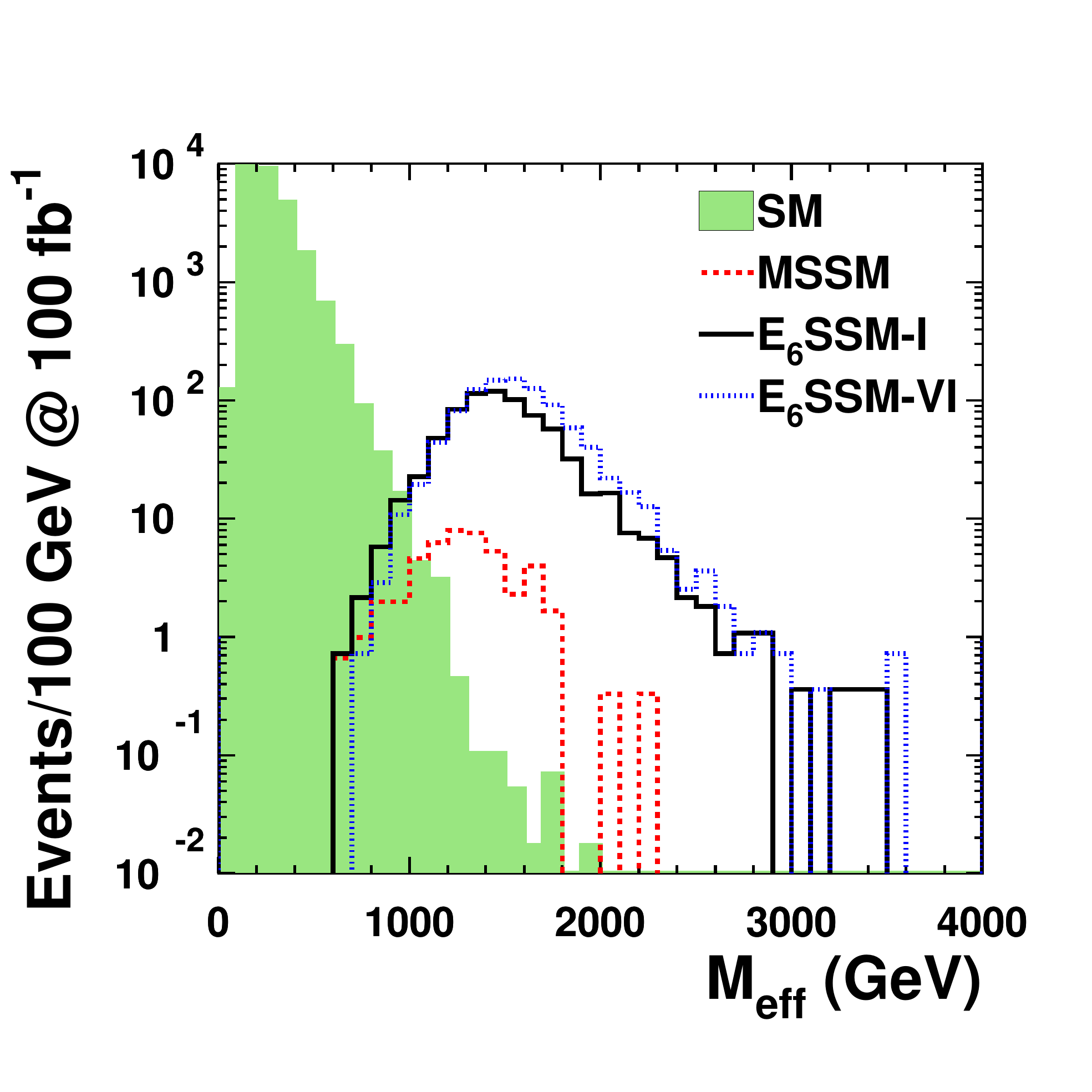}
\label{fig:meff-3lep14-900}
}
\subfigure[$m_{\tilde g}=1100$ GeV]{
	\includegraphics[width=.4\columnwidth]{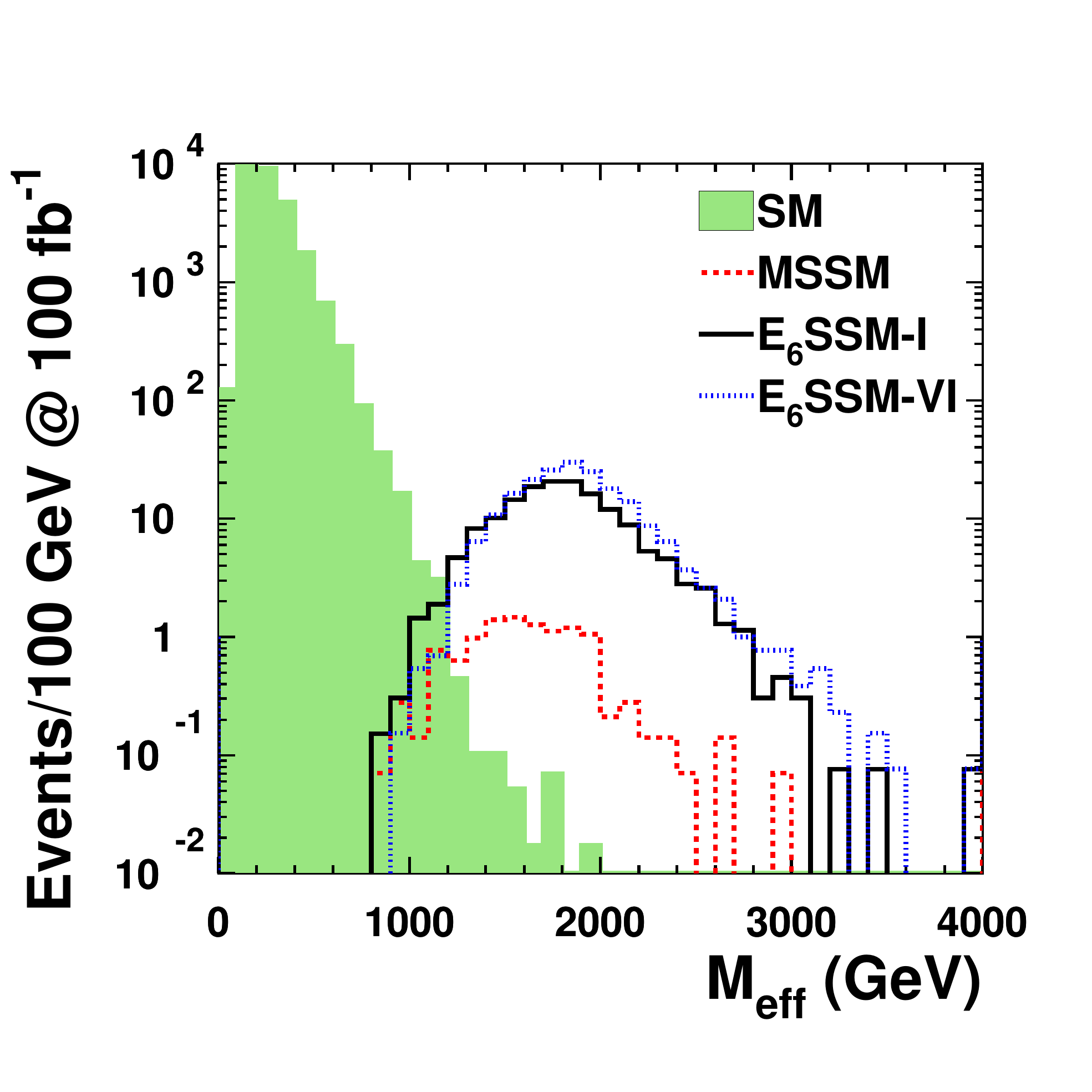}

\label{fig:meff-3lep14-1100}
}
\subfigure[$m_{\tilde g}=1300$ GeV]{
	\includegraphics[width=.4\columnwidth]{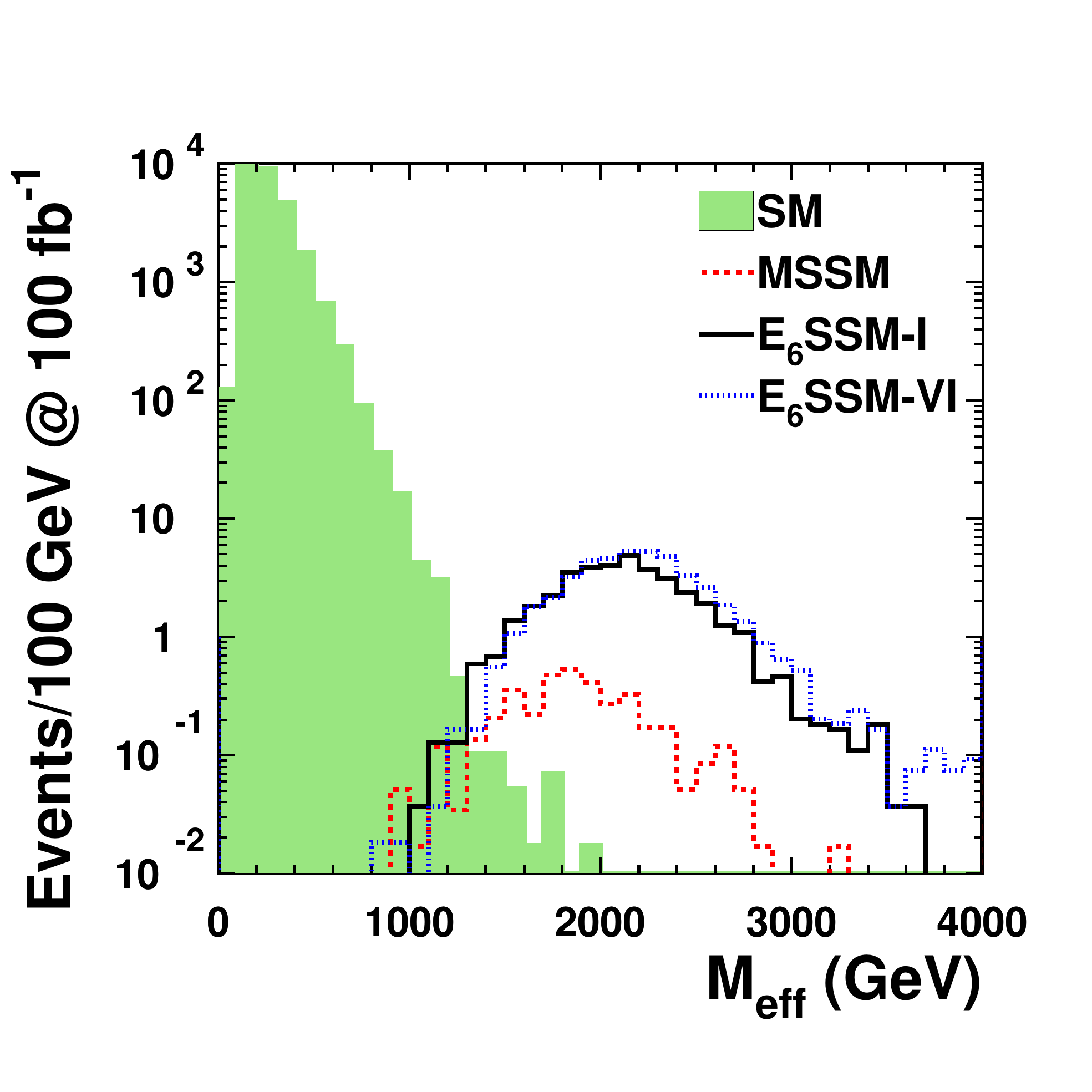}

\label{fig:meff-3lep14-1300}
}
\subfigure[$m_{\tilde g}=1500$ GeV]{
	\includegraphics[width=.4\columnwidth]{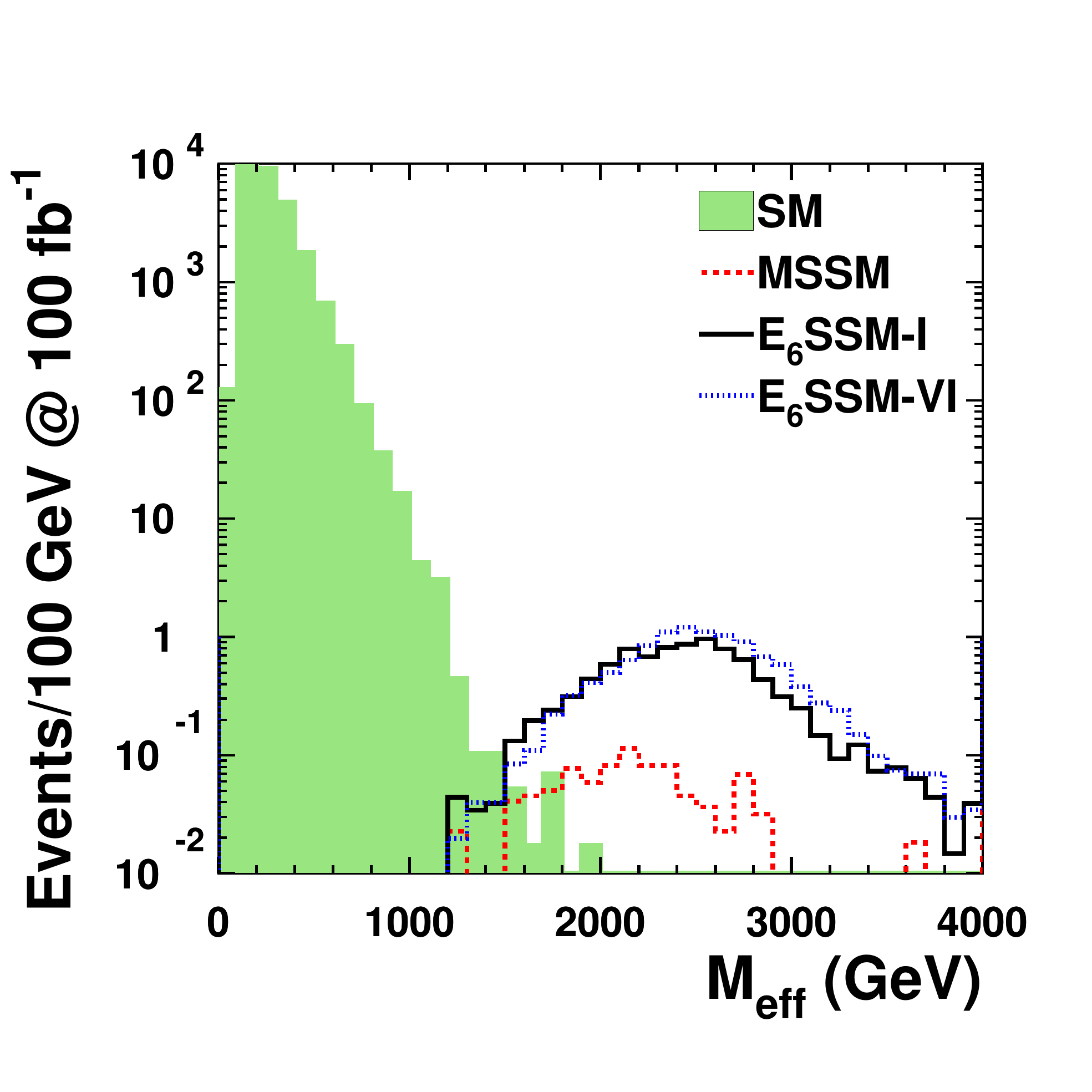}
\label{fig:meff-3lep14-1500}
}
\caption{
These plots show the evolution of the effective mass distribution after requiring at least three leptons for the benchmarks MSSM, E$_6$SSM-I, and E$_6$SSM-VI as the gluino mass changes.
In Figure \ref{fig:meff-3lep14-900} the gluino mass is 900~GeV and at this large integrated luminosity of 100~fb$^{-1}$ at $\sqrt{s}=14$~TeV the MSSM benchmark is almost discoverable.
 The E$_6$SSM benchmarks are both well above the background and clear signals are expected due to the good statistics.
}
\label{fig:3lep14}
\end{figure}

At higher collider energy the cross section for gluino production increases considerably. This causes both our MSSM and E$_6$SSM benchmarks to be clearly visible above the background, as can be seen in Fig.~\ref{fig:3lep14}. The figure shows the effective mass distribution for the MSSM, E$_6$SSM-I, and E$_6$SSM-VI benchmarks for different gluino masses in different subfigures, where a requirement of at least three leptons has been applied in the same way as in the 8 TeV analysis. For all gluino masses the E$_6$SSM-VI benchmark gives the largest signal, just as in the 8~TeV scenario discussed in section \ref{sec:3lep8}. The 14~TeV collider at such a large integrated luminosity as 100~fb$^{-1}$, which is used in Fig.~\ref{fig:3lep14}, allows for statistics needed for discovery of high mass gluinos. Heavier gluinos would push the effective mass distribution to higher values where there is essentially no background and where it is just a matter of acquiring enough statistics, but once that is done one expects a very clean signal.

\begin{figure}[htb]
\centering
\subfigure[Expected significance at 100 fb$^{-1}$, $\sqrt{s}=14$TeV]{
	\includegraphics[width=.4\columnwidth]{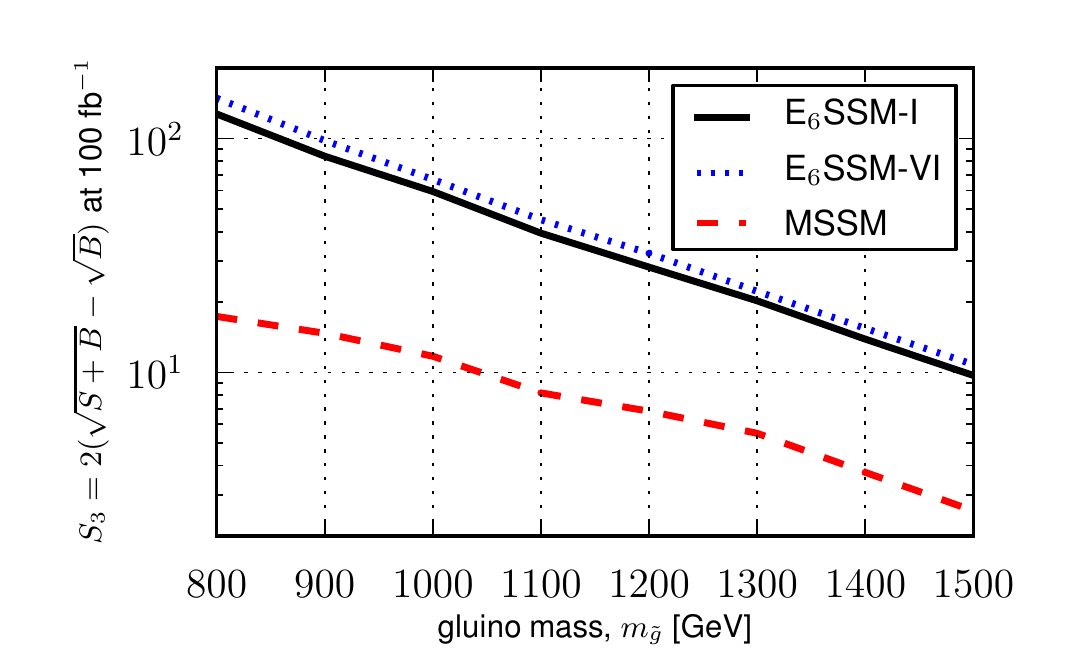}
\label{fig:3lep-significance-14TeV}
}
\subfigure[Luminosity required for 5$\sigma$ discovery and 2$\sigma$ exclusion, $\sqrt{s}=14$TeV]{
	\includegraphics[width=.4\columnwidth]{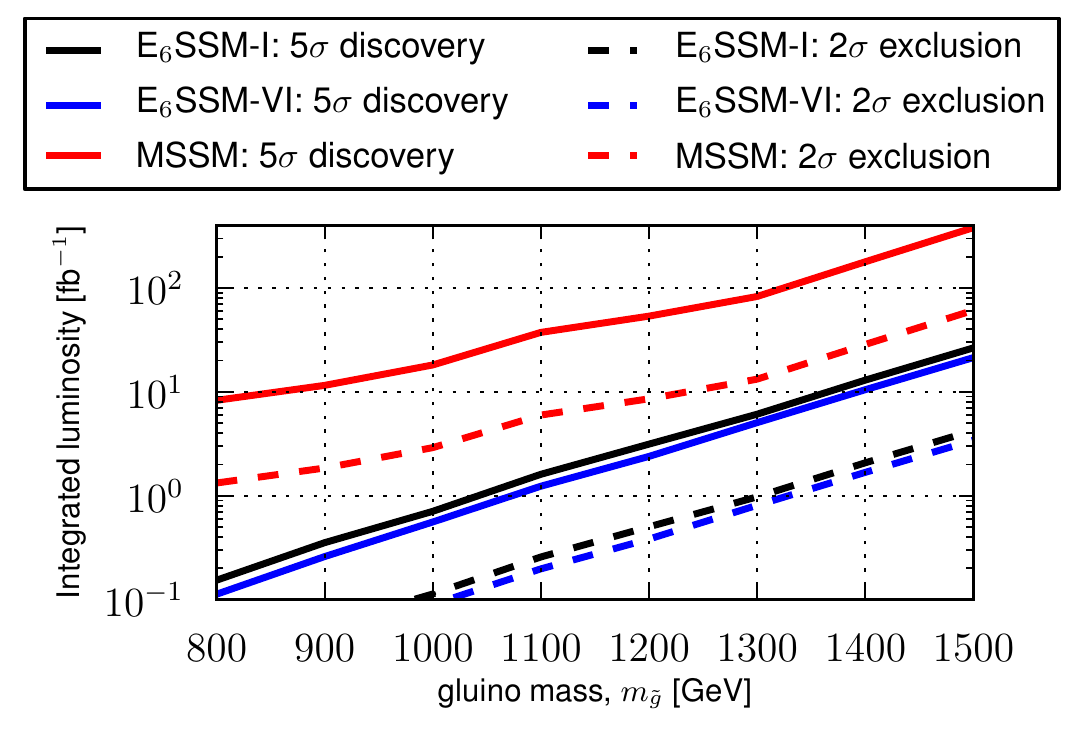}
\label{fig:3lep-discolumi-14TeV}
}

\caption{The gluino mass reach at $\sqrt{s}=$14~TeV for the three lepton channel. The gluino mass has been varied for the benchmarks E$_6$SSM-I,  E$_6$SSM-VI, and MSSM. The expected significance is calculated using the signal and background events surviving the cut $M_{\mathrm{eff}}>m_{\tilde g}$. A K-factor of 3 has also been applied to the signal. Here at the higher collider energy the MSSM benchmark also becomes discoverable through this 3 lepton channel.
The E$_6$SSM benchmarks are discoverable up to almost a 1500 GeV gluino mass.
}
\label{fig:3lep14-reach}
\end{figure}

In Fig.~\ref{fig:3lep14-reach} the gluino reach for the benchmarks at the 14 TeV collider is presented in the same way as was done for the 8 TeV collider in Fig. \ref{fig:3lep-reach}, again with a K-factor of 3 applied to the signal. The only difference is that the gluino mass dependent cut on the effective mass which defines the signal region is taken to be $M_{\mathrm{eff}}>m_{\tilde g}$ instead of $M_{\mathrm{eff}}>1.4m_{\tilde g}$. For the MSSM benchmark, which has become accessible at this energy, one will be able to exclude gluino masses up to $\sim$1400~GeV through this channel for an integrated luminosity of 100~fb$^{-1}$. The expected significance for the MSSM is however about an order of magnitude below the E$_6$SSM benchmarks as shown in Fig.~\ref{fig:3lep-significance-14TeV}. An order of magnitude difference in significance implies two orders of magnitude difference for the the integrated luminocity required for exclusion of a particular gluino mass, as can be seen in Fig.~\ref{fig:3lep-discolumi-14TeV}. For the E$_6$SSM-I and  E$_6$SSM-VI the 100~fb$^{-1}$ of data at 14 TeV allows not only for exclusion, but potentially for a 5$\sigma$ discovery for the whole gluino mass range up to 1.5~TeV.

It should be stressed once more that the results for the MSSM benchmark are different from other studies since we are not dealing with a GUT constrained model, but an electro-weak scale model and the spectrum is more general. Our study focuses on a specific scenario in which the squarks are two to three times heavier than the gluino. Therefore, typical results obtained in previous multi-lepton analyses for SUSY searches differ from our results. As an example, the three lepton signals derived for the mSUGRA points in \cite{baer2008} are larger than the signals for the benchmarks considered in this paper with the same gluino mass. This is mainly due to the lighter squark masses assumed in that study, giving rise to a larger cross section, but also partly due to the entire assumed spectrum being very different. These differences plays a crucial role in providing different $p_T$ distributions for the leptons arising from gluino cascade decays between the different models.

\section{Conclusions}
SUSY is an attractive candidate for new BSM physics, but so far it remains elusive at the LHC. A generic prediction of all SUSY models is the gluino, which can be produced with a large cross-section at the LHC due to its colour octet nature. Moreover in many SUSY models it may be the lightest coloured SUSY particle, which could make it the first SUSY particle to be discovered. However the gluino typically decays in cascade decays, via a chain of charginos and neutralinos, emitting jets or leptons at each stage of the decay chain, with the chain ending when the LSP is produced which is typically the lightest neutralino. 

In this paper we have discussed the gluino cascade decays in $E_6$ models, which include matter and Higgs filling out three complete 27 dimensional representations close to the TeV scale. In such models there are three families of Higgs doublets
of both $H_u$ and $H_d$ kinds, plus three Higgs SM-singlets of the NMSSM kind $S$. Only the third family acquire VEVs and the other two families do not and are called inert. All these Higgs states are accompanied by 
spin-1/2 SUSY partners, the Higgsinos. The extra inert charged and neutral Higgsinos, some of which are necessarily light, will mix with the usual MSSM charginos and neutralinos, yielding extra light states, providing extra links in the gluino cascade decays, and hence extra jets and leptons, with less missing energy.

The extra neutralinos and charginos generically appearing in a large class of $E_6$ inspired models lead to distinctive signatures from gluino cascade decays in comparison to those from the MSSM.
These signatures involve longer decay chains, more visible transverse energy, higher multiplicities of jets and leptons, and less missing transverse energy than in the MSSM. We have studied this effect in gluino cascade decays for the MSSM and $E_6$SSM and have shown that it can provide a characteristic and distinctive signature of the model, enabling an early discovery of the $E_6$SSM which may be discriminated from the MSSM.

In order to demonstrate this, we have first defined a rather large set of benchmark points within the E$_6$SSM. These benchmark points are chosen for the variety of ways in which the inert charginos and neutralinos can appear, and they all are chosen to have a nominal gluino mass of about 800 GeV, although this can be varied while keeping the inert chargino and neutralino masses unchanged. 
The results discussed below are remarkably robust, and apply to all of the $E_6$ benchmark points.
These $E_6$ benchmarks are also compared to an MSSM benchmark which is chosen to have similar conventional
(non-inert) chargino and neutralino masses to the $E_6$ ones, as well as mSUGRA points, in order to verify the model independence of the conclusions.

Given this set of benchmark points, we have then studied gluino pair production and decay at the LHC, first at 7~TeV,
then at 8~TeV, and eventually at 14~TeV using a Monte Carlo analysis. We already know that the gluino was not discovered at 7 TeV, which motivates the nominal choice of gluino mass of 800 GeV. For this gluino mass, we have first studied the missing transverse momentum and effective mass distributions for representative benchmark points, and seen that the former has a softer spectrum 
in the $E_6$ models when compared to the MSSM, as expected, while the latter has a similar or slightly harder distribution in the effective mass variable. Staying at 7 TeV, we then calculated the lepton and jet multiplicities in $E_6$ models and showed that they are significantly higher than in the MSSM. This motivated a study of lepton channels with increasing numbers of leptons, and decreasing statistics, where we showed that the 3 lepton channel provided a very good discriminator between the MSSM and the  $E_6$ models, although the statistics are rather too low at 7~TeV.

At 8~TeV we studied the 6 jet channel and showed that, in an effective mass analysis, a 1~TeV MSSM gluino may provide a similar limit to that of an 800~GeV $E_6$ gluino. Turning to the promising 3 lepton channels at 8~TeV,
we find increased statistics and possible observable signals in this channel for a range of gluino masses,
which would provide a striking confirmation of $E_6$ models. We calculate the required integrated luminosity in order to either discover or exclude the $E_6$ gluino in this channel at 8~TeV. We emphasise again that the MSSM gluino is unobservable in the 3~lepton channel.
Finally we have repeated the analysis for the promising 3 lepton channel at 14~TeV and 
found analogous results for required integrated luminosity to discover or exclude the $E_6$ gluino there as well.

In conclusion, the $E_6$ inspired models are clearly  distinguishable from the MSSM in gluino cascade decays
at the LHC with the full data set at 8 TeV, and certainly at 14 TeV, using the three lepton channel that we have proposed. 
Moreover, the present limits on the gluino mass, for example from a multijet analysis,  are 
weaker (and therefore not applicable) for $E_6$ models in comparison with  the MSSM, due to the longer decay chains with less missing transverse energy that is expected in $E_6$ models. 
We emphasise to the LHC experimental groups that the distinctive features present in gluino cascade decays,
resulting in different search strategies including the choice of the kinematical variables, 
such as those discussed here, not only represents a unique footprint of a particular model but also may provide the key  
to an earlier discovery of supersymmetry.

\acknowledgments{
We acknowledge partial support from the STFC Consolidated ST/J000396/1 and EU ITN grants UNILHC 237920 and INVISIBLES 289442. PS thanks the NExT institute and SEPnet for support.}

\cleardoublepage

\appendix

\section{Decay Diagrams for Benchmarks}
\label{ap:A}
\begin{figure}[ht]
\centering
	\subfigure[MSSM]{
	\raisebox{116pt}{\includegraphics{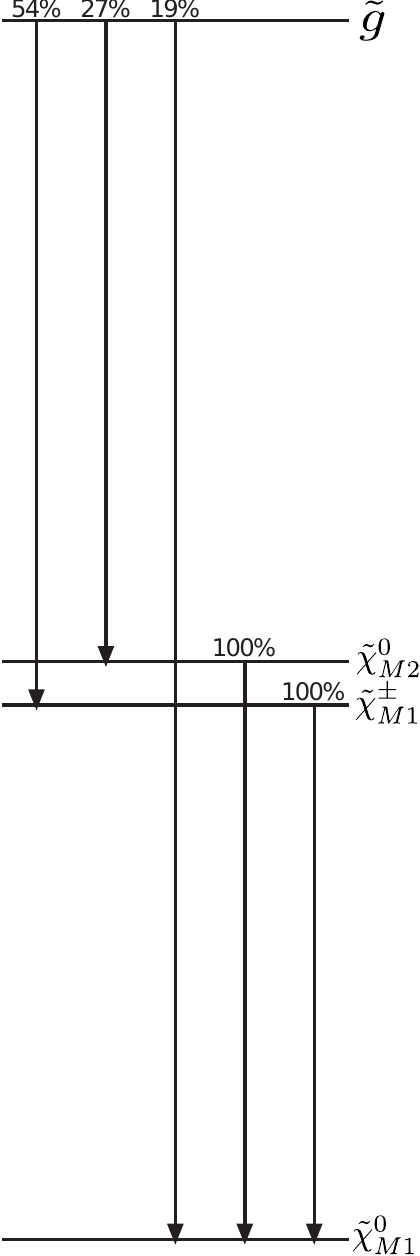}}
\label{fig:MSSM-decays}
}
	\subfigure[E$_6$SSM-I]{
\includegraphics{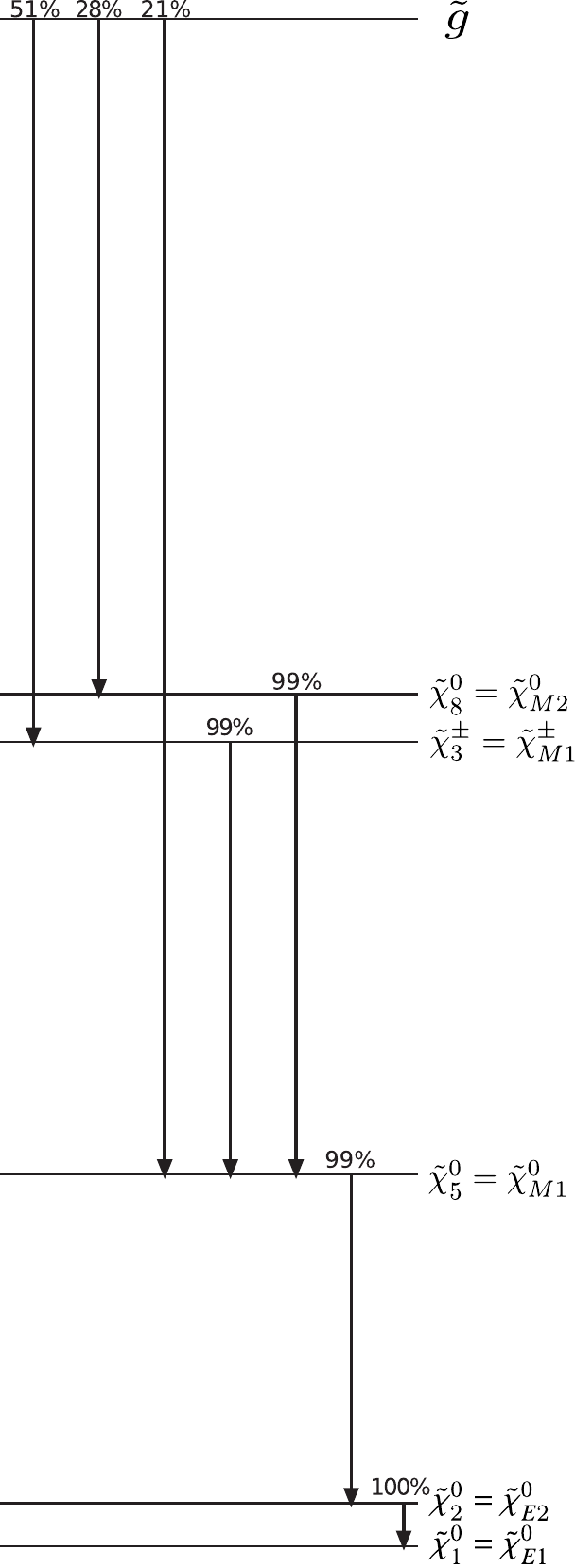}
\label{fig:E6SSM-I-decays}
	}
\end{figure}

\begin{figure}[ht]
	\subfigure[E$_6$SSM-II]{
\includegraphics[width=.865\linewidth]{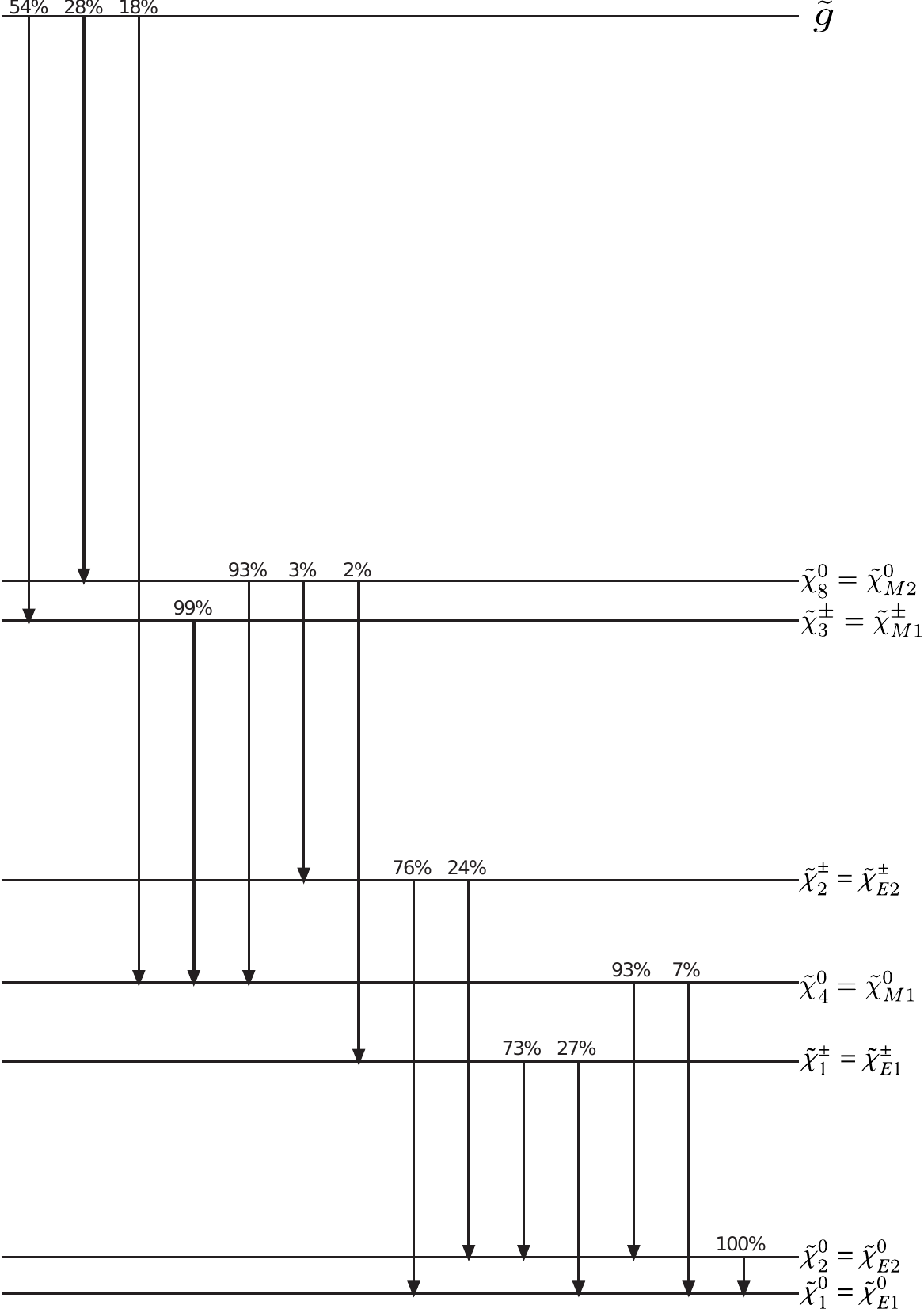}
\label{fig:E6SSM-II-decays}
	}
\end{figure}

\begin{figure}[ht]
	\subfigure[E$_6$SSM-III]{
\includegraphics[width=.53\linewidth]{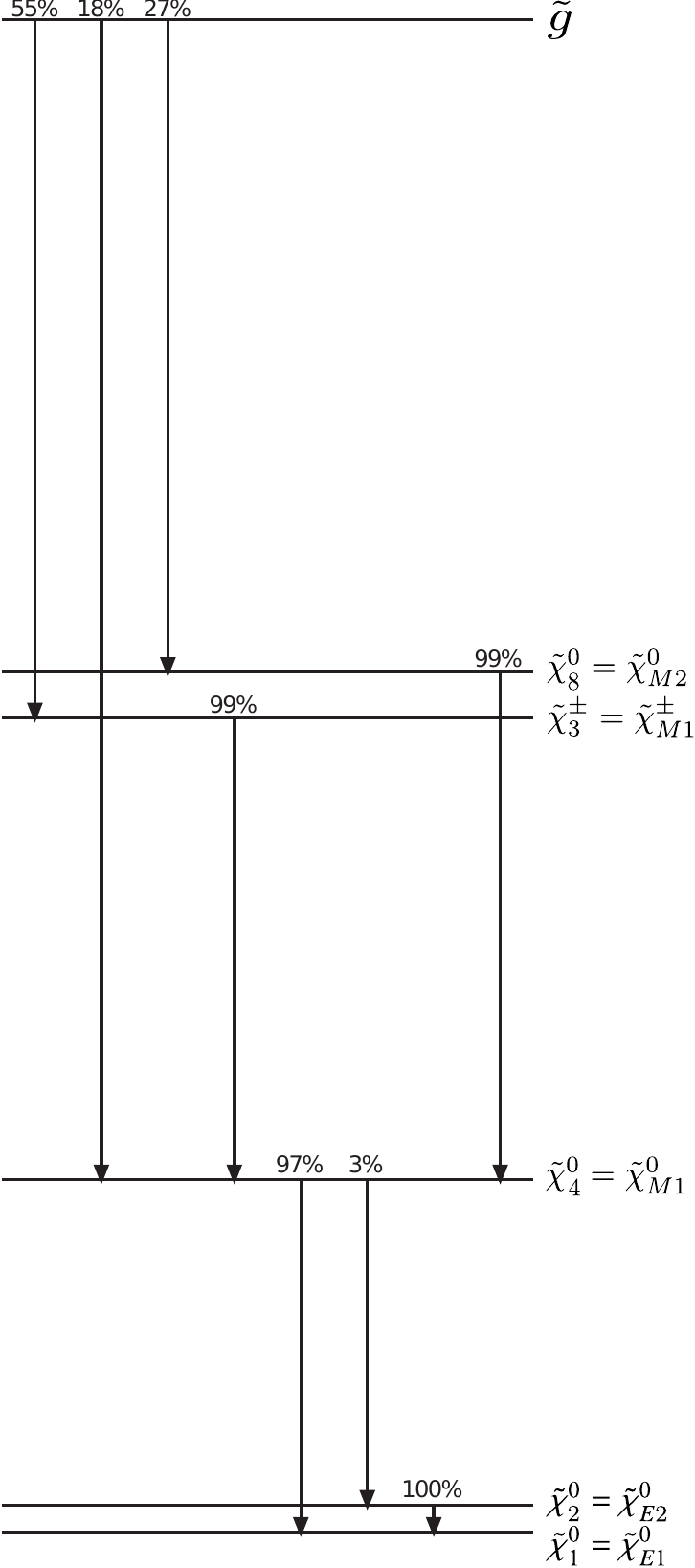}
\label{fig:E6SSM-III-decays}
	}
	\subfigure[E$_6$SSM-IV]{
	\raisebox{4.95cm}{\includegraphics[width=.365\linewidth]{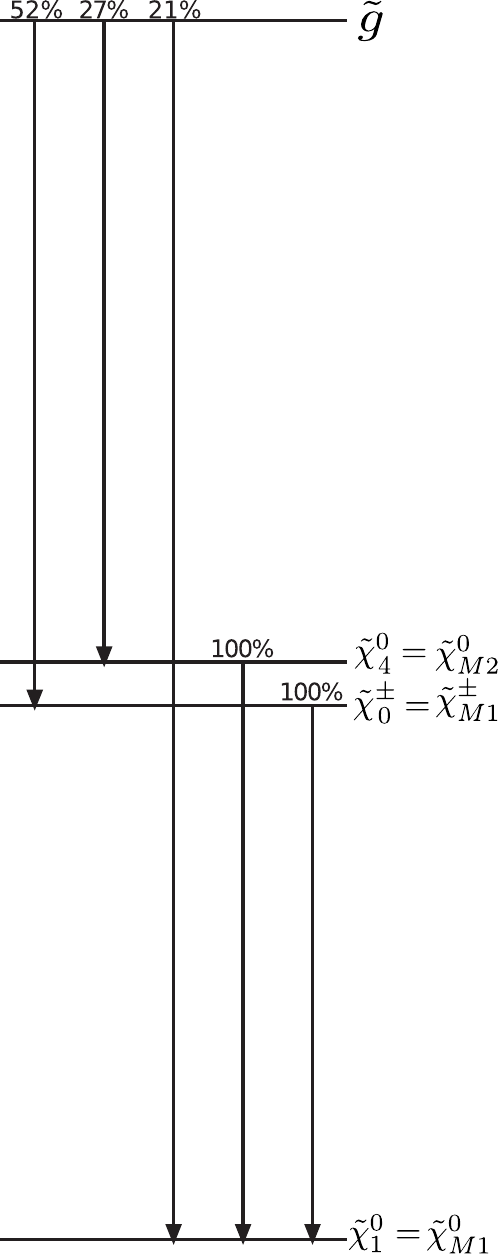}}
\label{fig:E6SSM-IV-decays}
	}
\end{figure}

\begin{figure}
	\subfigure[E$_6$SSM-V]{
	\raisebox{4.2cm}{\includegraphics[width=.4\linewidth]{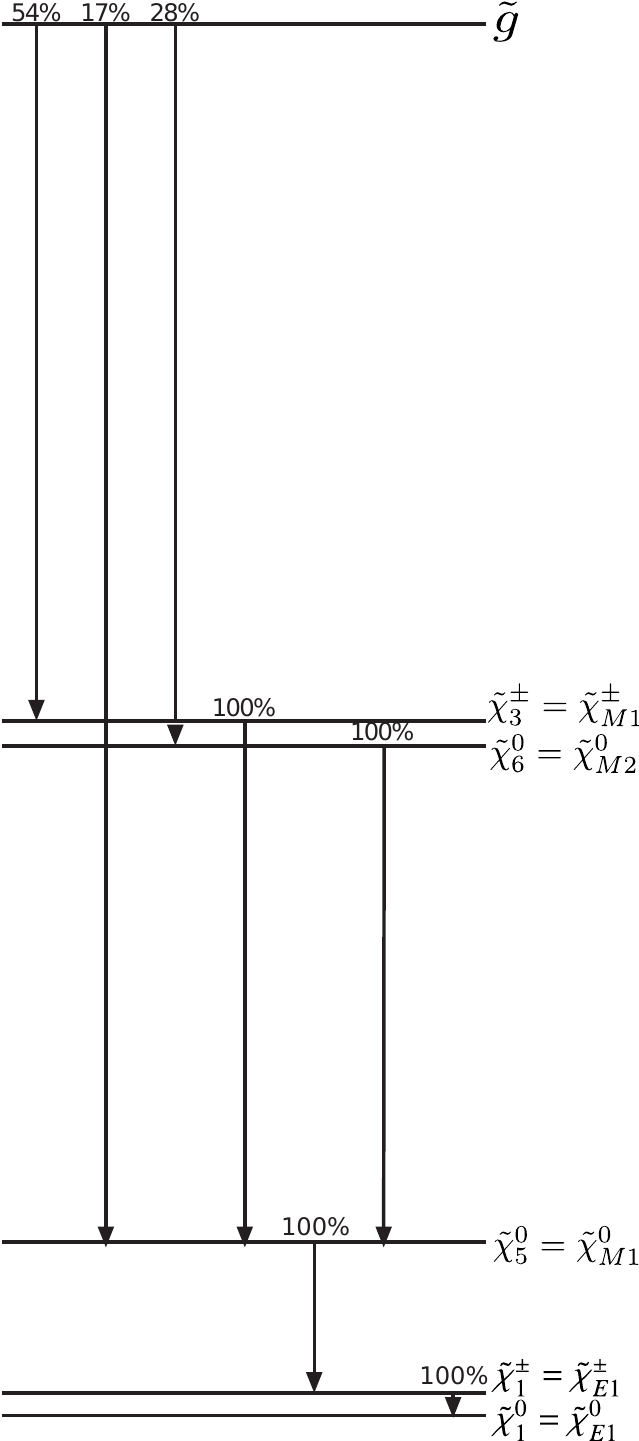}}
\label{fig:E6SSM-V-decays}
	}
	\subfigure[E$_6$SSM-VI]{
		\includegraphics[width=.4\linewidth]{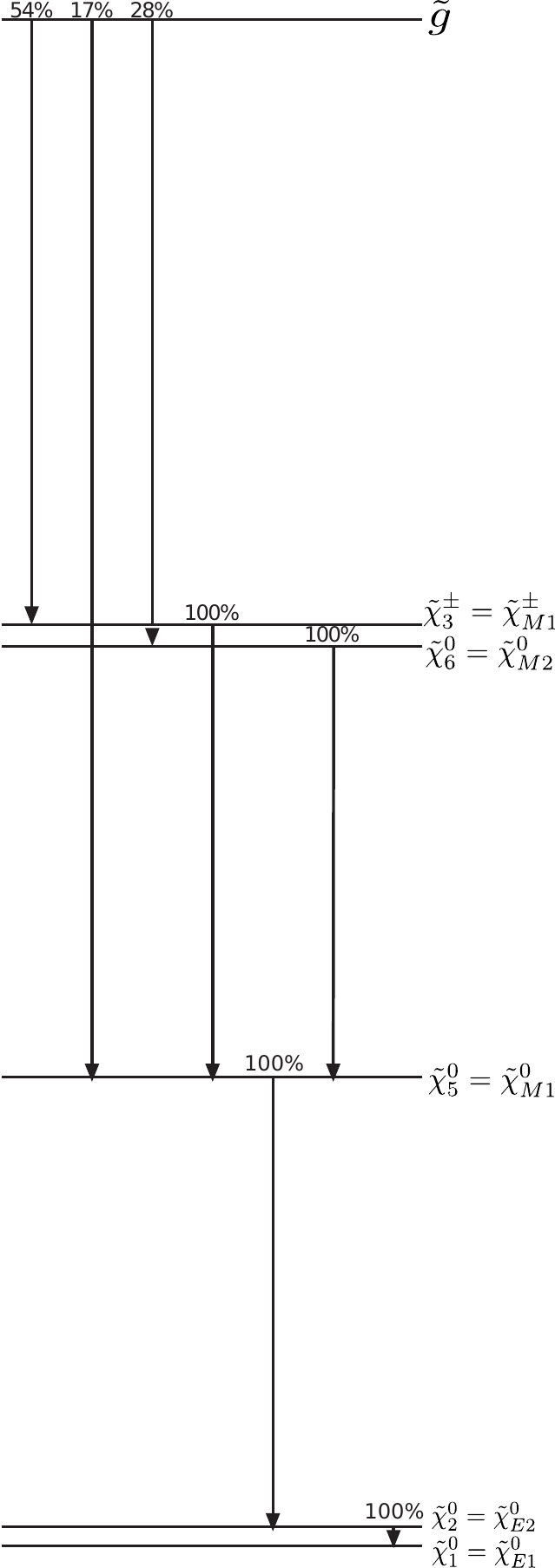}
\label{fig:E6SSM-VI-decays}
	}
\caption{Gluino decay diagrams for the MSSM and E$_6$SSM benchmarks, showing the leading decays (contributing more than 90\%) for the involved sparticles. The vertical line spacings are proportional to the mass splitting among the particles.}
\label{fig:gluino-decays}
\end{figure}

This section displays the possible gluino decays of each benchmark in a more complete fashion. Horizontal lines representing particle states are separated proportionally to their mass difference. Some exceptions are made where states are closely degenerate, in which case the lines have been separated more. An arrow then connects lines where possible decays occur with the corresponding branching ratio written above. The gluino decays of the MSSM benchmark is shown in Fig.~\ref{fig:MSSM-decays}, E$_6$SSM-I in Fig.~\ref{fig:E6SSM-I-decays}, E$_6$SSM-II in Fig.~\ref{fig:E6SSM-II-decays}, E$_6$SSM-III in Fig. \ref{fig:E6SSM-III-decays}, E$_6$SSM-IV in Fig.~\ref{fig:E6SSM-IV-decays}, E$_6$SSM-V in Fig.~\ref{fig:E6SSM-V-decays} and E$_6$SSM-VI in Fig.~\ref{fig:E6SSM-VI-decays}.

\section{Details about the CalcHEP Model E6SSM-12.02}
\label{ap:model}
In this section the contents and properties of the \verb+CalcHEP+ model \verb+E6SSM-12.02+ (hepmdb:1112.0106)\cite{E6SSM-12.02} are described in detail. The model files are accesible from HEPMDB \cite{HEPMDB} where one can either run it with \verb+CalcHEP+ on the IRIDIS cluster via the web interface or download it and run it on one's local \verb+CalcHEP+ installation. The model constitutes of four files; \verb+varsNN.mdl+, \verb+prtclsNN.mdl+, \verb+funcNN.mdl+, and \verb+lgrngNN.mdl+; for input variables, particle definitions, functions or constraints, and Feynman rules for vertices, respectively. 

\subsection{Particle content}
The model shares many features and particles with the MSSM and its extensions, e.g. the USSM or NMSSM, but has a greater particle content. This \verb+CalcHEP+ version of the E$_6$SSM, \verb+E6SSM-12.02+, includes particles from the three generations of 27 representations but not all. What is yet to be implemented is the full Higgs sector, including the \textit{inert} Higgs \textit{boson} states, and the coloured exotics. This is work in progress and will be added in a later version of the model. The extra particles included compared to the MSSM are a SM-singlet $S$, which mixes with the two MSSM-like CP-even Higgs particles; a $Z^\prime$-boson; two extra chargino states from inert Higgsinos; and 8 extra neutralino states, two from the bino$^\prime$ and the singlino and six from the inert neutral Higgsinos and singlinos. The  \verb+CalcHEP+ particle names and properties as used in the \verb+prtclsNN.mdl+ file are shown in Tab.~\ref{tab:prtcls}. The table shows the full name of each particle followed by the symbols used in \verb+CalcHEP+ for it and its antiparticle. The PDG code is a positive number assigned to each particle for referencing in the code, the antiparticles have PDG codes equal to the negative of the particles' PDGs. The spin, mass, width and color properties are also given in \verb+CalcHEP+ notation. Following the convention from earlier SUSY models in \verb+CalcHEP+, superparters, i.e. particles which are odd under $R$-parity, are denoted with a \ttilde~prefix. The conventions we use for the particle mass and width variables are \verb+M+$<\mbox{particle symbol}>$ and \verb+w+$<\mbox{particle symbol}>$. If the particle is charged the particle symbol is modified with a suffix -\verb+c+ for ``charged'' to avoid confusion, e.g. \verb+MHc+ for the charged Higgs \verb=H+=. In the case of superpartners the \ttilde~is not used in the mass or width variable. Instead a prefix \verb+S+- or suffix -\verb+o+  is attached to the particle name depending on whether the sparticle is a {\bf S}fermion or a boson{\bf o}, e.g. \verb+MGo+ for the gluino mass and \verb+MSe1+ for the first selectron mass. A number is added to the variable name whenever it refers to particles for which there are more than one of the same type, e.g. neutralinos or stops. We try to keep as close as possible to these conventions without departing too much from conventions used in earlier SUSY models. There are still some deviations from the set up conventions, e.g. the charged slepton sector, which we hope will not be too confusing until a more uniform way of naming the particles and variables is introduced. In future versions of the model the conventions will be used more strictly.

\begin{table}
	\centering
	\begin{tabular}{llllllll}
Full Name         & particle &antiparticle &  PDG ID&2$\times$spin& mass &width   &color\\
\hline
gluon             &G  &G  &21      &2     &0     &0       &8    \\
neutrino          &n1 &N1 &12      &1     &0     &0       &1    \\
electron          &e1 &E1 &11      &1     &Me1   &0       &1    \\
muon-neutrino     &n2 &N2 &14      &1     &0     &0       &1    \\
muon              &e2 &E2 &13      &1     &Me2   &0       &1    \\
tau-neutrino      &n3 &N3 &16      &1     &0     &0       &1    \\
tau-lepton        &e3 &E3 &15      &1     &Me3   &0       &1    \\
u-quark           &u  &U  &2       &1     &Mu    &0       &3    \\
d-quark           &d  &D  &1       &1     &Md    &0       &3    \\
c-quark           &c  &C  &4       &1     &Mc    &0       &3    \\
s-quark           &s  &S  &3       &1     &Ms    &0       &3    \\
t-quark           &t  &T  &6       &1     &Mt    &!wt     &3    \\
b-quark           &b  &B  &5       &1     &Mb    &0       &3    \\
light Higgs       &h1 &h1 &25      &0     &Mh1   &!wh1    &1    \\
heavier Higgs     &h2 &h2 &26      &0     &Mh2   &!wh2    &1    \\
heaviest Higgs    &h3 &h3 &27      &0     &Mh3   &!wh3    &1    \\
pseudoscalar Higgs&ha &ha &28      &0     &Mha   &!wha    &1    \\
charged Higgs     &H+ &H- &37      &0     &MHc   &!wHc    &1    \\
photon            &A  &A  &22      &2     &0     &0       &1    \\
Z-boson           &Z  &Z  &23      &2     &MZ    &!wZ     &1    \\
W-boson           &W+ &W- &24      &2     &MW    &!wW     &1    \\
Z-primed-boson    &Zb &Zb &32      &2     &MZb   &!wZb    &1    \\
chargino 1        &\ttilde 1+&\ttilde 1-&1000024 &1     &MCo1  &!wCo1   &1    \\
chargino 2        &\ttilde 2+&\ttilde 2-&1000037 &1     &MCo2  &!wCo2   &1    \\
chargino 3        &\ttilde 3+&\ttilde 3-&1000038 &1     &MCo3  &!wCo3   &1    \\
chargino 4        &\ttilde 4+&\ttilde 4-&1000039 &1     &MCo4  &!wCo4   &1    \\
neutralino 1      &\ttilde o1&\ttilde o1&1000022 &1     &MNo1  &0       &1    \\
neutralino 2      &\ttilde o2&\ttilde o2&1000023 &1     &MNo2  &!wNo2   &1    \\
neutralino 3      &\ttilde o3&\ttilde o3&1000025 &1     &MNo3  &!wNo3   &1    \\
neutralino 4      &\ttilde o4&\ttilde o4&1000026 &1     &MNo4  &!wNo4   &1    \\
neutralino 5      &\ttilde o5&\ttilde o5&1000027 &1     &MNo5  &!wNo5   &1    \\
neutralino 6      &\ttilde o6&\ttilde o6&1000028 &1     &MNo6  &!wNo6   &1    \\
neutralino 7      &\ttilde o7&\ttilde o7&1000029 &1     &MNo7  &!wNo7   &1    \\
neutralino 8      &\ttilde o8&\ttilde o8&1000030 &1     &MNo8  &!wNo8   &1    \\
neutralino 9      &\ttilde o9&\ttilde o9&1000031 &1     &MNo9  &!wNo9   &1    \\
neutralino 10     &\ttilde oA&\ttilde oA&1000032 &1     &MNoA  &!wNoA   &1    \\
neutralino 11     &\ttilde oB&\ttilde oB&1000033 &1     &MNoB  &!wNoB   &1    \\
neutralino 12     &\ttilde oC&\ttilde oC&1000034 &1     &MNoC  &!wNoC   &1    \\
gluino            &\ttilde g &\ttilde g &1000021 &1     &MGo   &!wGo    &8    \\
1st selectron     &\ttilde e1&\ttilde E1&1000011 &0     &MSe1  &!wSe1   &1    \\
2nd selectron     &\ttilde e4&\ttilde E4&2000011 &0     &MSe2  &!wSe2   &1    \\
1st smuon         &\ttilde e2&\ttilde E2&1000013 &0     &MSmu1 &!wSmu1  &1    \\
2nd smuon         &\ttilde e5&\ttilde E5&2000013 &0     &MSmu2 &!wSmu2  &1    \\
1st stau          &\ttilde e3&\ttilde E3&1000015 &0     &MStau1&!wStau1 &1    \\
2nd stau          &\ttilde e6&\ttilde E6&2000015 &0     &MStau2&!wStau2 &1    \\
e-sneutrino       &\ttilde n1&\ttilde N1&1000012 &0     &MSn1  &!wSn1   &1    \\
mu-sneutrino      &\ttilde n2&\ttilde N2&1000014 &0     &MSn2  &!wSn2   &1    \\
tau-sneutrino     &\ttilde n3&\ttilde N3&1000016 &0     &MSn3  &!wSn3   &1    \\
1st u-squark      &\ttilde u1&\ttilde U1&1000002 &0     &MSu1  &!wSu1   &3    \\
2nd u-squark      &\ttilde u2&\ttilde U2&2000002 &0     &MSu2  &!wSu2   &3    \\
1st d-squark      &\ttilde d1&\ttilde D1&1000001 &0     &MSd1  &!wSd1   &3    \\
2nd d-squark      &\ttilde d2&\ttilde D2&2000001 &0     &MSd2  &!wSd2   &3    \\
1st c-squark      &\ttilde c1&\ttilde C1&1000004 &0     &MSc1  &!wSc1   &3    \\
2nd c-squark      &\ttilde c2&\ttilde C2&2000004 &0     &MSc2  &!wSc2   &3    \\
1st s-squark      &\ttilde s1&\ttilde S1&1000003 &0     &MSs1  &!wSs1   &3    \\
2nd s-squark      &\ttilde s2&\ttilde S2&2000003 &0     &MSs2  &!wSs2   &3    \\
1st t-squark      &\ttilde t1&\ttilde T1&1000006 &0     &MSt1  &!wSt1   &3    \\
2nd t-squark      &\ttilde t2&\ttilde T2&2000006 &0     &MSt2  &!wSt2   &3    \\
1st b-squark      &\ttilde b1&\ttilde B1&1000005 &0     &MSb1  &!wSb1   &3    \\
2nd b-squark      &\ttilde b2&\ttilde B2&2000005 &0     &MSb2  &!wSb2   &3    \\
	\end{tabular}
	\caption{Particle content in the E6SSM-12.02 with CalcHEP naming conventions and properties.}
	\label{tab:prtcls}
\end{table}

\subsection{Input parameters}
The input parameters used in the model, with the exception of some SM parameters, are listed in Tab.~\ref{tab:vars}. The pseudoscalar Higgs mass is used as an input variable instead of the soft trilinear lambda coupling from the term $A_{\lambda}\lambda S H_d H_u$. The electroweak soft gaugino masses $M_1$, $M_2$, and $M_{1}^{\prime}$ are denoted \verb+ MG1+, \verb+MG2+, and \verb+MG1b+, respectively. The physical gluino mass is denoted \verb+MGo+. The soft squark masses for the first two generations are set by a common squark mass scale, \verb+MSq+, by default. It is however easy to modify the model files by moving the soft squark masses \verb+Mq1, Mq2, Mu1, Mu2, Md1, Md2+ from the function file \verb+funcNN.mdl+ to the input parameter file \verb+varsNN.mdl+ and assigning them separate values.

\begin{table}
	\centering
	\begin{tabular}{lll}
 Name  & Value       &  Comment                                   	\\
 \hline
g1b    &0.073        &U(1)-primed coupling with sqrt(1/40)\\
hL     &0.393        &Yukawa coupling for S  Hu  Hd\\
hL22   &-0.000357    &Yukawa coupling for S  Hd2 Hu2\\
hL21   &0.04         &Yukawa coupling for S  Hd2 Hu1\\
hL12   &0.0321       &Yukawa coupling for S  Hd1 Hu2\\
hL11   &0.000714     &Yukawa coupling for S  Hd1 Hu1\\
hFd22  &0.001        &Yukawa coupling for S2 Hd  Hu2\\
hFd21  &0.6844       &Yukawa coupling for S2 Hd  Hu1\\
hFd12  &0.65         &Yukawa coupling for S1 Hd  Hu2\\
hFd11  &0.001        &Yukawa coupling for S1 Hd  Hu1\\
hFu22  &0.001        &Yukawa coupling for S2 Hu  Hd2\\
hFu21  &0.67         &Yukawa coupling for S2 Hu  Hd1\\
hFu12  &0.64         &Yukawa coupling for S1 Hu  Hd2\\
hFu11  &0.001        &Yukawa coupling for S1 Hu  Hd1\\
hXd2   &0.000714     &Yukawa coupling for S  Hd  Hu2\\
hXd1   &0.000714     &Yukawa coupling for S  Hd  Hu1\\
hXu2   &0.000714     &Yukawa coupling for S  Hd2 Hu\\
hXu1   &0.000714     &Yukawa coupling for S  Hd1 Hu\\
hZ2    &0.001        &Yukawa coupling for S2 Hd  Hu\\
hZ1    &0.001        &Yukawa coupling for S1 Hd  Hu\\
Svev   &5180         &SM-singlet VEV\\
Hvev   &246          &SM Higgs VEV\\
MSq    &2000         &Common soft squark mass scale for gen. 1 and 2 = Mq1= Mq2= Mu1= Mu2= Md1= Md2\\
topA   &-2200        &soft trilinear A-coupling for top\\
botA   &-2200        &soft trilinear A-coupling for bottom\\
Mq3    &2000         &Soft squark mass for third gen. SU(2) doublet, q\\
Mu3    &2000         &Soft squark mass for third gen. right-handed u\\
Md3    &2000         &Soft squark mass for third gen. right-handed d\\
Ml1    &2000         &Soft slepton mass for 1st gen. SU(2) doublet, L\\
Ml2    &2000         &Soft slepton mass for 2nd gen. SU(2) doublet, L\\
Ml3    &2000         &Soft slepton mass for 3rd gen. SU(2) doublet, L\\
Mr1    &2000         &Soft slepton mass for 1st gen. right-handed e (selectron)\\
Mr2    &2000         &Soft slepton mass for 2nd gen. right-handed e (smuon)\\
Mr3    &2040         &Soft slepton mass for 3rd gen. right-handed e (stau)\\
tauA   &-2200        &soft trilinear A-coupling for tau\\
lsc    &165          &scale for Higgs loop corrections\\
lmt    &165          &top mass in Higgs loop corrections\\
Mha    &2736         &pseudoscalar Higgs mass\\
MG1    &150          &Soft gaugino mass for U(1) (hypercharge)\\
MG2    &300          &Soft gaugino mass for SU(2) \\
MG1b   &151          &Soft gaugino mass for U(1)' \\
Maux   &1            &mass of aux particles\\
tb     &1.5          &tangent beta\\
MGo    &800          &gluino mass	\\
	\end{tabular}
	\caption{Input parameters for the E6SSM-12.02 model in CalcHEP notation. Some SM parameters have been removed from this list. This is the format and content of the varsNN.mdl file that comes with the model. By default the parameter values of benchmark E$_6$SSM-I are given.}
	\label{tab:vars}
\end{table}

\subsection{Functions and dependent parameters}
In the file \verb+funcNN.mdl+ all dependent parameters are listed in terms of input parameters and dependent parameters above themselves in the list. The dependence can be given as simple algebraic expressions or as functions of external functions. As an example, expressions for mass matrix elements are given in this file. These matrix elements are then used as inputs for numerical diagonalisation routines from the \verb+SLHAplus+ library, which comes with \verb+CalcHEP+. \verb+SLHAplus+ then returns evaluated particle masses, which are defined in this file. As an example of what the contents of this file look likes, a few lines from it are presented in Tab.~\ref{tab:func}.

\begin{table}
	\centering
	\begin{tabular}{ll}
		Name	& Expression\\
		\hline
		$\,\,\vdots$	& $\,\,\vdots$\\
		MNE13  		& -MZ*SW*cb\\
		MNE14  		& MZ*SW*sb\\
		$\,\,\vdots$	& $\,\,\vdots$\\
		MNo1   		& MassArray(NeDiag, 1) \% Neutralino mass 1\\
		MNo2   		& MassArray(NeDiag, 2) \% Neutralino mass 2\\
		$\,\,\vdots$	& $\,\,\vdots$\\
	\end{tabular}
	\caption{Example lines from the model file funcNN.mdl, where dependent parameters are specified. Comments are allowed at the ends of lines after a \%. The lines shown are examples of simple expressions for matrix elements and masses defined by external numerical routines.}
	\label{tab:func}
\end{table}

\subsection{Feynman rules}
All of the vertices in the model are listed in the file \verb+lgrngNN.mdl+. Some of the vertices include auxiliary particles, which are included for technical reasons, e.g. to construct a four-gluon vertex. Examples of vertices and Feynman rules from this file are given in Tab.~\ref{tab:lgrng}.

\begin{table}
	\centering
	\begin{tabular}{llllll}
		P1   &P2   &P3   	&P4   		& Factor    			& dLagrangian/ dA(p1) dA(p2) dA(p3)  \\  
		\hline
		A    &H+   &H-   	&     		&-EE        			& m1.p2-m1.p3\\
		G    &W+   &\ttilde C1 	&\ttilde b1  	&-EE*GG*Sqrt2*Vcb*Zd33/SW	& m1.m2\\
		W+   &W+   &W-   	&W-   		&-EE\^{}2/SW\^{}2    		& m1.m4*m2.m3-2*m1.m2*m3.m4+m1.m3*m2.m4	\\
	\end{tabular}
	\caption{Example lines from the model file lgrngNN.mdl, where Feynman rules for all vertices in the model are listed.}
	\label{tab:lgrng}
\end{table}

\bibliographystyle{apsrev4-1}
\bibliography{bib}
\end{document}